\documentclass[11pt]{article}
\usepackage{axodraw}
\usepackage{epsfig}
\usepackage{amsfonts}
\usepackage{bbm}
\input pix.sty

\newcommand{\us}[1]{\frac{1}{d-x}}

\newcommand{\bmu}{\bar\mu}
\def\Ada(#1,#2)(#3,#4,#5){\DashCArc(#1,#2)(#3,#4,#5){3}}
\def\Lda(#1,#2)(#3,#4){\DashLine(#1,#2)(#3,#4){3}}

\def\ToprVBblob(#1,#2,#3,#4){\picb{#1(30,15)(15,-120,120)%
 #2(30,15)(15,120,240) #3(15,15)(15,60,300) #4(15,15)(15,-60,60)%
 \GCirc(45,15){2}{0}}}

\newcommand{\Nf}{N_{\rm f}}
\newcommand{\Nc}{N_{\rm c}}
\newcommand{\rmO}{{\mathcal{O}}}
\newcommand{\g}{D}
\newcommand{\CA}{C_A}

\def\lsi{\raise0.3ex\hbox{$<$\kern-0.75em\raise-1.1ex\hbox{$\sim$}}}
\def\gsi{\raise0.3ex\hbox{$>$\kern-0.75em\raise-1.1ex\hbox{$\sim$}}}

\newcommand{\bfx}{{\bf x}}

\def\counterA{\pic{\CArc(15,15)(15,0,360) \Text(21,10.5)[]{$\times$}}}
\def\counterB{\pic{\CArc(15,15)(15,0,360) \Text(0,10.5)[]{$\times$}%
\Text(21,10.5)[]{$\times$}}}
\def\counterC{\picc{\CArc(15,15)(15,0,360) \CArc(45,15)(15,0,360)%
\Text(42,10.5)[]{$\times$}}}
\def\counterD{\pic{\Photon(0,15)(30,15){1.5}{6} \CArc(15,15)(15,0,360)%
\Text(10.5,21)[]{\bf$\times$}}}
\def\counterE{\picc{\CArc(15,15)(15,0,360)%
\PhotonArc(45,15)(15,0,360){1.5}{16} \Text(0,10.5)[]{$\times$}}}
\newcommand{\hide}[1]{ }
\newcommand{\amE}{\hat m} 
\newcommand{\f}{\mbox{\sl f\,}}
\newcommand{\rmii}[1]{{\mbox{\tiny\rm{#1}}}}
\newcommand{\unit}{{\mathbbm{1}}} 

\newcommand{\ZZ}{{\mathbb{Z}}}

\newcommand{\Tc}{T_{\rm c}}

\makeatletter \@addtoreset{equation}{section} \makeatother
\renewcommand{\theequation}{\arabic{section}.\arabic{equation}}
\makeatletter
\renewcommand\section{\@startsection {section}{1}{\z@}%
                                   {-5.5ex \@plus -1ex \@minus -.2ex}
                                   {2.3ex \@plus.2ex}%
                                   {\normalfont\large\bfseries}}
\renewcommand\subsection{\@startsection{subsection}{2}{\z@}%
                                     {-3.25ex\@plus -1ex \@minus -.2ex}%
                                     {1.5ex \@plus .2ex}%
                                     {\normalfont\normalsize\bfseries}}
\renewcommand\thesection {\@arabic\c@section}
\renewcommand\thesubsection   {\thesection.\@arabic\c@subsection}
\renewcommand{\@seccntformat}[1]{%
\csname the#1\endcsname.\hspace{1.0em}}
\makeatother

\begin{document}

\begin{titlepage}
\begin{flushright}
BI-TP 2008/16
\\ arXiv:0808.0557
\end{flushright}

\begin{centering}
\vfill 

{\Large{\bf  Four-loop lattice-regularized vacuum energy density of \\[1mm] 
the three-dimensional SU(3) + adjoint Higgs theory}}

\vspace{0.8cm}

F.~Di Renzo$^{\rm a}$, 
M.~Laine$^{\rm b,c}$, 
Y.~Schr\"oder$^{\rm b}$, 
C.~Torrero$^{\rm d}$ 

\vspace{0.8cm}

{\em $^{\rm a}$%
Universit\`a di Parma \& INFN, 
I-43100 Parma, Italy \\}

\vspace{0.3cm}

{\em $^{\rm b}$%
Faculty of Physics, University of Bielefeld, 
D-33501 Bielefeld, Germany\\}

\vspace{0.3cm}

{\em $^{\rm c}$%
Department of Physics, University of Oulu, 
FI-90014 Oulu, Finland\\}

\vspace{0.3cm}


{\em $^{\rm d}$%
Institute for Theoretical Physics, University of Regensburg, 
D-93040 Regensburg, Germany\\}

\vspace*{0.8cm}

\mbox{\bf Abstract}

\end{centering}

\vspace*{0.3cm}
 
\noindent
The pressure of QCD admits at high temperatures a factorization
into purely perturbative contributions from ``hard'' thermal momenta, 
and slowly convergent as well as non-perturbative contributions
from ``soft'' thermal momenta. The latter can be related to various 
effective gluon condensates in a dimensionally reduced effective 
field theory, and measured there through lattice simulations. 
Practical measurements of one of the relevant
condensates have suffered, however, from difficulties in extrapolating 
convincingly to the continuum limit. In order to gain insight on 
this problem, we employ Numerical Stochastic Perturbation Theory
to estimate the problematic condensate up to 4-loop order in lattice 
perturbation theory. Our results seem to confirm the presence of ``large''
discretization effects, going like $a\ln(1/a)$, where $a$
is the lattice spacing. For definite conclusions, however, 
it would be helpful to repeat the corresponding part of 
our study with standard lattice perturbation theory techniques. 
%
\noindent
 

\vfill 

\noindent
September 2008

\vfill

\end{titlepage}

%
\section{Introduction}
\la{se:introduction}

Given possible applications in cosmology and in the phenomenology
of heavy ion collision experiments,
as well as the important theoretical role 
that the free energy density plays in 
understanding the properties of any 
finite-temperature system, the pressure (or minus the free energy density)
of Quantum Chromodynamics (QCD) is one of the central observables of
relativistic thermal field theory. 
In this paper we focus on its determination
at temperatures above a few hundred MeV, where 
the system is deconfined; 
there, if anywhere, it should eventually be 
possible to establish a quantitative first-principles description 
in terms of the temperature and the fundamental parameters of the theory.  

Given the importance of the problem, a huge amount of work
has already been carried out on the topic. 
(The references most directly related to the present work are cited
at the beginning of the next section.) 
In general, the approaches can be divided into numerical
(i.e.\ lattice Monte Carlo) and analytic (i.e.\ weak-coupling expansion
or various improvements thereof)
techniques. In addition, there is a strategy --- 
the one that we follow here --- which 
combines elements from both sides. The idea is 
to factorise the system into two parts: ``hard'' momenta, whose 
contribution is perturbative, and ``soft'' momenta, which need to 
be treated non-perturbatively. 
(One benefit of this approach is that dynamical quarks 
remain cheap even in the chiral continuum limit, since 
only gauge fields possess soft momenta.)
Our study concerns the non-perturbative soft part, 
but not directly its lattice measurement; rather, the point is that 
for such a factorization to work, both sides of the result need to be 
converted to the same regularization scheme, 
so that they can be added together. 
In some situations, like in ours, such a scheme conversion can turn out to be 
technically as demanding as the non-perturbative measurement itself, 
and this is the ultimate challenge that we try to tackle in this work. 

This paper is organised as follows. 
The general setup of factorising the pressure to  contributions 
from perturbatively computable terms, and ones that need to be 
estimated numerically, as well as the role that the present study
plays in this setup, is outlined in \se\ref{se:outline}. 
The outline is made quantitative in \se\ref{se:setup}, where
we define the precise quantities that we want 
to determine. The tool used for the computation, Numerical 
Stochastic Perturbation Theory, is reviewed in \se\ref{se:NSPT}. 
The numerical data is analysed in \se\ref{se:analysis}; our results
and conclusions comprise \se\ref{se:concl}.
In three appendices, we detail the perturbative expressions 
that we have worked out explicitly in lattice regularization. 

%
\section{Outline of setup}
\la{se:outline}

At a high temperature $T$ and a small gauge coupling constant $g$, 
there are parametrically three different momentum scales 
in hot QCD: $k \sim \pi T, gT, g^2T/\pi$~\cite{dr}. 
All the effects of the hard scale, $k\sim \pi T$, 
can be accounted for by a method called dimensional 
reduction~\cite{dr,generic}. In particular, the pressure, 
or minus the free energy density, can be written as~\cite{bn}
\ba
 p_\rmi{QCD}(T) & \equiv & p_\rmi{E}(T) 
 + \lim_{V\to\infty}\frac{T}{V}\ln 
 \int {\cal D}A_i^a \, {\cal D}A_0^a  \,
 \exp\Bigl( - S_\rmi{E} \Bigr)
 \;, \la{QCD_E} 
\ea
where $V$ is the volume, 
$A^a_i$ are gauge fields, $A_0^a$ are scalar
fields in the adjoint representation, 
and $S_\rmi{E}$ is a three-dimensional
effective field theory, to be specified presently
(the subscript may refer to ``Electrostatic QCD''~\cite{bn}).
The function $p_\rmi{E}(T)$ gets contributions only from the hard scale,
$k\sim \pi T$, and is computable in perturbation theory; 
the path integral with $S_\rmi{E}$ contains the contributions
of the soft modes, $k\sim gT, g^2T/\pi$, and should preferably be 
determined non-perturbatively (the contributions from the 
modes $k\sim g^2T/\pi$ are genuinely non-perturbative~\cite{linde,gpy};
those from the modes $k\sim gT$ are in principle still perturbative, 
but in general 
slowly convergent~\cite{bn}, \cite{az}--\cite{chm},
although some observables with possibly faster 
convergence have also been found~\cite{conv}). 
The same description applies also in the presence of a small
quark chemical potential~\cite{mu}, allowing to compute 
further quantities such as quark number susceptibilities~\cite{ahkr}.
At least when treated non-perturbatively, 
Electrostatic QCD appears to yield 
a quantitative description of the full four-dimensional theory
up from about $T\sim 1.5 \Tc$, where $\Tc$ denotes the 
pseudocritical temperature of the QCD crossover~
(see, e.g.,\ ref.~\cite{rbc} and references therein). 

Now, to give a precise meaning to \eq\nr{QCD_E}, requires
the specification of a regularization scheme. Though a purely 
perturbative challenge, the determination of $p_\rmi{E}(T)$ is fairly 
complicated in practice~\cite{az}. Therefore, it is preferable to use 
dimensional regularization for the algebra: all the 3-loop 
and 4-loop results available today have been obtained 
in the $\msbar$ scheme~\cite{bn}, \cite{az}--\cite{gsixg},
\cite{nspt_mass,phi4}. 

Let us, correspondingly, denote the $\msbar$ scheme  
``vacuum energy density'' of the theory defined by $S_\rmi{E}$, with
\be
 \f_\tinymsbar \equiv - \biggl\{ \lim_{V \to \infty} \frac{1}{V} 
 \ln \int \! {\cal D} A_i^a {\cal D} A_0^a  
 \, \exp\Bigl( -S_\rmi{E} 
 \Bigr) \biggr\}_\tinymsbar \;,
 \la{fMSdef}
\ee
where $V = \int\! {\rm d}^d\vec{x}$ is the $d$-dimensional volume.  
Then \eq\nr{QCD_E} becomes 
\be
 p_\rmi{QCD}(T) = \Bigl\{ p_\rmi{E}(T) \Bigr\}_\tinymsbar
                - T \f_\tinymsbar
 \;. 
\ee
Though each part is scheme-dependent, the expression as a whole is not. 
In the following, we concentrate exclusively on the determination
of $\f_\tinymsbar$.

To take further steps, we need to specify 
the effective action $S_\rmi{E}$.
It reads
\ba 
 S_\rmi{E} & = &  
 \int \! {\rm d}^d x\, {\cal L}_\rmi{E}
 \;, \la{seqcd} \\
 {\cal L}_\rmi{E} & = & \fr12 \tr [F_{ij}^2 ]+ \tr [D_i,A_0]^2 + 
 m_3^2\tr [A_0^2] 
 +\lambda_{3} (\tr [A_0^2])^2
 + ... 
 \; . 
 \hspace*{0.5cm} \la{eqcd}
\ea
Here $i=1,...,d$, $d=3-2\epsilon$, 
$F_{ij} = (i/g_{3}) [D_i,D_j]$, 
$D_i = \partial_i - i g_{3} A_i$, 
$A_i = A^a_i T^a$, 
$A_0 = A^a_0 T^a$, 
and $T^a$ are hermitean generators of SU($\Nc$), 
normalised as $\tr[T^aT^b] = \delta^{ab}/2$.\footnote{%
 In the present paper we concentrate on $\Nc = 3$, but 
 lattice measurements have previously been carried out
 also for $\Nc = 2$~\cite{adjoint}. For $\Nc \ge 4$, 
 another independent quartic coupling should be included
 in \eq\nr{eqcd}. 
 }
The dimensionalities of $g_3^2$ and $\lambda_3$ are GeV$^{1+2\epsilon}$.
There are also higher order operators, classified in ref.~\cite{sc}, 
whose parametric importance has been analysed in ref.~\cite{gsixg}; 
even though some of them contribute at the same parametric order
as some of the effects that we are after, it is still a well-defined problem
to start by determining the full non-perturbative effect of the truncated 
form of the theory in \eq\nr{eqcd}. Therefore we will ignore all higher
order operators in the following.

Now, $\f_\tinymsbar$ may include a part which is independent
of $m_3^2$. However, this part can be evaluated by sending $m_3^2\to\infty$, 
whereby the field $A_0$ can be integrated out. Thereby the problem
reduces to that already considered in refs.~\cite{plaq,nspt_mass}.
In the following, we concentrate on the part of $\f_\tinymsbar$
which does depend on $m_3^2$. 

The part of $\f_\tinymsbar$ depending on $m_3^2$ can be isolated
through a partial derivative, which in turn yields a condensate:\footnote{%
 This condensate is analogous to the Polyakov loop 
 condensate, playing a role
 in various attempts at improved effective theories of hot QCD
 (see, e.g., refs.~\cite{Pol,Z3} and references therein), 
 but it appears difficult to promote the 
 analogy to a precise relation. 
 } 
\be 
 \partial_{m_3^2} \f_\tinymsbar
  =
 \Bigl\langle \tr [A_0^2] \Bigr\rangle_\tinymsbar
 \;.
\ee
Thus, if we are able to measure the condensate 
$\langle \tr [A_0^2] \rangle$ in lattice regularization, 
and convert the result to the $\msbar$ scheme, we are able to 
determine the $m_3^2$-dependent part of $\f_\tinymsbar$ non-perturbatively. 
In principle this is, indeed, doable: the relation of the two regularization
schemes, which is exact in the continuum limit due to the 
super-renormalizability of the theory in \eq\nr{eqcd}, can be found 
in refs.~\cite{framework,contlatt}
(see also \eq\nr{a0msbar} below). 

Let us be a bit more specific about what we would like to achieve 
with the lattice simulations. Note first that in the $\msbar$ scheme, 
the $m_3^2$-dependent part of $\f_\tinymsbar$ is known analytically 
up to 4-loop 
order~\cite{aminusb}. This corresponds to an expansion of the form
\be
 \Bigl\langle \tr [A_0^2] \Bigr\rangle_\tinymsbar
 = m_3 + g_3^2 \ln\frac{\bmu}{m_3} + \frac{g_3^4}{m_3} + \frac{g_3^6}{m_3^2}
 + \rmO\biggl(\frac{g_3^8}{m_3^3}\biggr)
 \;, \la{a02ms}
\ee
where $\bmu$ is the $\msbar$ scale parameter, and 
we have for simplicity omitted all numerical coefficients, 
as well as terms containing $\lambda_3$. What we would
like to determine is the ``remainder'', i.e. the sum of terms beyond
the level that is already known analytically. Denoting by 
$
 \langle \tr [A_0^2] \rangle_a
$
the condensate in lattice regularization,
and by $a$ the lattice spacing, the remainder is 
\be
 \Bigl\langle \tr [A_0^2] \Bigr\rangle_\tinymsbar^{\mathcal{R}}
 \equiv 
 \lim_{a\to 0} 
 \Bigl\{ \Bigl\langle \tr [A_0^2] \Bigr\rangle_a 
        - \frac{1}{a} - g_3^2 \ln \frac{1}{a\bmu}
 \Bigr\} 
 - m_3 - g_3^2 \ln\frac{\bmu}{m_3} - \frac{g_3^4}{m_3} - \frac{g_3^6}{m_3^2}
 \;, \la{fit1}
\ee
where the limit inside the curly brackets
takes us to the $\msbar$ scheme~\cite{framework,contlatt}, and the 
subsequent continuum expression subtracts the known terms
in \eq\nr{a02ms}. 

The problem with this procedure is that in practice the limit 
$a\to 0$ in \eq\nr{fit1} cannot be taken exactly, 
but it introduces systematic errors. To carry out the limit requires
a fit ansatz; in three dimensions, discretization effects go like $\rmO(a)$. 
These dominant errors could in principle be removed through an improvement 
program~\cite{moore_a}, but even if it had been completed (which is presently
not the case), one would still need an ansatz for the subsequent terms, 
and at this point it is for instance not clear whether logarithms
like $a^2\ln(1/a)$ should be included.  
Yet, this may have a noticeable impact on the results. Moreover, the ansatz 
is necessarily of a finite order; this means that there will remain  
residual 1-loop discretization errors in the result, while the subsequent
subtraction of the continuum terms is attempting to take us to the 5-loop
level and beyond. Evidently, this situation is unsatisfactory; for 
a demonstration of the problems encountered, see ref.~\cite{ahkr}.

The goal of the present paper is, then, to determine the terms 
corresponding to those in \eq\nr{a02ms} ``exactly'' in lattice 
regularization. That is, we do not carry out any expansion in 
$a m_3$; only one in the loop order. The corresponding result
contains the counterterms needed for the limit in \eq\nr{fit1}; 
the finite terms in \eq\nr{fit1}; but also an infinite number 
of higher order corrections, starting at $\rmO(a m_3)$. 
Then, we can write the remainder as 
\be
 \Bigl\langle \tr [A_0^2] \Bigr\rangle_\tinymsbar^{\mathcal{R}}
 = 
 \lim_{a\to 0} 
 \Bigl\{ \Bigl\langle \tr [A_0^2] \Bigr\rangle_a 
 - m_3\, f_0(a m_3) - g_3^2\, f_1(am_3)  
 - \frac{g_3^4}{m_3} f_2(am_3) - \frac{g_3^6}{m_3^2} f_3(am_3)
 \Bigr\} 
 \;, \la{fit2}
\ee
where $f_i$ are the functions to be determined. We of course still need to 
take the limit $a\to 0$ at the end, as indicated by \eq\nr{fit2}, 
but we have gained in that we do not
need to be as worried about discretization errors as before: 
any number yielded by the difference inside 
the curly brackets in \eq\nr{fit2} is already
an approximation for the remainder, and 
there is no danger of 1-loop discretization errors overtaking
an interesting 5-loop continuum effect.

%
\section{Precise setup}
\la{se:setup}

Let us now add the missing details to the formulation
of the problem. The 4-loop computation of $\f_\tinymsbar$
in dimensional regularization
has been described in detail in ref.~\cite{aminusb}. The mass
parameter needs renormalisation, 
\ba
 m_\rmi{3,bare}^2 & = & m_3^2(\bmu) + \delta m_\tinymsbar^2 
 \;, \\ 
 \delta m_\tinymsbar^2     & = & 
 \frac{1}{(4\pi)^2} \frac{\mu^{-4\epsilon}}{4\epsilon}
 2 (d_A + 2) \Bigl( -g_3^2 \lambda_3 C_A + \lambda_3^2 \Bigr)
 \;,
\ea
where
$
 \bmu^2 \equiv 4\pi \mu^2 e^{-\gamma_\rmii{E}}
$,
$C_A \equiv \Nc$,
and $d_A \equiv \Nc^2-1$.
Consequently the renormalised mass parameter satisfies the equation
\be
 \bmu \frac{{\rm d}}{{\rm d}\bmu} m_3^2(\bmu) = 
 \frac{1}{8\pi^2} (d_A + 2) \Bigl( -g_3^2 \lambda_3 C_A + \lambda_3^2 \Bigr)
 \;. \la{mrunning}
\ee
Here and in the following, we assume $\epsilon\to 0$ 
taken in all finite quantities. 
It is convenient to define the dimensionless ratios 
\ba
 x  & \equiv & \frac{\lambda_3}{g_3^2}
 \;, \la{xdef} \\ 
 y & \equiv & \frac{m_3^2(\bmu=g_3^2)}{g_3^4}
 \;. \la{ydef}
\ea
As indicated, we choose $\bmu \equiv g_3^2$ for defining $y$. 

Now, it is thought that the 
perturbative series for the pressure of QCD converges very
slowly~\cite{az,zk}. The reason for this can be traced
back to the slow convergence of the perturbative 
series for $\partial_{m_3^2} \f_\tinymsbar$~\cite{bn,a0cond,gsixg}.
Making use of the results of ref.~\cite{aminusb}, and inserting
the choice $\bmu = g_3^2$, 
the known 4-loop result can be written as
\ba
 \Bigl\langle \tr[A_0^2/g_3^2] \Bigr\rangle_{\tinymsbar,\bmu=g_3^2}
 & = &  
 -\frac{d_A}{(4\pi)} \frac{y^{\fr12}}{2} +
 \nn \!\! & + & \!\!
 \frac{d_A}{(4\pi)^2} \fr14
 \biggl\{
 {C_A} \biggl[ 
 1 - 2\ln(4y) 
 \biggr] + 
 x (d_A+2) 
 \biggr\} +
 \nn \!\! & + & \!\!
 \frac{d_A}{(4\pi)^3} \frac{1}{4 y^{\fr12}}
 \biggl\{ 
 C_A^2 
 \biggl[ 
 \frac{89}{12}  + \frac{\pi^2}{3} -\frac{11}{3}\ln 2
 \biggr]
 + x \, C_A (d_A+2)  
 \biggl[ 
 \ln(4y) + \fr32
 \biggr] +
 \nn &&{} ~~~~
 + x^2 \, (d_A+2) 
 \biggl[ 
 1 -\ln(16 y) 
 \biggr]
 - \frac{x^2}{4} (d_A + 2)^2 
 \biggr\} +
 \nn \!\! & + & \!\!
 \frac{d_A}{(4\pi)^4} \frac{1}{4 y}
 \biggl\{ 
 2 C_A^3
 \biggl[
 \frac{43}{4} - \frac{491}{768} \pi^2  
 \biggr]
 + 10 \, x \, C_A^2 
 \biggl[ 
 1 - \frac{\pi^2}{8}
 \biggr]  +
 \nn &&{} ~~~~
 + x \, C_A^2 (d_A + 2) 
 \biggl[ 
 2 \ln(4 y) - 1
 \biggr] +
 \nn &&{} ~~~~
 + 2 x^2 C_A (d_A+2) 
 \biggl[
 \frac{36-\pi^2}{8}  -\ln(4 y) 
 \biggr]
 -x^2 C_A (d_A+2)^2  -
 \nn &&{} ~~~~
 - x^3 (d_A + 2)(d_A+8)  \frac{\pi^2}{12} 
 + x^3 (d_A+2)^2 
 \biggr\}
 \;. \la{dyFpert}
\ea 
The terms beyond \eq\nr{dyFpert} 
die away as $y^{-\fr32}$ at large $y$, modulo possible 
logarithms, but realistic values of $y$ are not that large
(for $\Nc=3,\Nf=3$, 
$
 y\simeq 0.39\, [\log_{10}(T/\Lambdamsbar) + 1.0]
$).
The goal of a non-perturbative determination is therefore
to sum the whole series beyond these known terms.

We then move to the lattice side. 
To this effect, let us define the lattice action, 
$S_a$, corresponding 
to \eq\nr{seqcd}. The standard Wilson discretization yields
\ba
 S_a & = &  
 \beta \sum_{\bf x} \sum_{i < j}
 \biggl\{ 1 - \frac{1}{C_A} \mathop{\mbox{Re}} \tr [P_{ij}({\bf x}) ] 
 \biggr\} 
 + \nn & + & 
 {2a}\sum_{\bf x} \sum_i
 \left\{ \tr [A_0^2({\bf x})]-
 \tr [A_0({\bf x})U_i({\bf x}) A_0({\bf x}+i)U_i^\dagger({\bf x})]\right\} + 
 \nn 
 & + & a^3 \sum_{\bf x} \Bigl\{ 
 m_\rmi{3,bare}^2\tr [A_0^2(\bfx)]+
 \lambda_3\Bigl(\tr [A_0^2(\bfx)]\Bigr)^2 \Bigr\} \;, \la{SAH}
\ea
where $a$ is the lattice spacing,  
$U_i({\bf x})$ is a link matrix, 
${\bf x}+i\equiv  {\bf x}+a\hat\e_i$, where $\hat\e_i$ 
is a unit vector, $P_{ij}({\bf x})$ is the plaquette, and
\be
 \beta \equiv \frac{2 C_A}{g_3^2 a}
 \;. \la{beta}
\ee
The bare mass parameter of \eq\nr{SAH} reads~\cite{contlatt}
\ba
 m_\rmi{3,bare}^2 \!\! &  = & \!\! m_3^2(\bmu) + \delta m^2_a 
 \;, \\
\delta m^2_a
 \!\! &=& \!\!
 - \Bigl[2g_3^2C_A + \lambda_3(d_A+2)\Bigr] \fr\Sigma{4\pi a}+ \nn
 & & \!\!
 +\fr1{(4\pi)^2} \biggl\{  2\lambda_3(d_A+2)(\lambda_3-g_3^2C_A)
\Bigl( \ln\fr6{a\bmu} +\zeta\Bigr) -2g_3^2C_A\lambda_3(d_A+2)
\Bigl( \fr{\Sigma^2}4-\delta\Big) - \nn 
 & & \!\!
 -g_3^4 C_A^2 \biggl[
 \frac{5}{8}\Sigma^2+\biggl(\frac{1}{2}-\frac{4}{3C_A^2}\biggr)\pi\Sigma
 -4(\delta+\rho)+2\kappa_1-\kappa_4
 \biggr] \biggr\} \;, \la{dmassL}
\ea
where $\Sigma\approx 3.175911535625$  
is a three-dimensional hybercubic lattice integral which can be expressed as
$
 \Sigma = (\sqrt{3}-1)\Gamma^2[\fr1{24}]\Gamma^2[\fr{11}{24}]/ 48 \pi^2
$~\cite{math}\footnote{%
 We thank D.~Broadhurst for bringing these references 
 to our attention. 
 };  
$\zeta$, $\delta$, $\rho$, $\kappa_1$, $\kappa_4$ are further
lattice integrals which are only known numerically~\cite{contlatt};
and $\bmu \equiv g_3^2$. This bare mass parameter guarantees
the existence of a continuum limit for any fixed 
$m_3^2(\bmu)$ (but  
$\rmO(a)$ discretization effects remain~\cite{moore_a}).

The derivative of the vacuum energy density, $\f_a$,  
with respect to the mass 
parameter (bare or renormalized)
yields then the quadratic condensate in lattice regularization, 
\be
 \partial_{m_3^2} \f_a = \Bigl\langle \tr[A_0^2] \Bigr\rangle_a
 \;. \la{latt_cond}
\ee
This can be related to the $\msbar$ condensate by~\cite{framework,contlatt} 
\ba
 \Bigl\langle \tr [A_0^2/g_3^2] \Bigr\rangle_\tinymsbar
 & = & 
 \nn & & \hspace*{-3cm}
 \lim_{\beta\to\infty}
 \biggl\{
 \Bigl\langle
 \tr[ 
 {A_0^2}/{g_3^2}
  ] 
 \Bigr\rangle_a
 - \biggl[ \frac{d_A \Sigma\beta}{16\pi\CA} 
 + \frac{d_A \CA}{(4\pi)^2} 
 \biggl(\ln\beta + 
 \zeta + \frac{\Sigma^2}{4} - \delta - \ln 
 \frac{\CA\bmu}{3 g_3^2} 
 \biggr) 
 \biggr] \biggr\}
 \;. \hspace*{0.5cm} \la{a0msbar}
\ea
Writing the renormalised mass parameter in lattice units as 
\be
 \amE \equiv a m_3(\bmu=g_3^2) = \frac{2C_A y^{\fr12}}{\beta}
 \quad \Leftrightarrow \quad
 y^{\fr12} = \frac{\beta \amE}{2C_A}
 \;, \la{amE} \la{lnb}
\ee
where $y$ was defined in \eq\nr{ydef},
we can express the lattice condensate as
\ba
  \frac{1}{d_A} \left\langle 
   \tr \bigl[ 
          {A_0^2}/{g_3^2}
       \bigr]
  \right\rangle_a
  & = & 
  \frac{1}{d_A g_3^2} \partial_{m_3^2} \f_a 
  \\
  & \equiv & 
  \biggl( \frac{\beta\amE}{2C_A} \biggr)^{+1} 
  \phi_{00}(\amE) 
  + \nn & + & 
  \biggl( \frac{\beta\amE}{2C_A} \biggr)^{0}
  \biggl\{
  \phi_{10}(\amE) + 
  x\, \phi_{11}(\amE) 
  \biggr\}
  + \nn & + & 
  \biggl( \frac{\beta\amE}{2C_A} \biggr)^{-1}
  \biggl\{
    \phi_{20}(\amE) + 
    \sum_{n=1}^{2} x^n 
    \biggl[
       \phi_{2n}(\amE) + \tilde \phi_{2n}(\amE) 
       \ln\biggl( \frac{\beta\amE}{2C_A} \biggr)
    \biggr] 
  \biggr\} 
 + \nn & + & 
  \biggl( \frac{\beta\amE}{2C_A} \biggr)^{-2}
  \biggl\{
    \phi_{30}(\amE) + 
    \sum_{n=1}^{3} x^n 
    \biggl[
       \phi_{3n}(\amE) + \tilde \phi_{3n}(\amE) 
       \ln\biggl( \frac{\beta\amE}{2C_A} \biggr)
    \biggr] 
  \biggr\} 
 + \nn & + & 
  \rmO\biggl( \frac{\beta\amE}{2C_A} \biggr)^{-3}
  \;. \la{phi}
\ea
Regarding the structure of this equation, we note that
higher powers of logarithms than the terms shown do not need 
to be considered, as will be explained below. 

Now, super-renormalizability guarantees that only a finite set
among the functions $\phi_{mn}$, $\tilde \phi_{mn}$ diverge
in the continuum limit. In fact, as can be 
deduced from \eq\nr{a0msbar}, 
$\phi_{00}$ diverges as $1/\amE$
and $\phi_{10}$ diverges as $\ln(1/\amE)$, but all the others
are finite in the limit $\amE\to 0$~\cite{framework}. In this limit,
they then agree with the corresponding $\msbar$ scheme expressions, 
readily extracted from \eq\nr{dyFpert}, after taking note of \eq\nr{lnb}. 

For $\amE\neq 0$, the functions $\phi_{mn}, \tilde \phi_{mn}$
can be computed in lattice perturbation theory. The first one, 
$\phi_{00}$, follows from a 1-loop computation, 
while $\phi_{10}, \phi_{11}$ require 2-loop computations
(details are given in appendix~A). 
The functions 
$\tilde \phi_{21}$, 
$\tilde \phi_{22}$, 
$\tilde \phi_{31}$,
$\tilde \phi_{32}$,
$\tilde \phi_{33}$
can be deduced from the fact that $\langle \tr[A_0^2] \rangle_a$ 
is independent of $\bmu$; on the other hand
$\phi_{00}$, 
$\phi_{10}$, 
$\phi_{11}$
have $\bmu$-dependence, 
emerging through the running of the $\msbar$ 
mass parameter according to \eq\nr{mrunning}.  
This dependence must cancel against explicit 3-loop and 4-loop
logarithms, containing $\ln(6/a\bmu)$; these logarithms arise
exclusively from the mass counterterm in \eq\nr{dmassL}.
Therefore, the 3-loop and 4-loop coefficients 
$\tilde \phi_{21}$, 
$\tilde \phi_{22}$, 
$\tilde \phi_{31}$,
$\tilde \phi_{32}$,
$\tilde \phi_{33}$
can be deduced from mass derivatives
of the 1-loop and 2-loop expressions. 

As far as the ``genuine'' 3-loop coefficients are concerned,  
we have computed explicitly only $\phi_{21}, \phi_{22}$
(details are given in appendix~A). 
These arise from graphs containing at least one quartic 
coupling, which means that most of them 
(with one exception) factorise into products 
of lower-order graphs. 

To display the results, we use the notation
of basic lattice integrals
($\hat J_a$, $\hat I_a$, $\hat H_a$, $\hat G_a$, $\hat B_a$) 
explained in appendix~B. Denoting furthermore 
\ba
 \mathcal{K}_1(\amE^2) & \equiv & 
 2 \Bigl[ \hat I_a(\amE^2) + \hat I_a(0) - \frac{\Sigma}{2\pi} \Bigr] 
 \hat I_a'(\amE^2) 
 + \hat I_a(0) 
 \Bigl[ 1 + \amE^2 \partial_{\amE^2}\Bigr] \hat I_a(\amE^2)
 + \nn & + & 
 4 \Bigl[ 1 + \amE^2 \partial_{\amE^2}\Bigr] \hat H_a(\amE^2)
 + \hat G_a'(\amE^2)
 \;, \la{K1} \\ 
 \mathcal{K}_2(\amE^2) & \equiv & 
 \partial_{\amE^2} \mathcal{K}_1(\amE^2) 
 \nn  & = & 
 2 \Bigl[ \hat I_a(\amE^2) + \hat I_a(0) - \frac{\Sigma}{2\pi} \Bigr] 
 \hat I_a''(\amE^2) 
 + \hat I_a(0) 
 \Bigl[ 2 + \amE^2 \partial_{\amE^2}\Bigr] \hat I_a'(\amE^2)
 + \nn & + & 
 2 \Bigl[ \hat I_a'(\amE^2) \Bigr]^2
 + 4 \Bigl[ 2 + \amE^2 \partial_{\amE^2}\Bigr] 
 \hat H_a'(\amE^2)
 + \hat G_a''(\amE^2)
 \;, \hspace*{0.5cm} \la{K2} \\
 \mathcal{K}_3(\amE^2) & \equiv & 
 \hat I_a'(\amE^2)
 \biggl[
   \Bigl( 1 + \amE^2 \partial_{\amE^2}\Bigr) \hat H_a(\amE^2)
 + \fr14 \hat G_a'(\amE^2)
 - 
 \frac{1}{(4\pi)^2}
 \biggl(
   \ln\frac{6}{\amE} + \zeta + \frac{\Sigma^2}{4} - \delta 
 \biggr)   
 \biggr]
 + \nn & + & 
 \Bigl[ \hat I_a(\amE^2) - \frac{\Sigma}{4\pi} \Bigr] 
 \biggl[
 \Bigl( 2 + \amE^2 \partial_{\amE^2}\Bigr) 
 \hat H_a'(\amE^2)  
 + \fr14 \hat G_a''(\amE^2)
 \biggr]
 + \nn & + &
 \fr12 
 \Bigl[ \hat I_a(\amE^2) - \frac{\Sigma}{4\pi} \Bigr]
 \Bigl[ \hat I_a(\amE^2) 
 + \Bigl( 1 + \frac{\amE^2}{2} \Bigr)\hat I_a(0) - \frac{\Sigma}{2\pi} \Bigr]
 \hat I_a''(\amE^2)
 + \nn & + &   
 \Bigl[ \hat I_a(\amE^2) 
 + \fr12 \Bigl( 1 + \frac{\amE^2}{2} \Bigr) \hat I_a(0) 
 - \frac{3\Sigma}{8\pi} \Bigr]  
 \Bigl[ \hat I_a'(\amE^2) \Bigr]^2 
 + \nn & + &  
 \fr34 \Bigl[ \hat I_a(\amE^2) - \frac{\Sigma}{6\pi} \Bigr] 
 \hat I_a(0) 
 \hat I_a'(\amE^2) 
 \;, \la{K3}
\ea
we obtain from appendix~A, as well as from the continuum
values in \eq\nr{dyFpert}: 
\ba
 \phi_{00} & = & 
 \frac{1}{2\amE} \hat I_a(\amE^2) 
 \la{def:f00} \\ & \approx &
 \frac{\Sigma}{8\pi \amE} - \frac{1}{8\pi} + \rmO(\amE)
 \;, \\ 
 \phi_{10} & = & \fr14 C_A\, \mathcal{K}_1(\amE^2)
 \\ & \approx & 
 \frac{C_A}{(4\pi)^2}
 \biggl[
   \ln\frac{3}{\amE} + \zeta + \frac{\Sigma^2}{4} - \delta + \fr14 + 
    \rmO(\amE) 
 \biggr]
 \;, \\ 
 \phi_{11} & = & 
 \fr12 (d_A + 2)
 \Bigl[ \hat I_a(\amE^2) - \frac{\Sigma}{4\pi} \Bigr]
 \hat I_a'(\amE^2)
 \\ & \approx & 
 \frac{d_A+2}{(4\pi)^2}\Bigl[ \fr14 + \rmO(\amE) \Bigr]
 \;, \\
 \phi_{20} & \approx &
 \frac{C_A^2}{(4\pi)^3} 
 \biggl[ 
   \fr{89}{48}-\fr{11}{12}\! \ln2 +\fr{\pi^2}{12} + \rmO(\amE)
 \biggr] 
 \;, \la{phi20} \\ 
 \phi_{21} & = & 
 (d_A+2)C_A \amE\, \mathcal{K}_3(\amE^2)
 \\ & \approx & 
 \frac{(d_A+2)C_A}{(4\pi)^3}\biggl[
  \fr12 \! \ln2 + \fr38 + \rmO(\amE) 
 \biggr] 
 \;, \\ 
 \tilde \phi_{21} & = & 
 -\frac{(d_A+2)C_A}{(4\pi)^2} \amE \, \hat I_a'(\amE^2)
 \\ & \approx &
 \frac{(d_A+2)C_A}{(4\pi)^3}\biggl[
  \fr12 + \rmO(\amE) 
 \biggr] 
 \;, \\
 \phi_{22} & = & 
 (d_A+2) \amE \biggl[
 \frac{1}{(4\pi)^2} \hat I_a'(\amE^2)
 \biggl(\ln\frac{6}{\amE} + \zeta \biggr)
 - \fr14 \hat B_a'(\amE^2)
 \biggr]
 + \nn & + &
 \fr12 (d_A+2)^2 \amE \Bigl[ \hat I_a(\amE^2) - \frac{\Sigma}{4\pi} \Bigr]
 \times \nn & & \times
 \biggl\{ 
  \Bigl[ \hat I_a'(\amE^2) \Bigr]^2 +
  \fr12 \Bigl[ \hat I_a(\amE^2) - \frac{\Sigma}{4\pi} \Bigr]
  \hat I_a''(\amE^2)
 \biggr\}
 \\ & \approx & 
 \frac{d_A+2}{(4\pi)^3}\biggl[
  \fr14 - \ln2 -\frac{d_A+2}{16} + \rmO(\amE) 
 \biggr] 
 \;, \\ 
 \tilde \phi_{22} & = & 
 \frac{d_A+2}{(4\pi)^2} \amE \hat I_a'(\amE^2)
 \\ & \approx &
 \frac{d_A+2}{(4\pi)^3}\biggl[
  - \fr12 + \rmO(\amE) 
 \biggr] 
 \;, \\
 \phi_{30} & \approx & 
 \frac{C_A^3}{(4\pi)^4}
 \biggl[
   \frac{43}{8} - \frac{491}{1536} \pi^2  
   + \rmO(\amE)
 \biggr]
 \;, \la{phi30} \\  
 \phi_{31} & \approx & 
 \frac{C_A^2}{(4\pi)^4}
 \biggl[ 
   \fr52 \biggl( 1 - \frac{\pi^2}{8} \biggr)
   + (d_A + 2)
   \Bigl(
     \ln2 - \fr14  
   \Bigr) 
   + \rmO(\amE)
 \biggr]
 \;, \la{phi31} \\    
 \tilde \phi_{31} & = &
 -\fr12 \frac{(d_A+2) C_A^2}{(4\pi)^2} \amE^2 \mathcal{K}_2(\amE^2)
 \\ & \approx & 
 \frac{(d_A+2) C_A^2}{(4\pi)^4}
 \Bigl[ 1 + \rmO(\amE) \Bigr]
 \;, \\  
 \phi_{32} & \approx & 
 \frac{(d_A+2)C_A}{(4\pi)^4}
 \biggl[ 
   \fr94  - \frac{\pi^2}{16}  -\ln2 
   -\frac{d_A + 2}{4} 
   + \rmO(\amE)
 \biggr]
 \;, \la{phi32} \\    
 \tilde \phi_{32} & = &
 \frac{(d_A+2) C_A}{(4\pi)^2}  
 \fr12 \amE^2 \mathcal{K}_2(\amE^2)
 - \nn & - & 
 \frac{(d_A+2)^2 C_A}{(4\pi)^2} \amE^2 
 \biggl\{ 
   \Bigl[ \hat I_a'(\amE^2) \Bigr]^2 + 
   \Bigl[ \hat I_a(\amE^2) - \frac{\Sigma}{4\pi} \Bigr] 
   \hat I_a''(\amE^2)
 \biggr\}
 \\ & \approx & 
 \frac{(d_A+2) C_A}{(4\pi)^4}
 \Bigl[ -1 + \rmO(\amE) \Bigr]
 \;, \\  
 \phi_{33} & \approx & 
 \frac{d_A+2}{(4\pi)^4}
 \biggl[ 
   \fr14 (d_A + 2) -  \frac{\pi^2}{48} (d_A+8) 
   + \rmO(\amE)
 \biggr]
 \;, \la{phi33} \\    
 \tilde \phi_{33} & = &
 \frac{(d_A+2)^2}{(4\pi)^2} \amE^2 
 \biggl\{
   \Bigl[ \hat I_a'(\amE^2) \Bigr]^2 + 
   \Bigl[ \hat I_a(\amE^2) - \frac{\Sigma}{4\pi} \Bigr] 
   \hat I_a''(\amE^2)
 \biggr\}
 \la{def:tf33} \\ & \approx & 
 \frac{(d_A+2)^2}{(4\pi)^4}
 \Bigl[ 0 + \rmO(\amE) \Bigr]
 \;.
\ea
Note that at infinite volume, $\hat I_a(0)=\Sigma/4\pi$, 
so that the functions $\mathcal{K}_1$, $\mathcal{K}_2$, 
$\mathcal{K}_3$ defined in \eqs\nr{K1}--\nr{K3} can be simplified; however, 
in a finite volume, $\Sigma/4\pi$ appearing in the mass counterterm
is kept fixed, while $\hat I_a(0)$,
emerging from loops, gets modified (cf.\ appendix~B).

%
\section{Numerical Stochastic Perturbation Theory}
\la{se:NSPT}

In order to estimate numerically the coefficients 
$\phi_{20}$,
$\phi_{30}$,
$\phi_{31}$,
$\phi_{32}$,
$\phi_{33}$,
for which only the continuum values ($\hat m\to 0$ limits)
are known exactly
(cf.\ \eqs\nr{phi20}, \nr{phi30}, \nr{phi31}, \nr{phi32}, \nr{phi33}), 
we find it convenient to rewrite the 
action in \eq\nr{SAH} as
\ba
 S_\rmii{latt}&=&
 \beta\sum_\rmii{$\vec{x},~\!i<j$}
 \Bigl(1-\frac{1}{3}\mathop{\rm Re}\tr\big[P_{ij}(\vec{x})\big]\Bigr) - 
 \nn  &-&
 2\sum_\rmii{$\vec{x},~\!i$}
 \tr\big[\Phi(\vec{x}) U_i(\vec{x}) 
         \Phi(\vec{x}+i) U_i^{\dagger}(\vec{x})\big] 
 + \nn &+&
 \sum_\rmii{$\vec{x}$}\bigg\{\alpha(\beta,\lambda, \amE)
 \tr\big[\Phi^2(\vec{x})\big]+
 \lambda\Big(\!\tr\big[\Phi^2(\vec{x})\big]\Big)^2 \bigg\}
 \;,
\ea
where $\Phi\equiv \sqrt{a} A_0$,
$\lambda \equiv a \lambda_3$, 
$\hat m \equiv a m_3(\bmu = g_3^2)$ and, 
for $\Nc = 3$, \eq\nr{dmassL} implies that
\ba
 \alpha(\beta,\lambda,\amE)
 \!\! &=& \!\!
 6\bigg\{1+\frac{\amE^2}{6} -\bigg(6+\frac{5}{3}\lambda\beta\bigg)
 \frac{3.175911525625}{4\pi\beta}-
 \nn 
 \!\! &-& \!\! \frac{3}{8\pi^{2}\beta^{2}}~\!
 \bigg[\bigg(10\lambda\beta-\frac{5}{9}\lambda^{2}\beta^{2}\bigg)
 \bigg(\ln\beta+0.08849\bigg)+
 \frac{34.768}{6}\lambda\beta+36.130\bigg]\bigg\}
 \;. \hspace*{5mm} \la{alpha}
\ea
We write the expansion of the lattice condensate now as 
\ba
 \langle~\!\! \tr~\!\![~\!\!\Phi^2]~\!\rangle&=&
 d_{00}+d_{10}~\!\frac{1}{\beta}+d_{11}\lambda+
 d_{20}~\!\frac{1}{\beta^{2}}+
 d_{21}~\!\frac{\lambda}{\beta}+d_{22}\lambda^2
 +\nn &+&
    d_{30}~\!\frac{1}{\beta{^{3}}}+
    d_{31}~\!\frac{\lambda}{\beta{^{2}}}+
    d_{32}~\!\frac{\lambda^2}{\beta}+
    d_{33}\lambda^3+{O}\bigg(~\!\frac{\lambda^n}{\beta{^{4-n}}}\bigg)
 \;. \la{Phi_exp}
\ea
The coefficients here are related to those in \eq\nr{phi} through
\ba
 d_{00} & = & 
  {d_A \amE}\, \phi_{00} 
 \;, \la{def:d00} \\[2mm] 
 d_{10} & = & 
  {2 d_A C_A}\phi_{10}
 \;, \\[1mm] 
 d_{11} & = &  {d_A} \phi_{11}
 \;, \\[1mm]
 d_{20}  & = & \frac{4 d_A C_A^2}{\amE}  \phi_{20}  
 \;, \\ 
  d_{21} & = & 
 \frac{2 d_A C_A}{\amE} 
 \biggl[
 \phi_{21} + 
 \tilde \phi_{21} 
 \ln \biggl( \frac{\beta \amE}{2C_A} \biggr)
 \biggr]
 \;, \la{def:d21} \\ 
 d_{22}  & = & 
 \frac{d_A}{\amE} \biggl[ 
 \phi_{22}
 + 
 \tilde \phi_{22}
 \ln \biggl( \frac{\beta \amE}{2C_A} \biggr)
 \biggr]
 \;, \la{def:d22} \\ 
 d_{30} & = & 
 \frac{8 d_A C_A^3}{\amE^2} \phi_{30}
 \;, \\  
 d_{31} & = & 
 \frac{4 d_A C_A^2}{\amE^2} \biggl[ 
 \phi_{31}
 + 
 \tilde \phi_{31}
 \ln \biggl( \frac{\beta \amE}{2C_A} \biggr)
 \biggr]
 \;, \la{def:d31} \\    
 d_{32} & = & 
 \frac{2 d_A C_A}{\amE^2} \biggl[ 
  \phi_{32}
  + 
  \tilde\phi_{32}
 \ln \biggl( \frac{\beta \amE}{2C_A} \biggr)
 \biggr]
 \;, \la{def:d32} \\    
 d_{33} & = & 
 \frac{d_A}{\amE^2} \biggl[ 
  \phi_{33}
  + 
  \tilde\phi_{33}
 \ln \biggl( \frac{\beta \amE}{2C_A} \biggr)
 \biggr]
 \;. \la{def:d33}
\ea

The perturbative study is concretely carried out by means of 
\emph{Numerical Stochastic Perturbation Theory (NSPT)}~\cite{FDR,drs}.
(It would certainly also be interesting to pursue 
the same computation with standard techniques~\cite{HP}; 
we comment on this in more detail in \se\ref{se:concl}.)
Its origins lie in {Stochastic Quantization}~\cite{Par}, 
based on introducing an extra coordinate $t$
and {an evolution equation of the Langevin type}, namely
\be
 \partial_t\Phi(\vec{x},t)= -\delta_{\Phi} S[\Phi]+\eta(\vec{x},t)
 \;, \la{Phi_evol}
\ee
where $\eta(\vec{x},t)$ is a Gaussian noise.
The usual Feynman-Gibbs path integral is recovered 
by averaging over the stochastic time,
\be
 Z^{-1}\!\!\int \mathcal{D}\Phi\, 
 O[\Phi(\vec{x})]e^{-S[\Phi(\vec{x})]}=\lim_{t\rightarrow\infty}
 \frac{1}{t} \int_0^t \! {\rm d} t' \,
 \big\langle O[\Phi_{\eta}(\vec{x},t')]\big\rangle_{\eta} 
 \;. 
\ee
In the case of gauge degrees of freedom  
the Langevin equation reads
\be
 \partial_{\tilde t}U_{\tilde\eta} = -i\Bigl( \nabla S[U_{\tilde\eta}]
 +\tilde\eta \Bigr)
 U_{\tilde\eta} \;, \la{U_evol}
\ee
where $\tilde\eta(\vec{x},\tilde t)$ is another Gaussian noise, 
and $\tilde t$ is another fictitious time coordinate.
Both time coordinates are dimensionless, being effectively
measured in spatial lattice units. 

In practice, the time coordinates $t$, $\tilde t$ need to be discretized
as well: $t = n \epsilon$, $\tilde t = n \tilde\epsilon$, 
$n \in \ZZ$, with $\epsilon,\tilde\epsilon\to 0$ 
in the end. The discretized version of the scalar evolution, 
\eq\nr{Phi_evol}, reads
\be
 \Phi(\vec{x},(n+1)\epsilon) 
 = 
 \Phi(\vec{x},n \epsilon)
 - \epsilon \delta_{\Phi(\vec{x})} S + 
 \sqrt{\epsilon} \eta(\vec{x},n\epsilon) 
 \;, \la{Phi_evol_eps}
\ee
while the discretized version of the link evolution, \eq\nr{U_evol}, becomes
\be
 U_k(\vec{x},(n+1)\tilde\epsilon) 
 = \exp\Bigl\{
  - i \Bigl[ 
 \tilde \epsilon \nabla_{k,\vec{x}} S + \sqrt{\tilde\epsilon} 
 \tilde\eta_k(\vec{x}, n \tilde\epsilon)
 \Bigr]
 \Bigr\}\, U_k(\vec{x}, n \tilde\epsilon)
 \;. \la{U_evol_eps}
\ee
Here $\eta \equiv T^a \eta^a$, $\tilde\eta_k \equiv T^a \tilde\eta_k^a$; 
we have rescaled the noise fields by a factor $\sqrt{\epsilon}$,
$\sqrt{\tilde\epsilon}$ ;
$\nabla_{k,\vec{x}} \equiv T^a \nabla^a_{k,\vec{x}}$, 
where $T^a$ are the generators of SU(3), normalised as
$\tr[T^a T^b] = \delta^{ab}/2$; 
and the covariant derivative is defined as 
\be
 \nabla^a_{k,\vec{x}} S \equiv 
 \lim_{\delta\to 0}
  \frac{1}{\delta}
 \Bigl\{
   S[e^{i \delta T^a} U_k(\vec{x})] - S[U_k(\vec{x})]  
 \Bigr\} 
 \;.
\ee
To be explicit, the expressions for the functional derivatives
in \eqs\nr{Phi_evol_eps}, \nr{U_evol_eps} read 
\ba
 \delta_{\Phi(\vec{x})} S & = & 
 - \sum_{k}
 \Bigl[ 
  U_k(\vec{x}) \Phi(\vec{x} + k) U^\dagger_k(\vec{x}) + 
  U^\dagger _k (\vec{x} - k) \Phi(\vec{x} - k) U_k(\vec{x} - k)
 \Bigr]
 + \nn & + &
 \alpha \Phi(\vec{x}) 
 + 2 \lambda \Phi(\vec{x}) \tr [\Phi^2(\vec{x})]
 \;, \\
 i \nabla_{k,\vec{x}} S & = &  
 \frac{\beta}{12} 
 \sum_{|l|\neq k}
 \biggl\{  P_{kl}(\vec{x}) - P_{kl}^\dagger(\vec{x}) 
         - \frac{\unit}{3} \tr \Bigl[ P_{kl}( \vec{x}) 
        - P_{kl}^\dagger(\vec{x}) \Bigr]
 \biggr\}
 + \nn  & + & 
  \Bigl[ U_k(\vec{x}) \Phi(\vec{x}+ k) U_k^\dagger(\vec{x}), 
    \Phi(\vec{x})  \Bigr]
 \;.
\ea
Furthermore we write the gauge-field time-step in the form
$
 \tilde \epsilon \equiv 10^{-3} {\tau} / {\beta}
$,
while
$
 \epsilon \equiv 10^{-3} \tau
$.

To now introduce NSPT, we expand the variables as
\be
 \Phi(\vec{x} )\longrightarrow\sum_{i}g_0^i\Phi^{(i)}(\vec{x})\;,
 \quad
  U_k(\vec{x}) = 
 \unit + \sum_{i=1} \, \beta^{-\frac{i}{2}} \; U_k^{(i)}(\vec{x}) 
 \;,
\ee
where $g_0$ is some small coupling; in our case, this role
is played by two expansion parameters, $\beta^{-1/2}$ and $\lambda$.
This results in a hierarchical system of difference equations 
that can be numerically solved, to obtain the series in 
\eq\nr{Phi_exp} for $\langle~\!\! \tr~\!\![~\!\!\Phi^{2}]~\!\rangle$, 
for each $\tau$. Subsequently, we need to extrapolate to $\tau = 0$.

Finally, we recall that the gauge field equation of motion
possesses a zero-mode solution. When constructing the gauge 
field propagator, we omit this contribution; its effects are, 
in any case, insignificant in the infinite-volume limit needed
for constructing \eq\nr{Phi_exp}.

%
\section{Data analysis}
\la{se:analysis}

%
\begin{table}[t]

\begin{center}
\begin{tabular}{llrr} \hline
 $a m_3$ & $\ln\beta$ &  $N$~~~ & $\tau$~~~~~~~~~ \\ \hline
 0.25 &  $\ln24$   & 11 -- 22 &  5$^\rmi{(a)}$,10,15,20,25   \\
 0.30 &  $\ln24$   & 10 -- 19 &  5$^\rmi{(b)}$,10,15,20,25   \\
 0.40 &  $\ln24$   & 7 -- 16  &  5$^\rmi{(c)}$,10,15,20,25   \\ 
 0.50 &  $\ln24$   & 7 -- 16  &  10,15,20,25   \\
 0.60 &  $\ln24$   & 10 -- 15 &  10,15,20,25   \\
 0.80 &  $\ln24$   & 4 -- 15  &  10,15,20,25   \\
 1.00 &  $\ln24$   & 5 -- 13  &  10,15,20,25   \\
 1.00$^*$ &  $\ln80$   & 10 -- 13 &  10,15,20,25   \\ \hline
\end{tabular}
\end{center}

\caption[a]{The masses $am_3 \equiv a m_3(\bmu = g_3^2)$, 
counter term parts $\ln\beta$ 
(cf.\ \eq\nr{alpha}), 
box sizes $N$ ($V = a^3 N^3$), and time discretizations $\tau$ studied.
The box sizes were increased in unit steps within the intervals shown.
The time step $\tau = 5$ was only used for the box sizes 
(a) $N=16,19,22$;
(b) $N=16,19$;
(c) $N=16$.
In total, our sample consists of 298 lattices. 
}

\la{table:stats}
\end{table}
%

Our approach involves three different extrapolations / interpolations 
in total: first, the above-mentioned extrapolation to $\tau\to 0$; 
second, an extrapolation to infinite volume ($N\to\infty$); 
third, an interpolation between the different 
$\amE = a m_3(\bmu = g_3^2)$ simulated. 
We discuss these steps one by one. 
The complete data sample is listed in Table~\ref{table:stats}.

%
\subsection{Extrapolation $\tau\to 0$}

\begin{figure}[p]


\centerline{%
\epsfysize=5.0cm\epsfbox{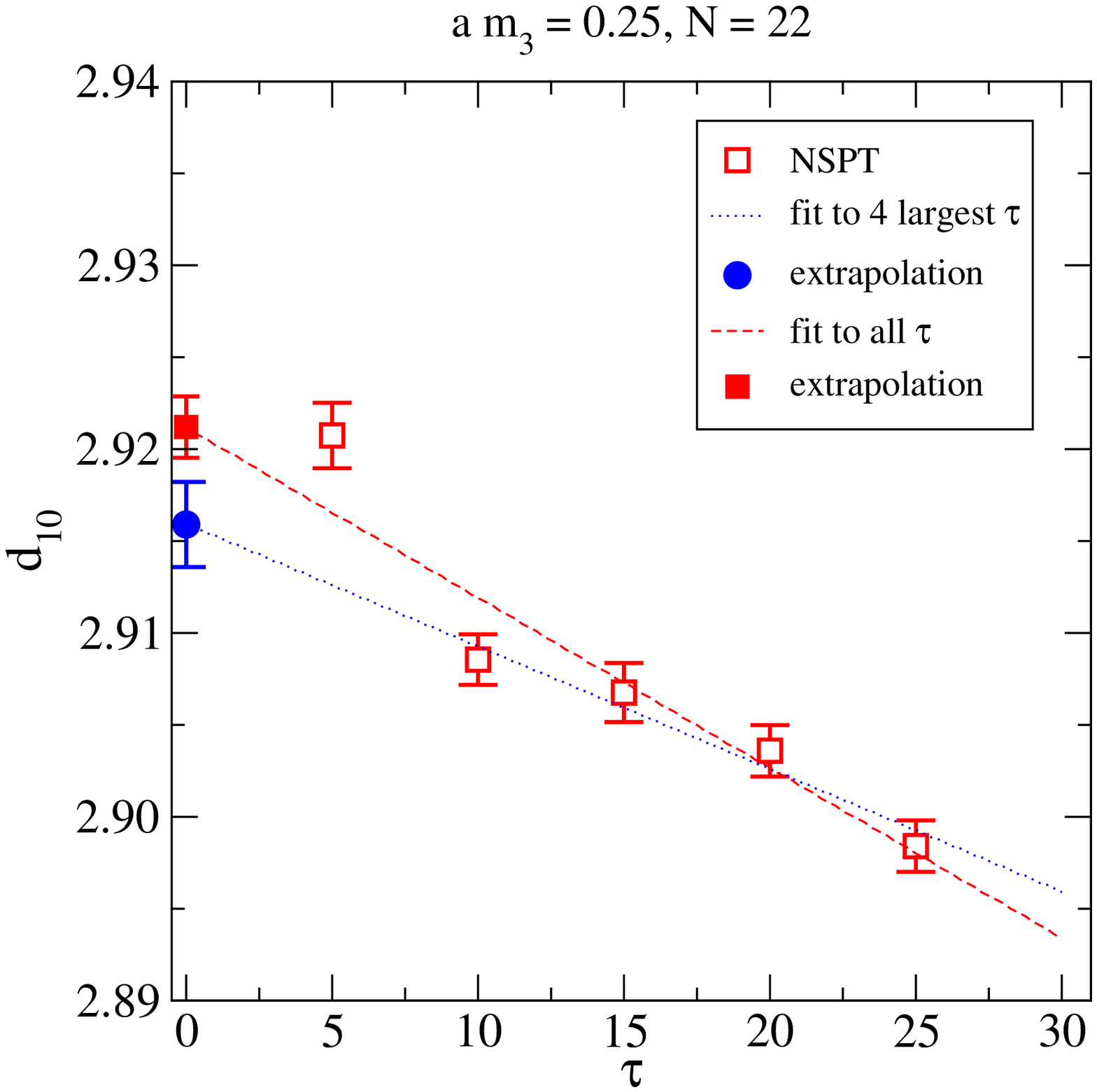}%
~~\epsfysize=5.0cm\epsfbox{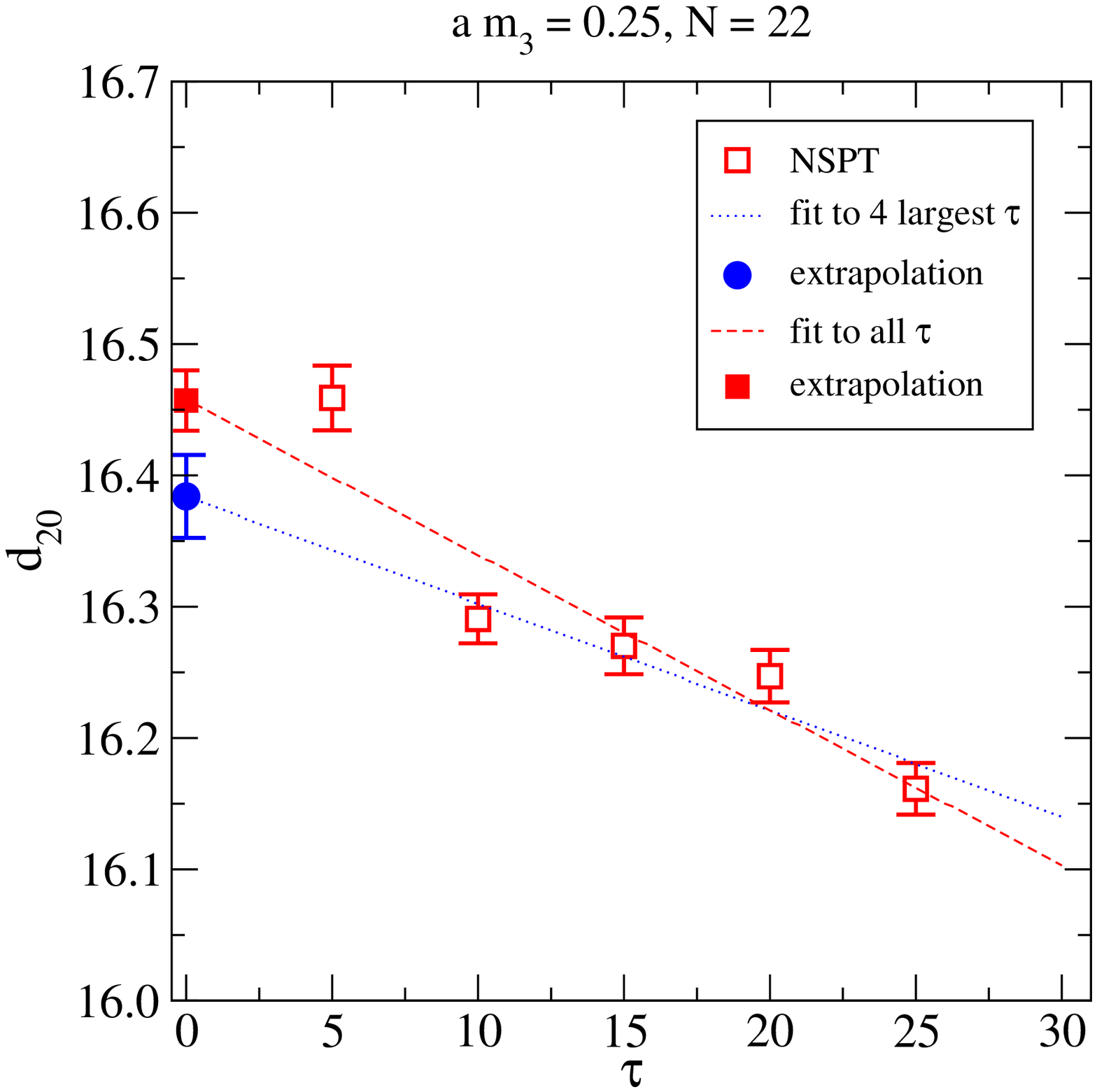}%
~~\epsfysize=5.0cm\epsfbox{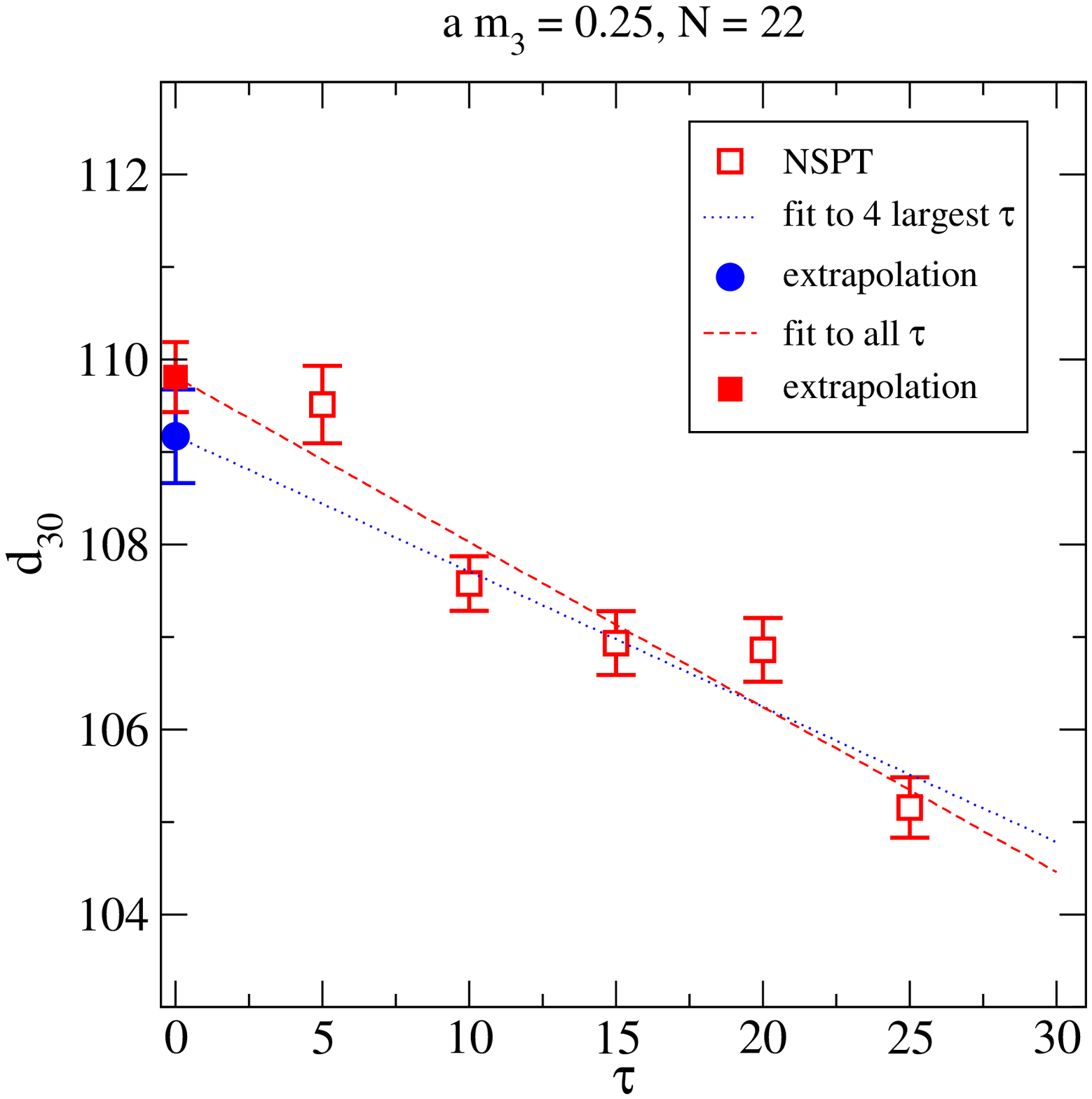}%
}

\vspace*{0.5cm}

\centerline{%
\epsfysize=5.0cm\epsfbox{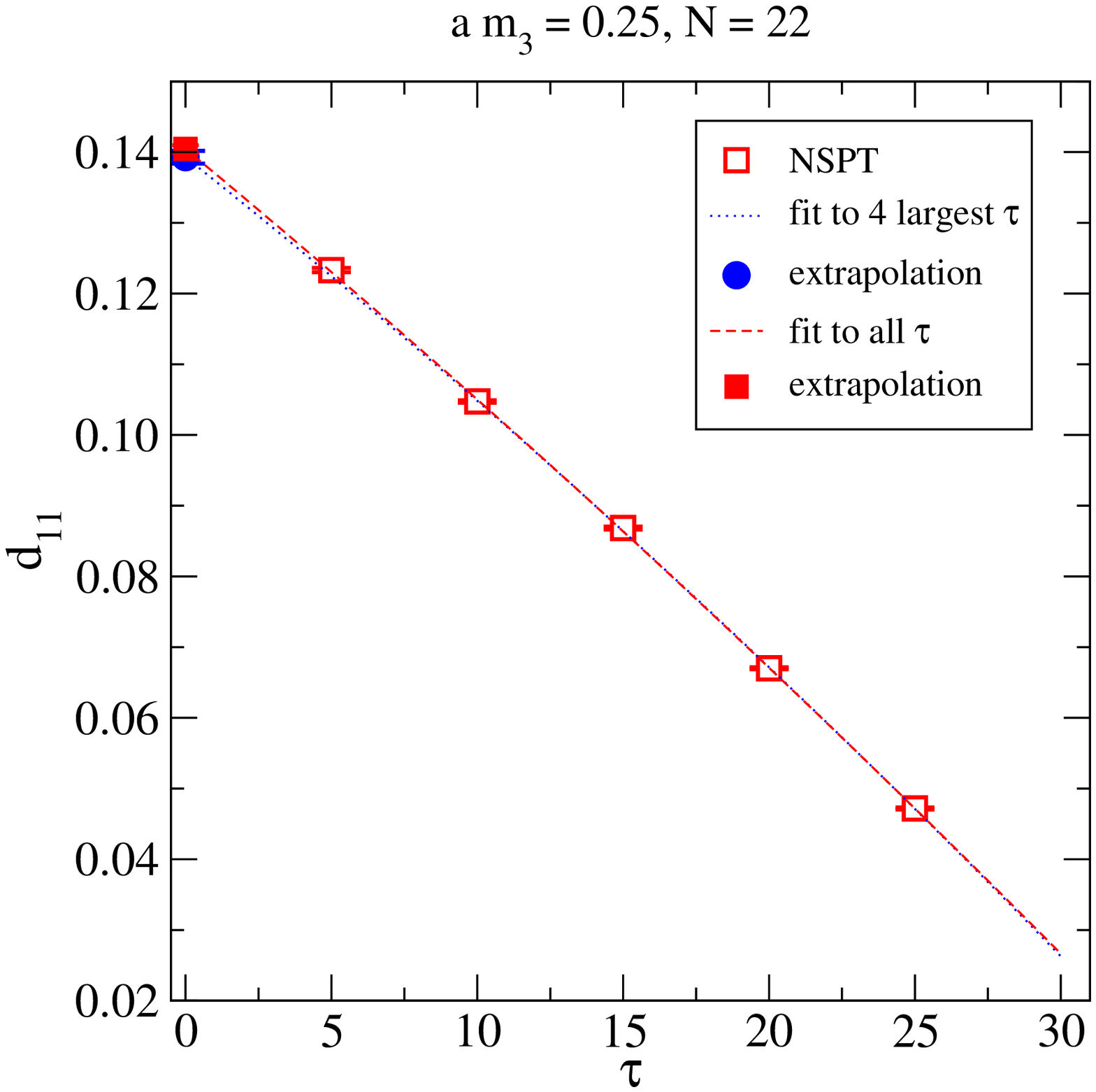}%
~~\epsfysize=5.0cm\epsfbox{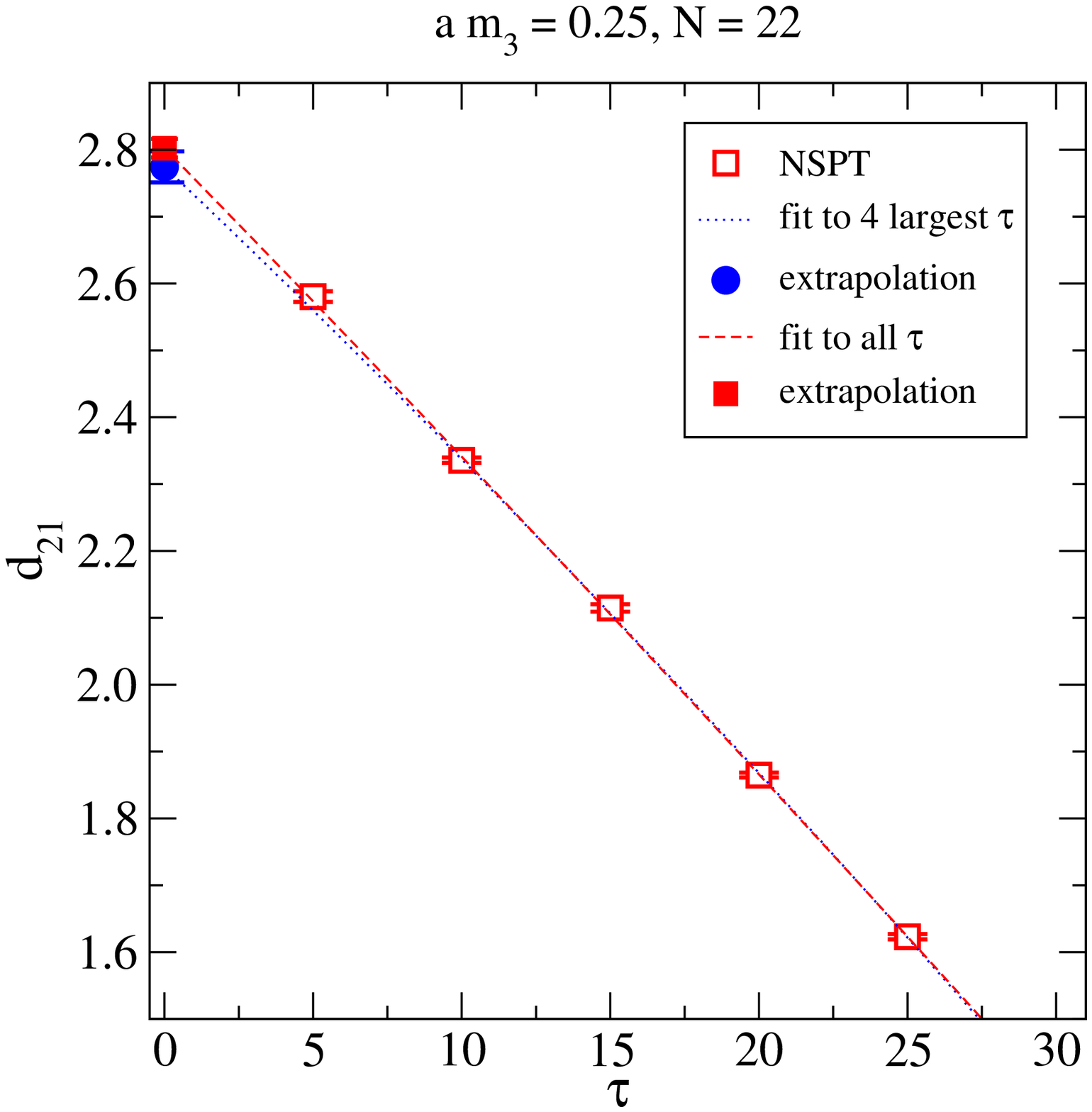}%
~~\epsfysize=5.0cm\epsfbox{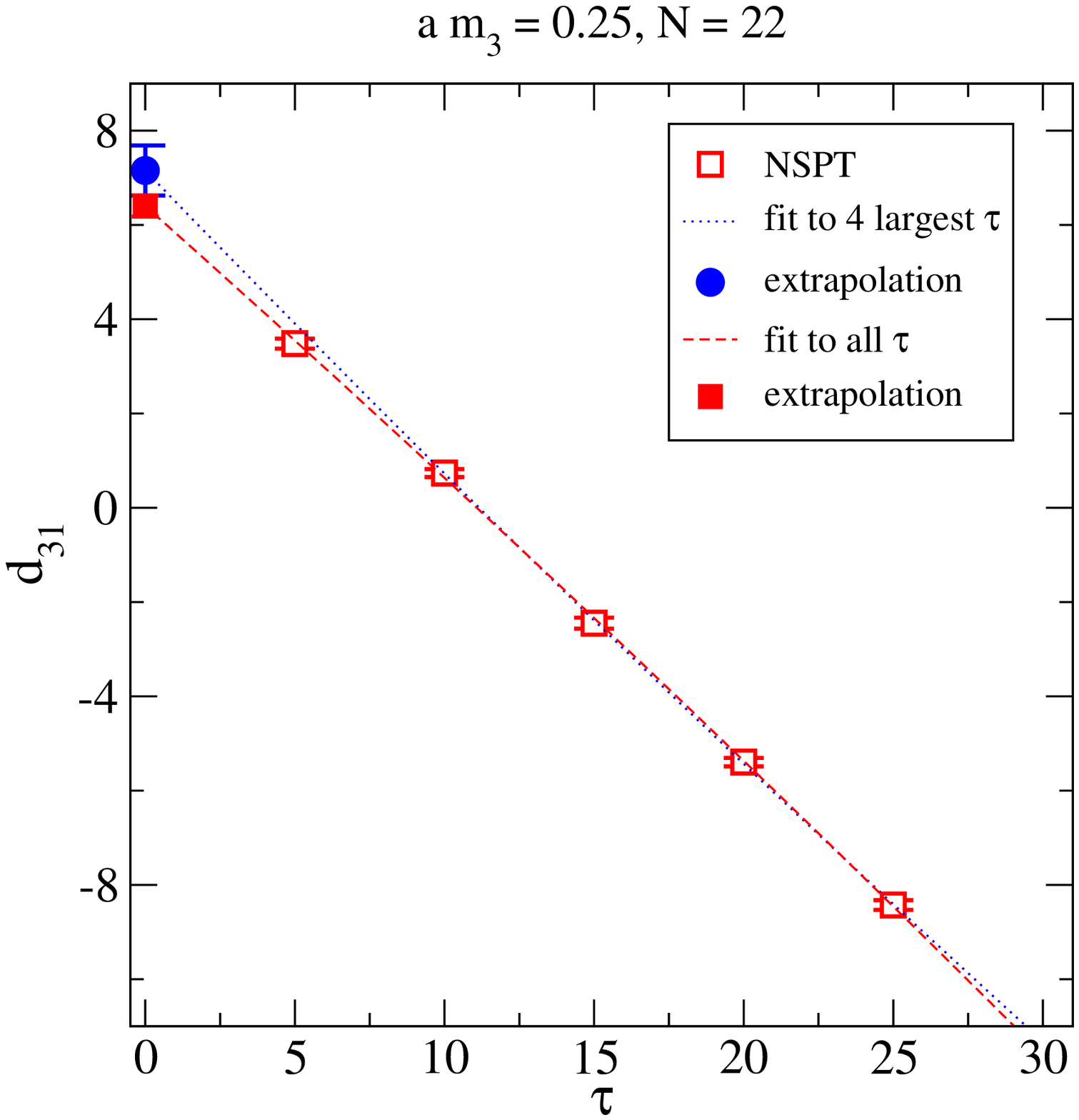}%
}

\vspace*{0.5cm}

\centerline{%
\epsfysize=5.0cm\epsfbox{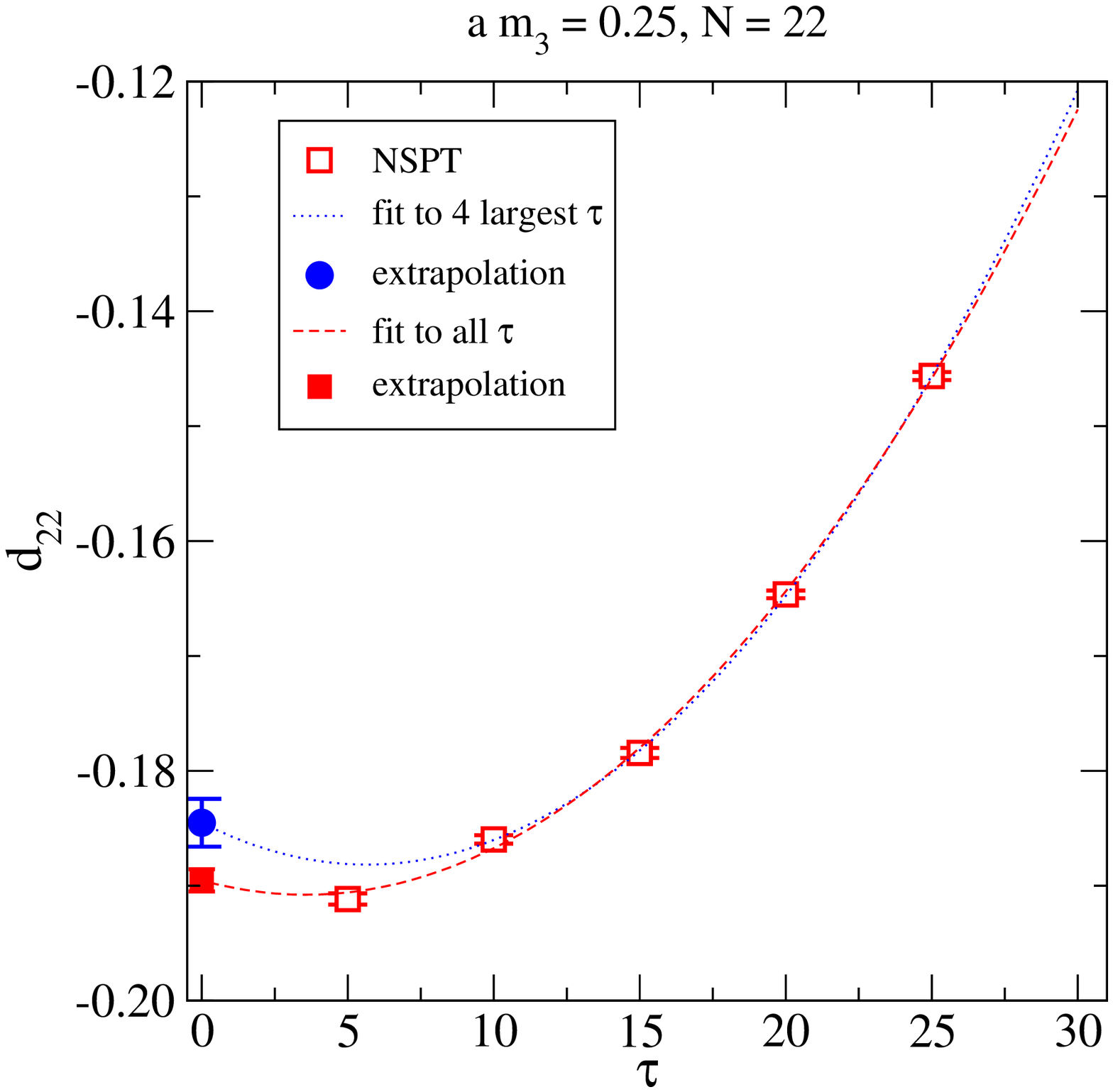}%
~~\epsfysize=5.0cm\epsfbox{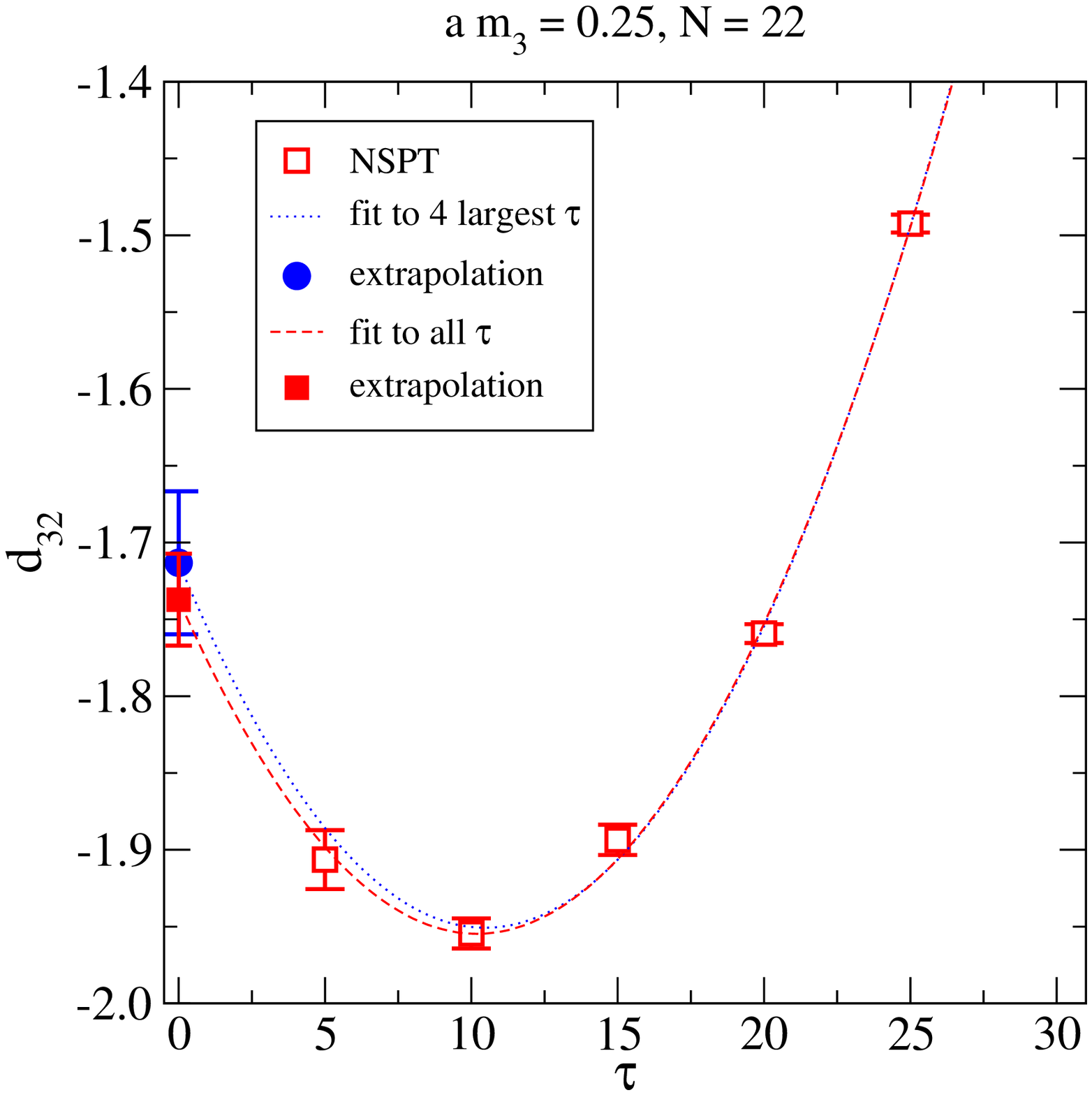}%
~~\epsfysize=5.0cm\epsfbox{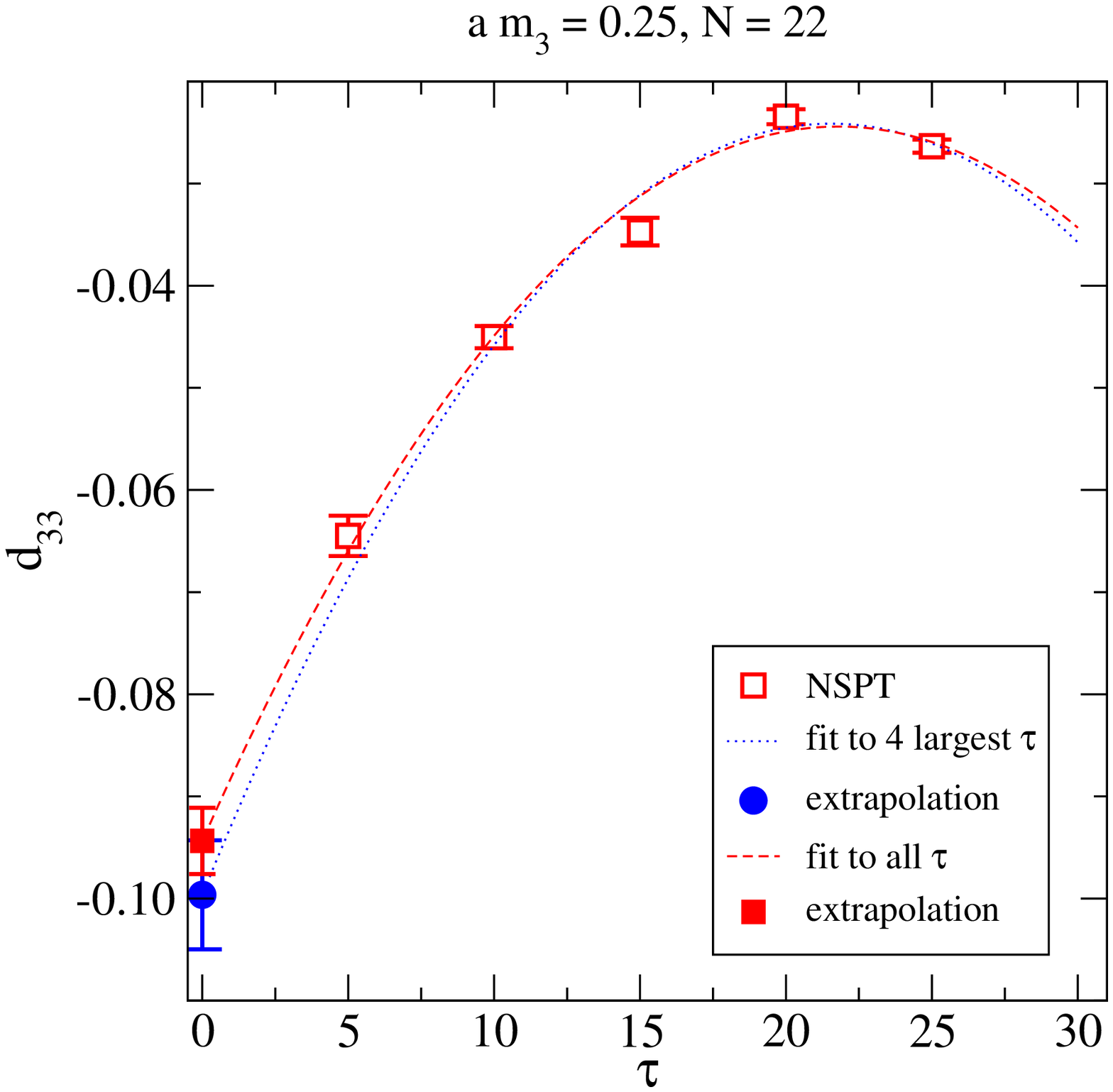}%
}


\caption[a]{The results for $d_{ij}$ 
as a function of $\tau$, at $a m_3 = 0.25$, $N=22$.
The curves show the results of linear or quadratic fits. 
}

\la{fig:dij_tau}
\end{figure}

Examples of the $\tau\to 0$ extrapolations
are shown in \fig\ref{fig:dij_tau}, at $a m_3 = 0.25$, 
$N=22$. (We omit, for layout-reasons, the simplest coefficient $d_{00}$.)
The data immediately lead to the important observation that the 
shapes of the curves are practically 
the same for $d_{10}$, $d_{20}$ and $d_{30}$; 
for $d_{11}$, $d_{21}$ and $d_{31}$; 
and for $d_{22}$ and $d_{32}$. 
In other words, the behaviour as a function of $\tau$ is dictated
by the number of scalar couplings $\lambda$ that are associated
with the coefficient. (This is at least partly due to the simple
way in which we chose the time steps related to gauge and scalar
field time evolutions; with some tuning, it might have been possible 
to optimise the time steps such that the time evolutions would have been 
more balanced~\cite{sw}.)

We have used the data at $\tau = 5$, which were the most expensive ones
to produce and are only available for a subset of the parameter values, 
as a probe for the type of extrapolation that should be used for 
obtaining the $\tau \to 0$ limits. Indeed, 
we choose the order of the polynomial 
fit in $\tau$ low enough so that the results remain more or less stable 
in the inclusion of the $\tau = 5$ points. 
For $d_{10}$, $d_{20}$ and $d_{30}$, this requires linear extrapolations; 
for the other coefficients, we use quadratic fits. Nevertheless, we note
that in some cases the results of the extrapolations do change by 
a statistically significant amount in the inclusion of the 
points with $\tau = 5$, indicating that our systematic 
errors may be non-negligible. 

\begin{figure}[t]


\centerline{%
\epsfysize=5.0cm\epsfbox{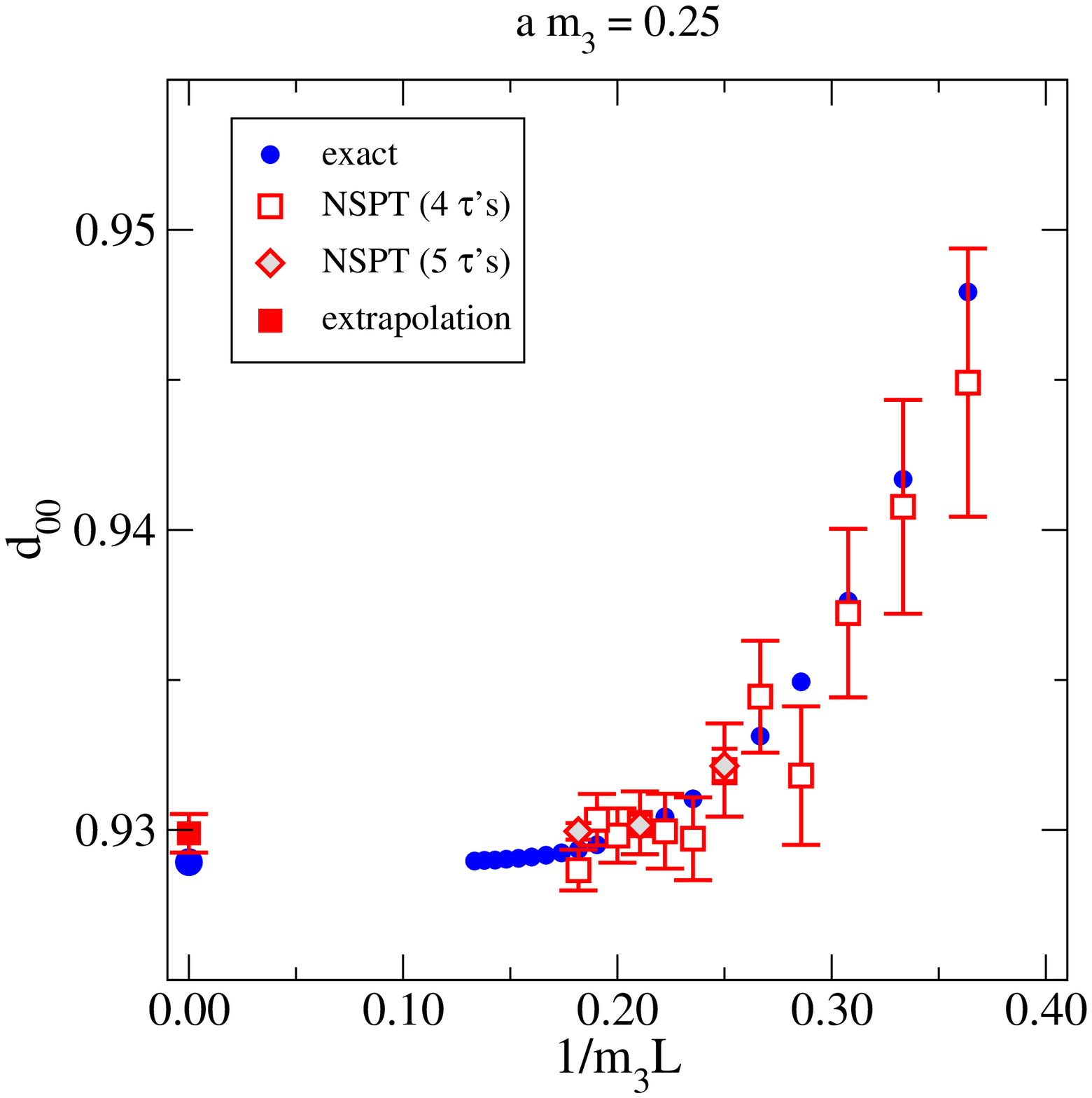}%
~~\epsfysize=5.0cm\epsfbox{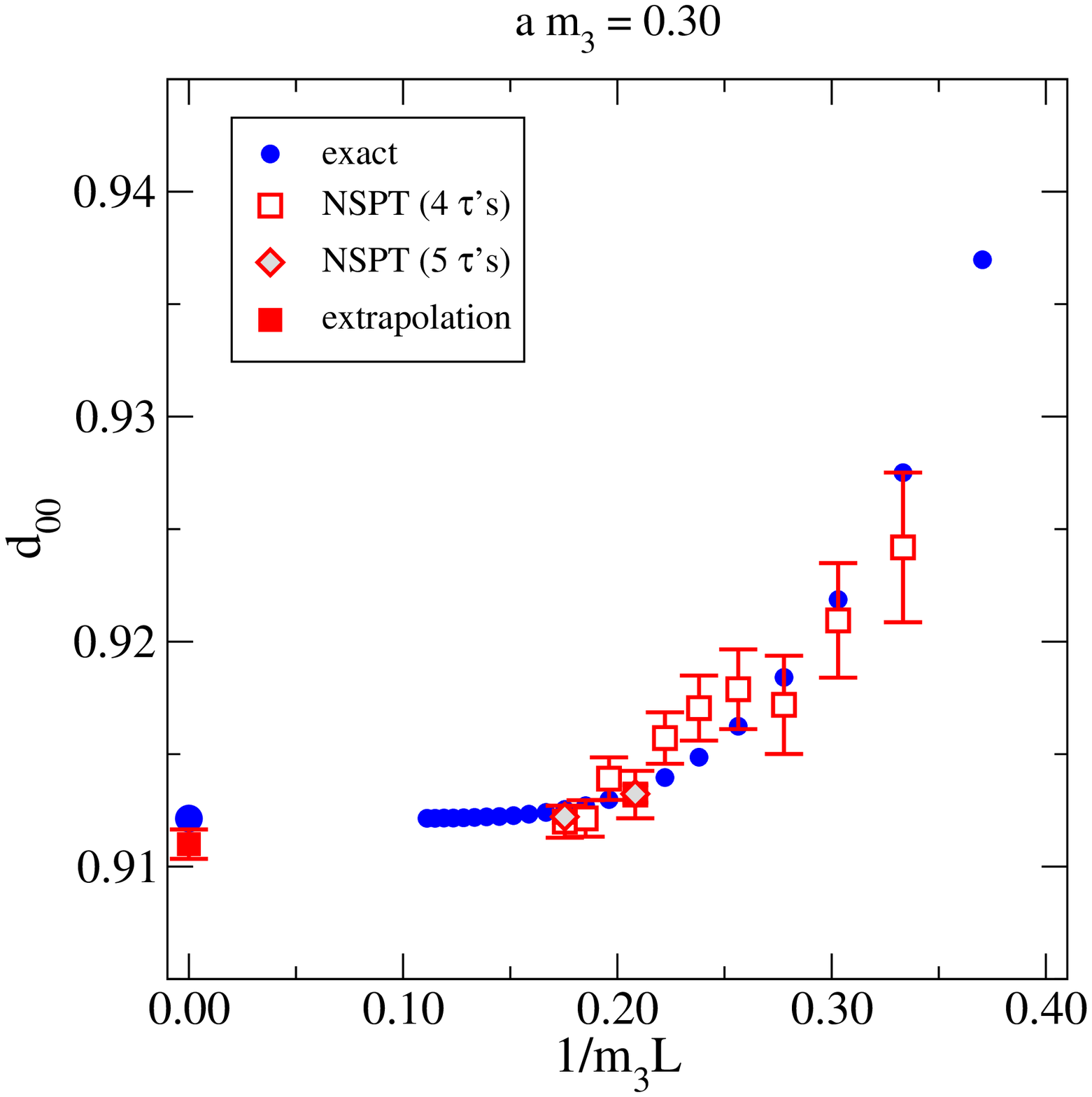}%
~~\epsfysize=5.0cm\epsfbox{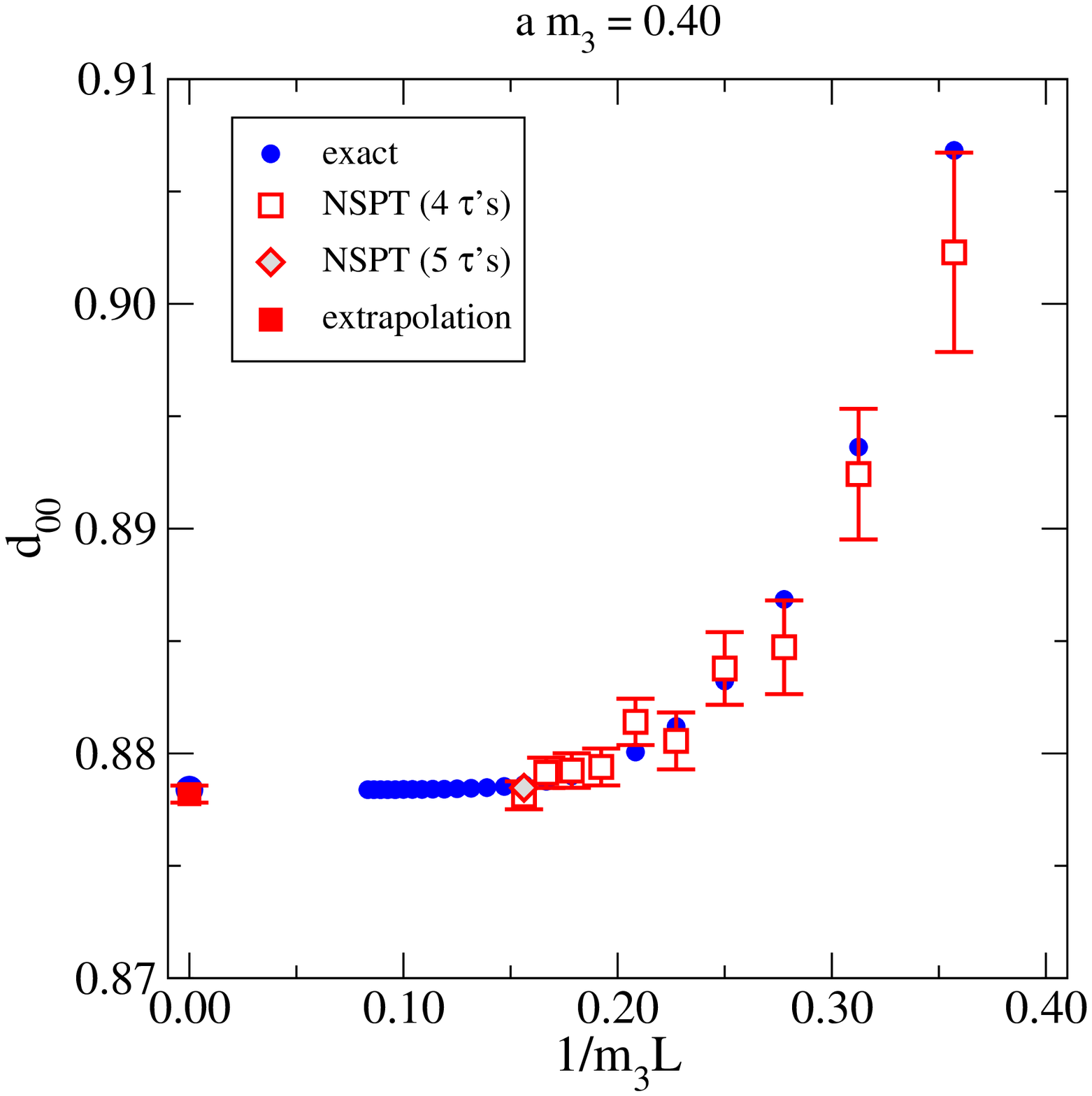}%
}

 \vspace*{0.5cm}

\centerline{%
\epsfysize=5.0cm\epsfbox{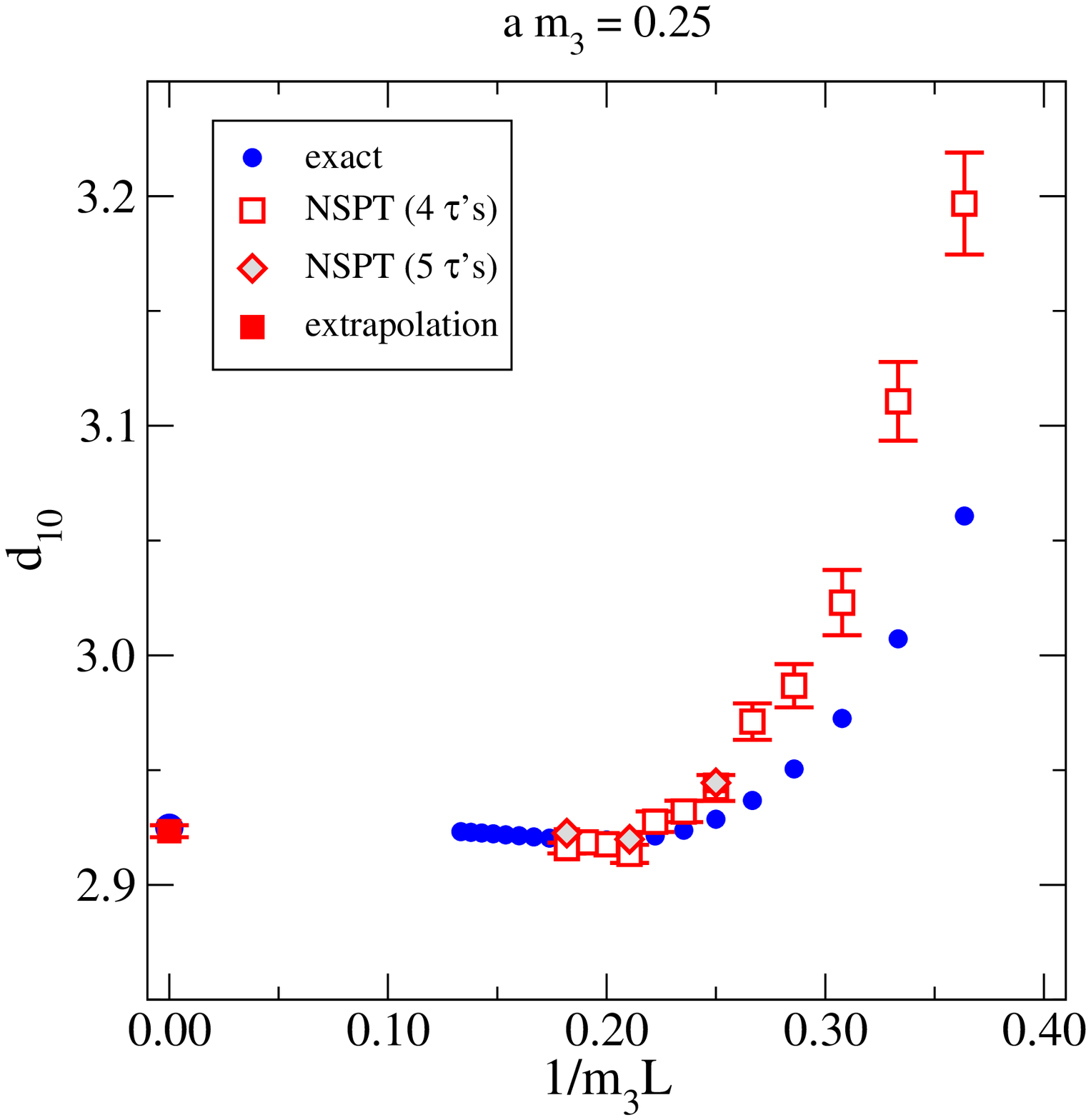}%
~~\epsfysize=5.0cm\epsfbox{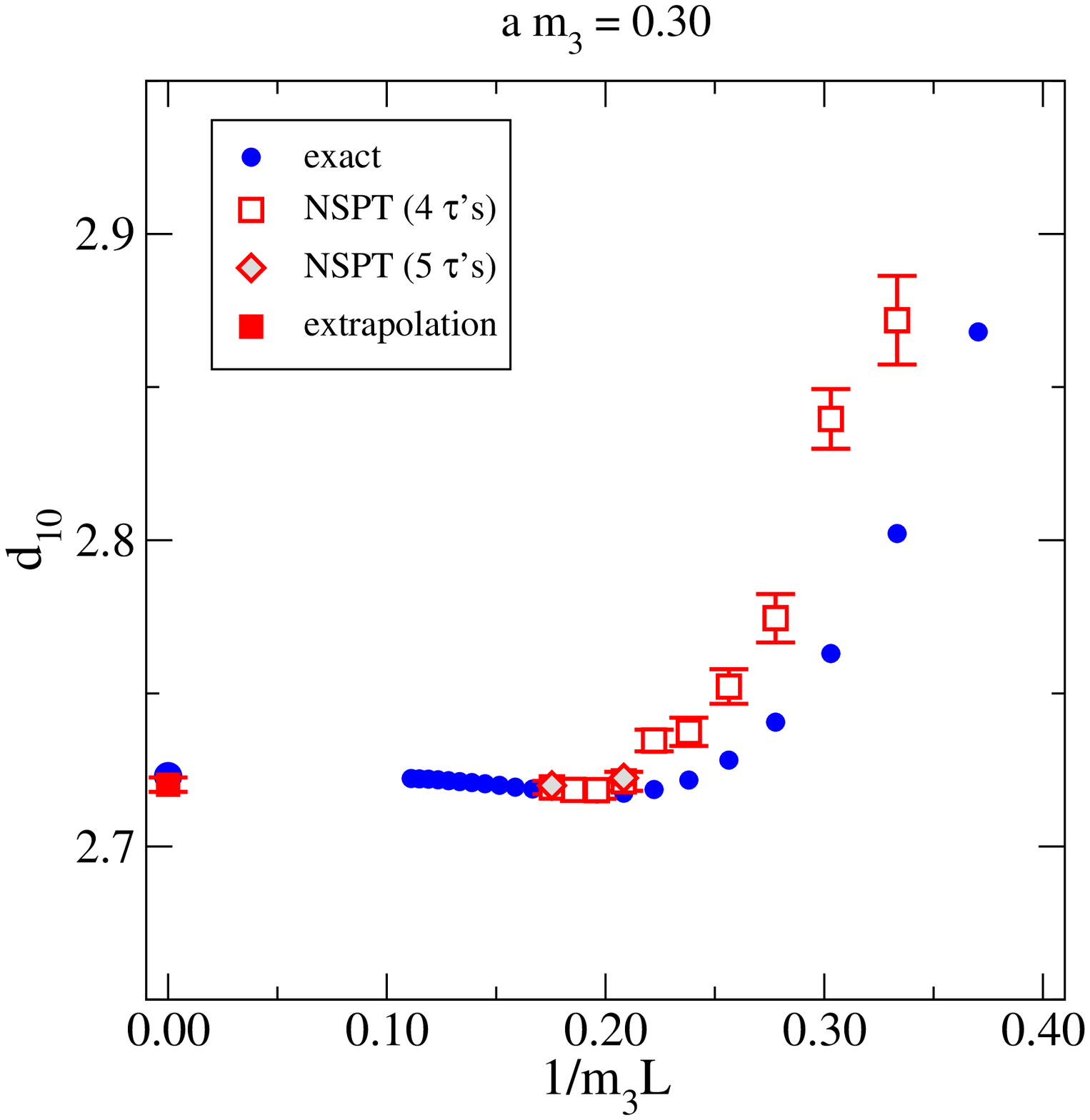}%
~~\epsfysize=5.0cm\epsfbox{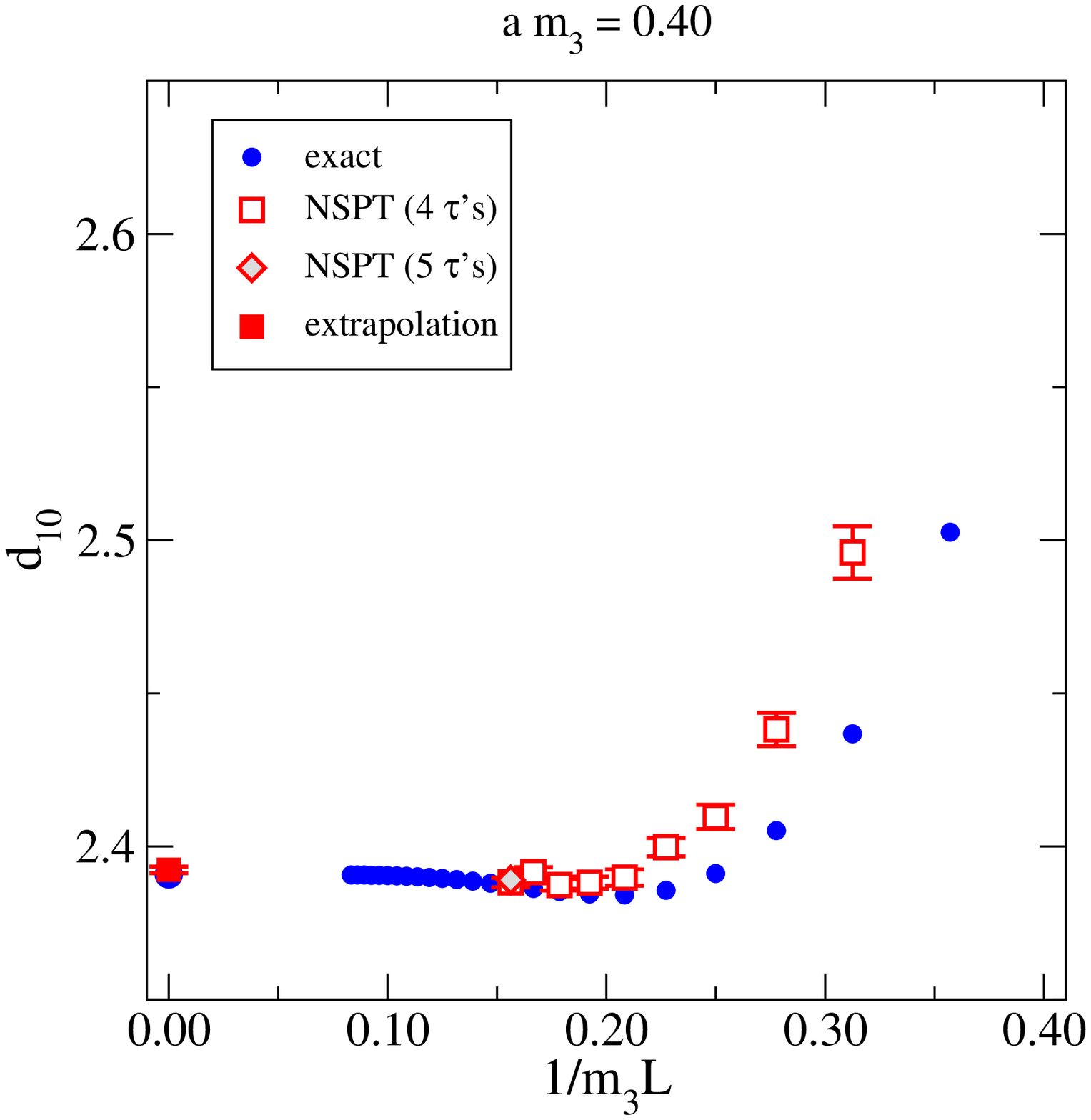}%
}


\caption[a]{The results for $d_{00}$ and $d_{10}$
 after the $\tau\to 0$ extrapolation, 
 as a function of the inverse box size in ``physical'' units 
($m_3 L = a m_3 N$), 
together with the exact values. 
The (infinite volume) ``extrapolation'' denoted with 
the closed square refers to the procedure defined around \eq\nr{dij_prime}.}

\la{fig:precise}
\end{figure}


\begin{figure}[p]


\centerline{%
\epsfysize=5.0cm\epsfbox{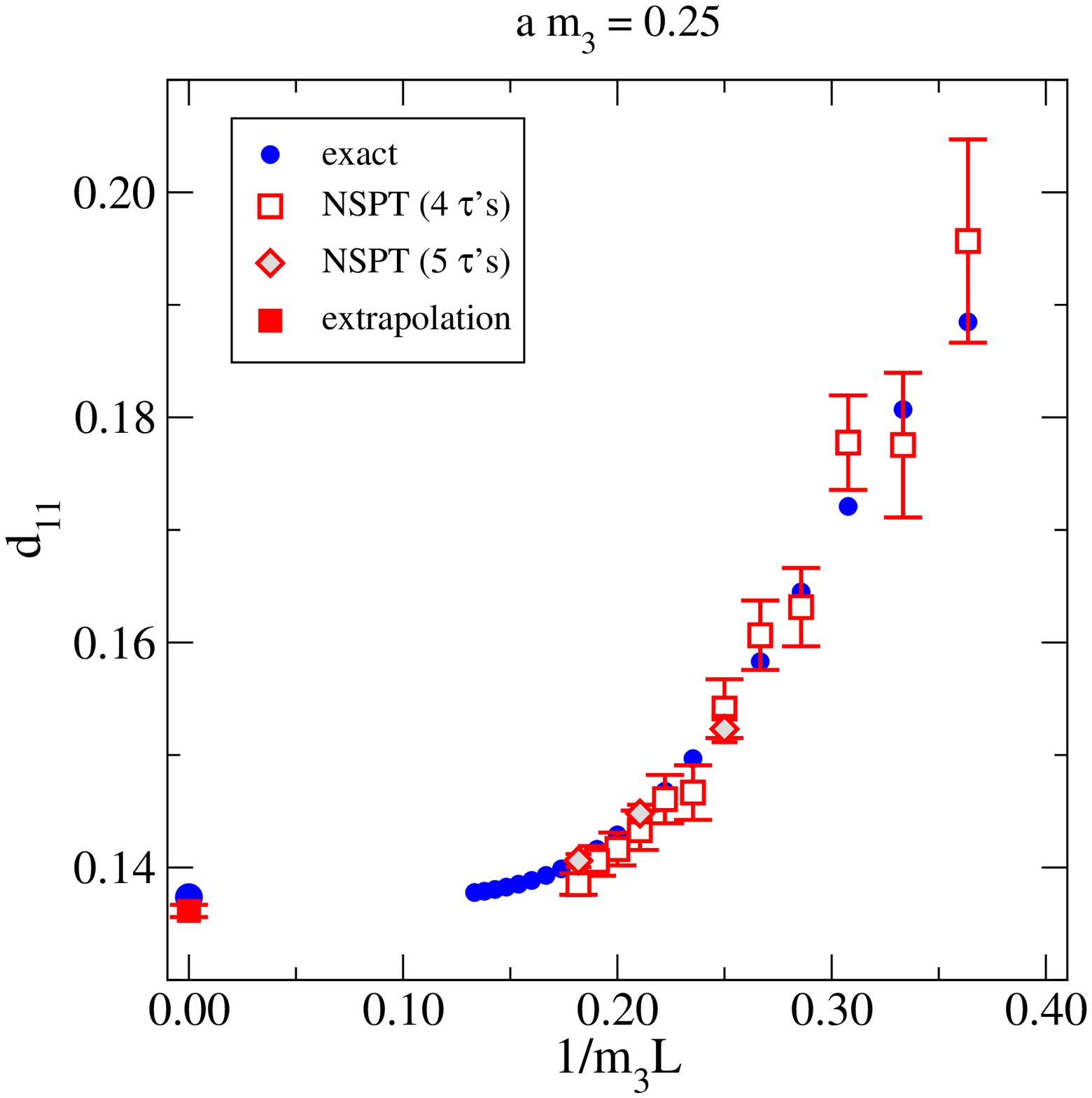}%
~~\epsfysize=5.0cm\epsfbox{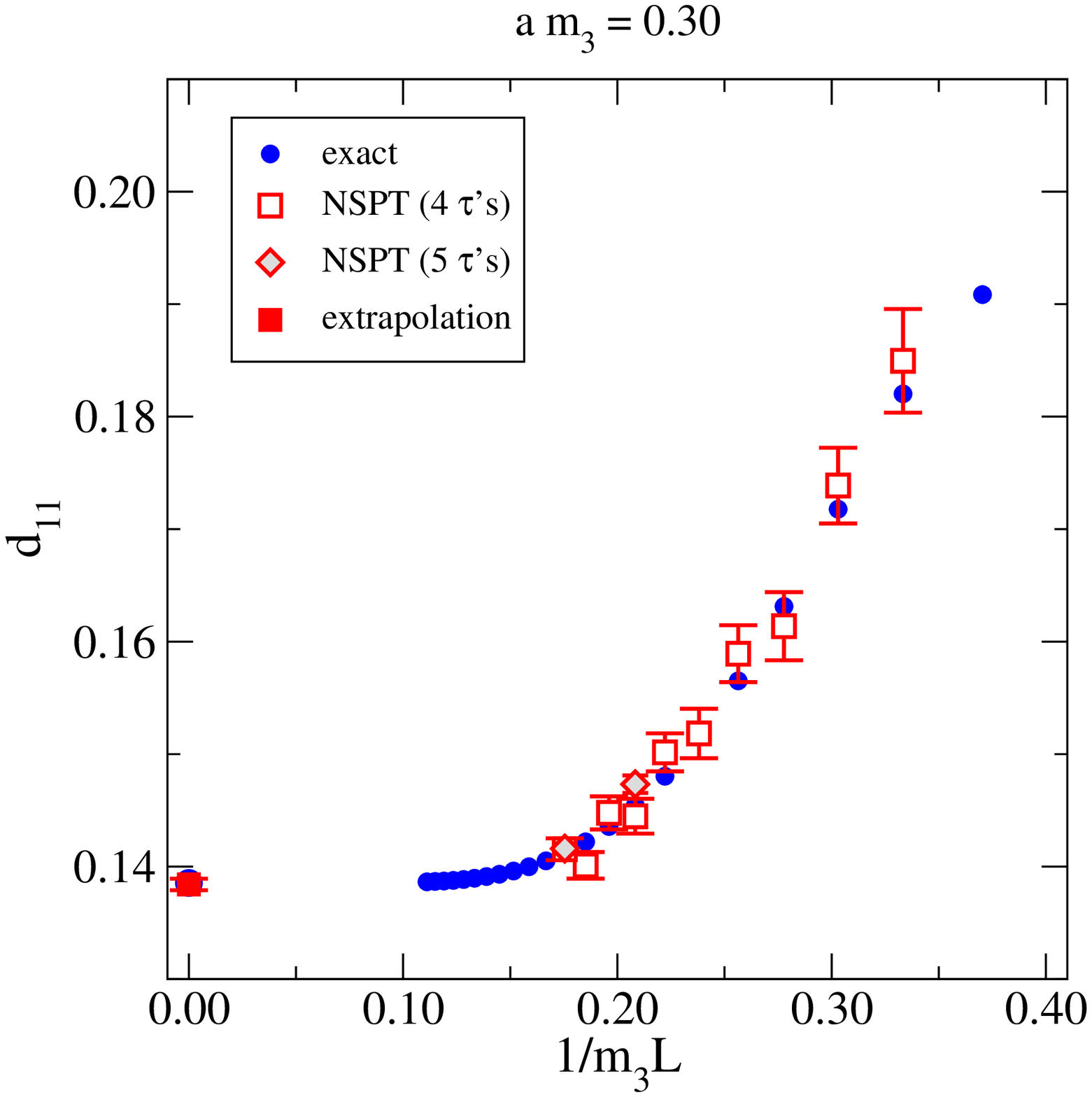}%
~~\epsfysize=5.0cm\epsfbox{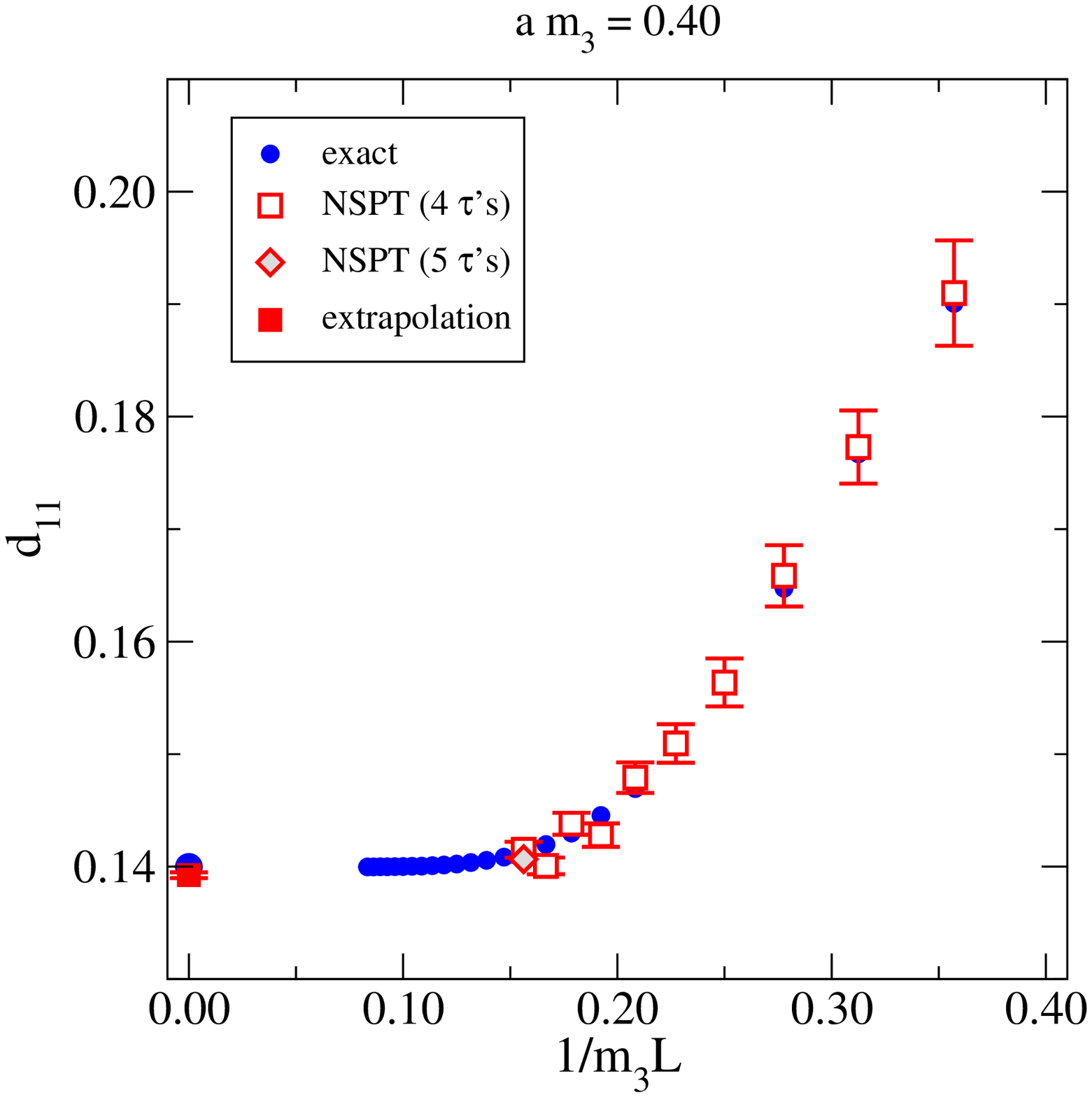}%
}

 \vspace*{0.5cm}

\centerline{%
\epsfysize=5.0cm\epsfbox{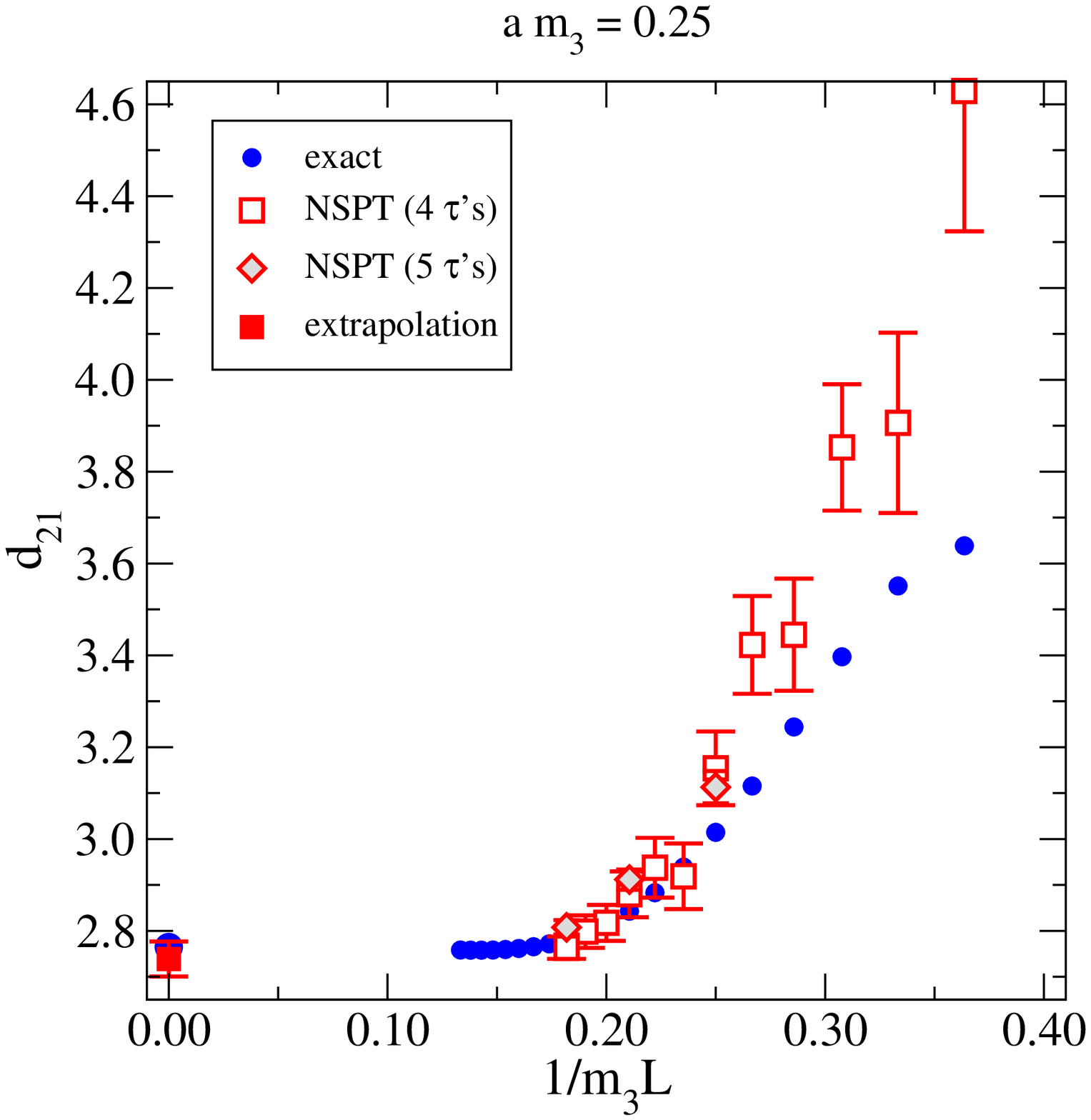}%
~~\epsfysize=5.0cm\epsfbox{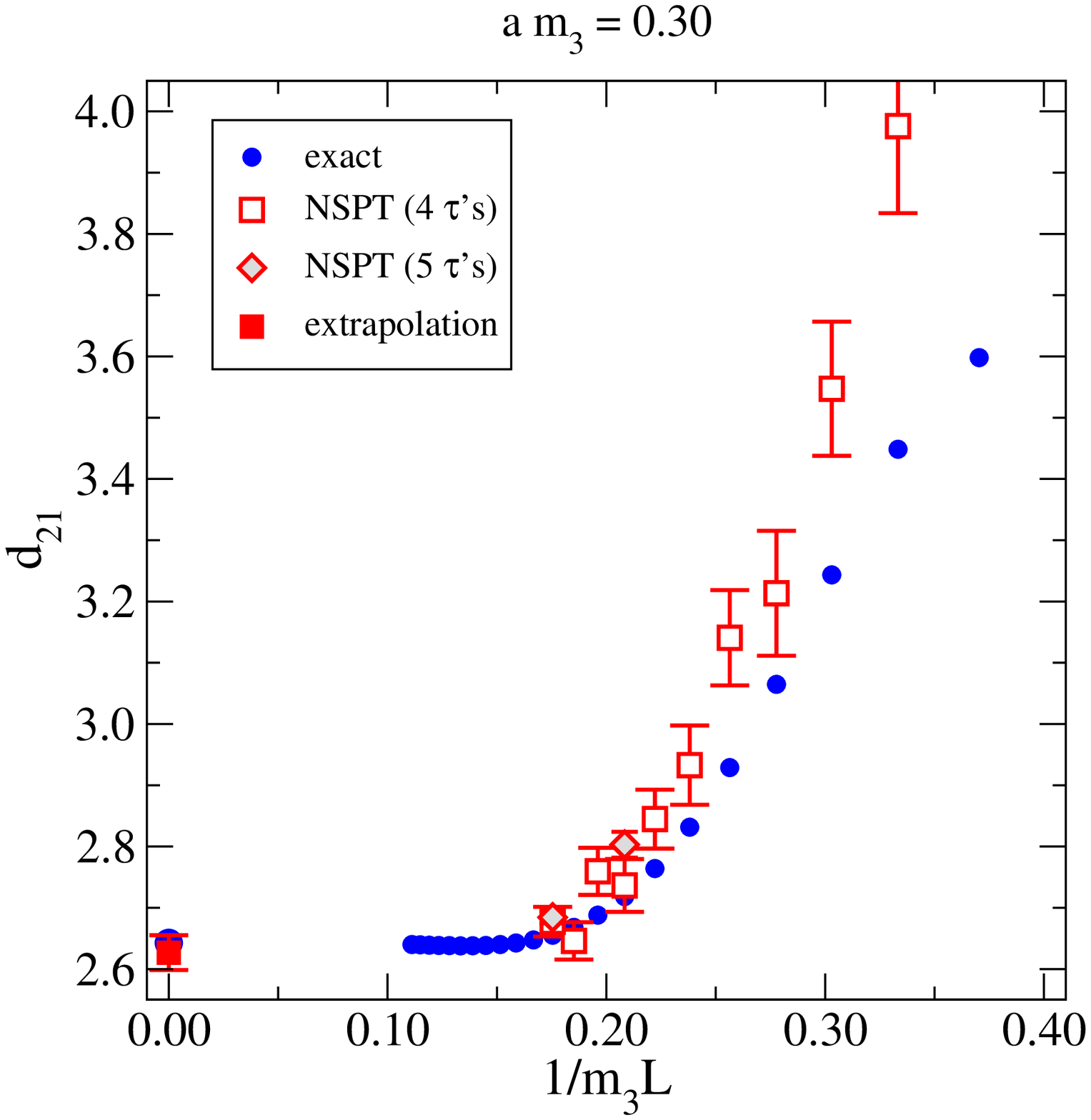}%
~~\epsfysize=5.0cm\epsfbox{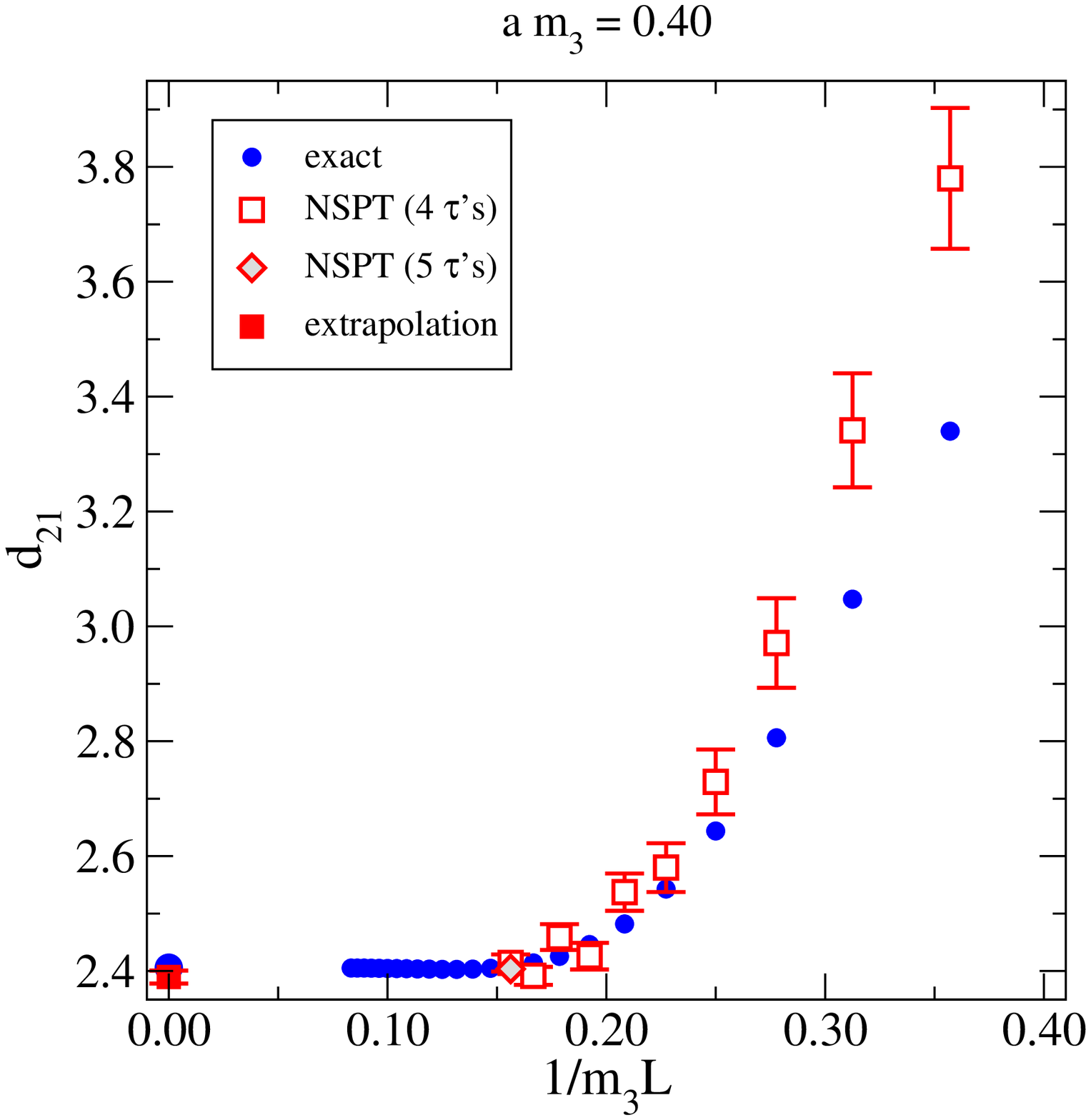}%
}

 \vspace*{0.5cm}

\centerline{%
\epsfysize=5.0cm\epsfbox{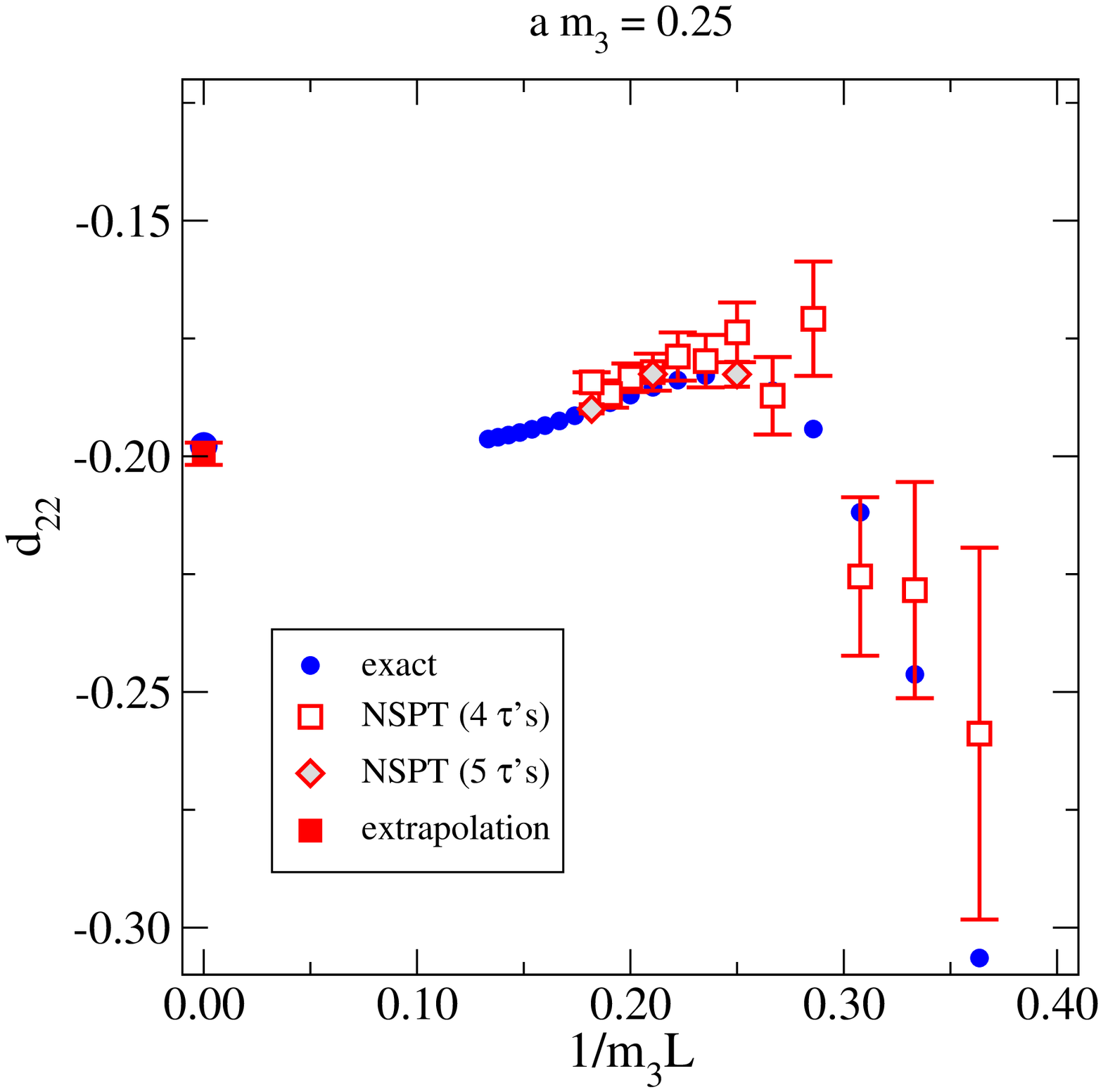}%
~~\epsfysize=5.0cm\epsfbox{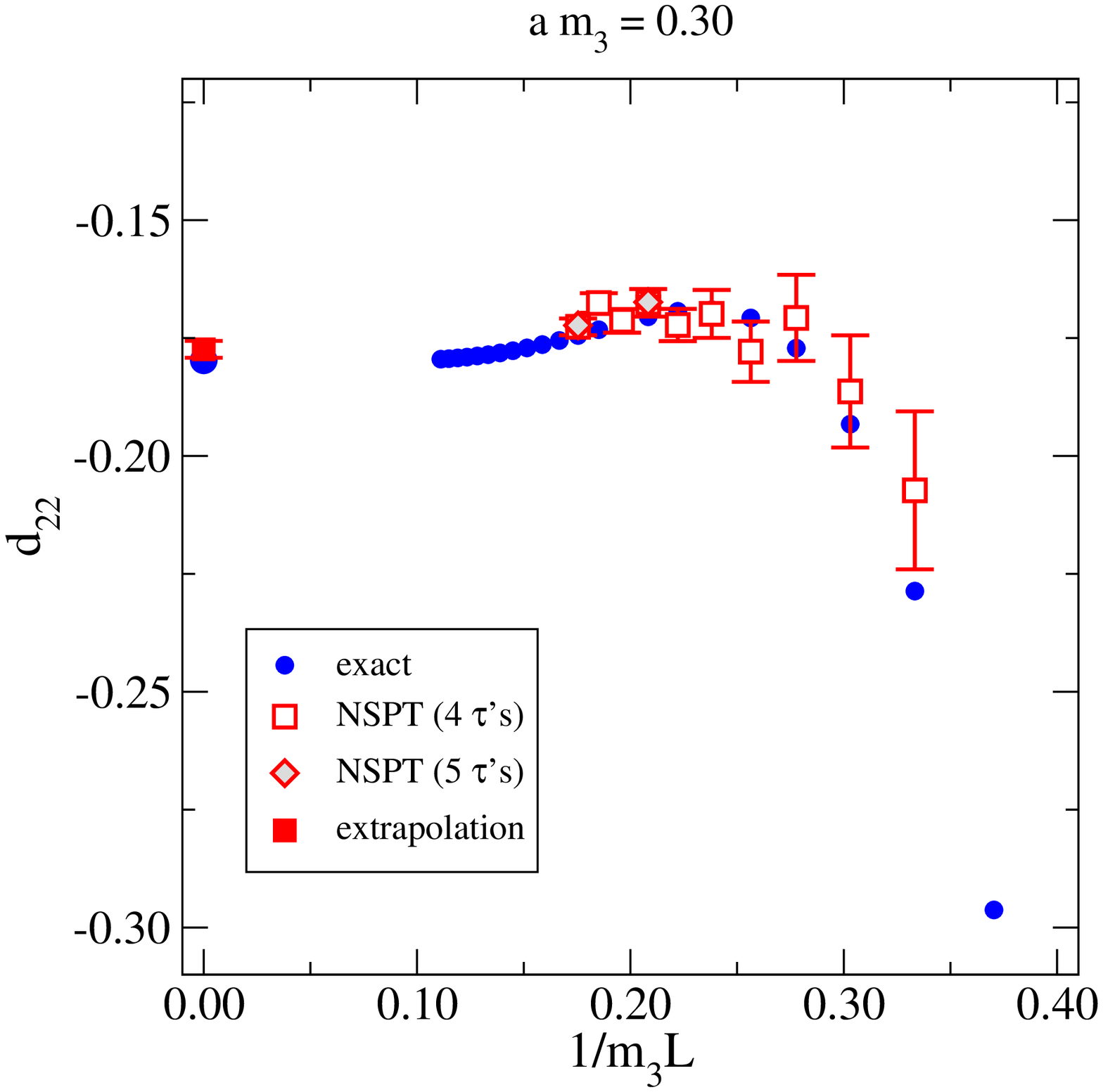}%
~~\epsfysize=5.0cm\epsfbox{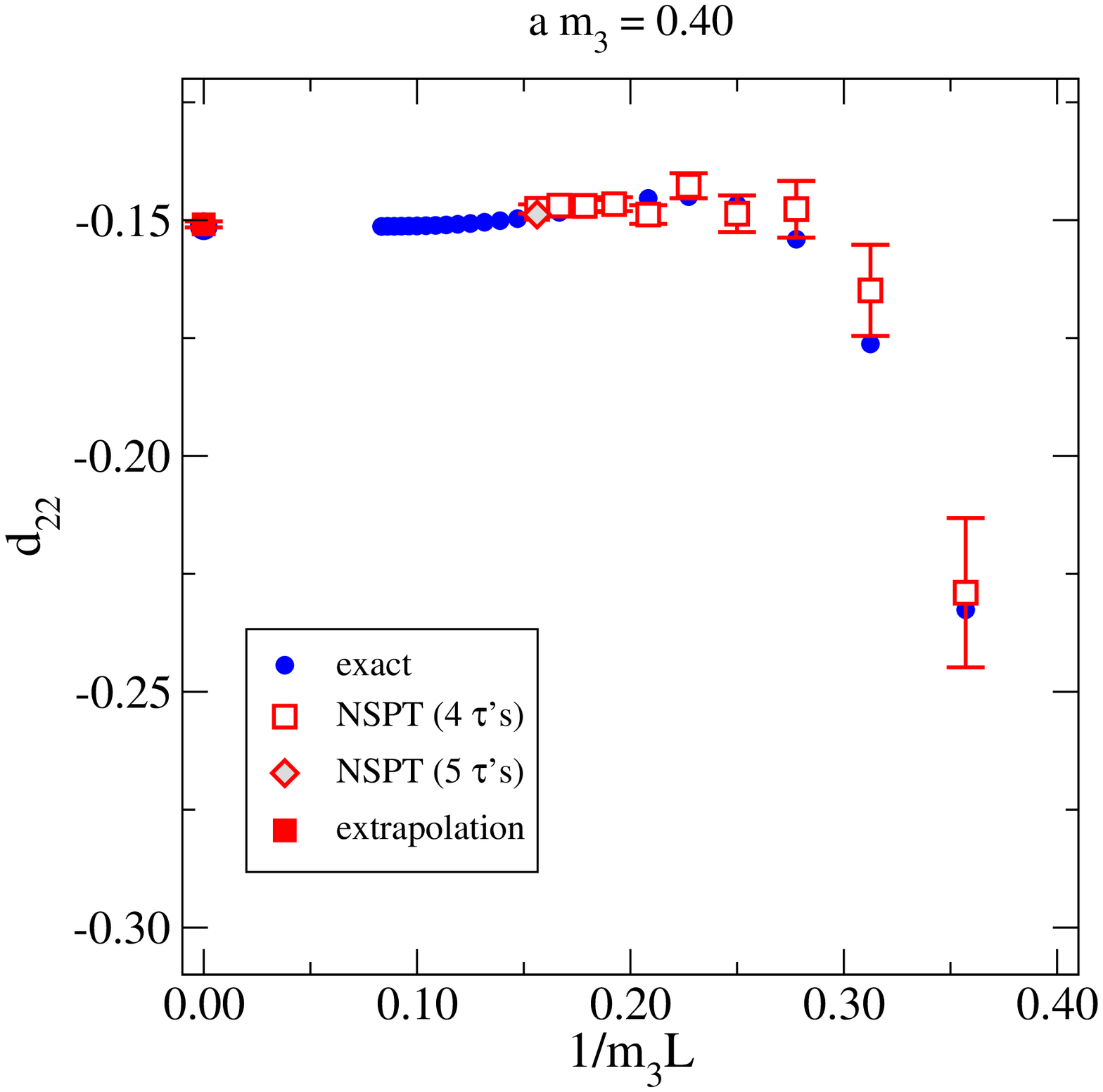}%
}


\caption[a]{Like \fig\ref{fig:precise} but for 
$d_{11}$, $d_{21}$, $d_{22}$. 
}

\la{fig:precise_2}
\end{figure}

Given the possible existence of systematic errors,  
it is important to crosscheck the results in a number of known cases. 
As has been discussed in \se\ref{se:setup}, we do have
exact results available, for any given $a m_3$ and box size $N$, 
for the coefficients $d_{00}, d_{10}, d_{11}, d_{21}, d_{22}$.  
In \figs\ref{fig:precise}, \ref{fig:precise_2}
we compare the $\tau\to 0$ extrapolations, 
based on 4 $\tau$'s in most cases, and on 5 $\tau$'s where
available, with the exact values. 

We observe that, in general, the results of the $\tau\to 0$ extrapolations
{\em do} scatter around the correct values. The exception is 
$d_{10}$, and to a lesser extent $d_{21}$, at small volumes. We suspect
that the reason for this 
is related to the way in which the zero-modes are subtracted 
in standard lattice perturbation theory (i.e.\ ``exact values'', 
cf.\ appendix~B) 
and in NSPT, respectively. 
Nevertheless, given that the discrepancy rapidly disappears with 
increasing volume, there does not appear to be serious reason for concern.

Moreover, we note that the extrapolations including $\tau = 5$
are in general closer to the exact values than those excluding it. 
In a few cases, the correct value is between the extrapolations
based of 4 and 5 $\tau$'s;  in other words, the data point at 
$\tau = 5$ ``overcorrects'' the result of the extrapolation. 
Based on these tests, we have decided to 
always include extrapolations based on 4 and, where available, 
5 $\tau$'s, as independent estimates of the intercepts at $\tau = 0$.
The extrapolations based on 5 $\tau$'s have smaller error bars, and thus
more weight in the subsequent fits; at the same time, the inclusion 
of the extrapolations based on 4 $\tau$'s allows
us to correct for the mentioned overshooting in the cases
where it does take place.

%
\subsection{Extrapolation $N \to \infty$}

\begin{figure}[p]

\vspace*{-1cm}

\centerline{%
\epsfysize=5.0cm\epsfbox{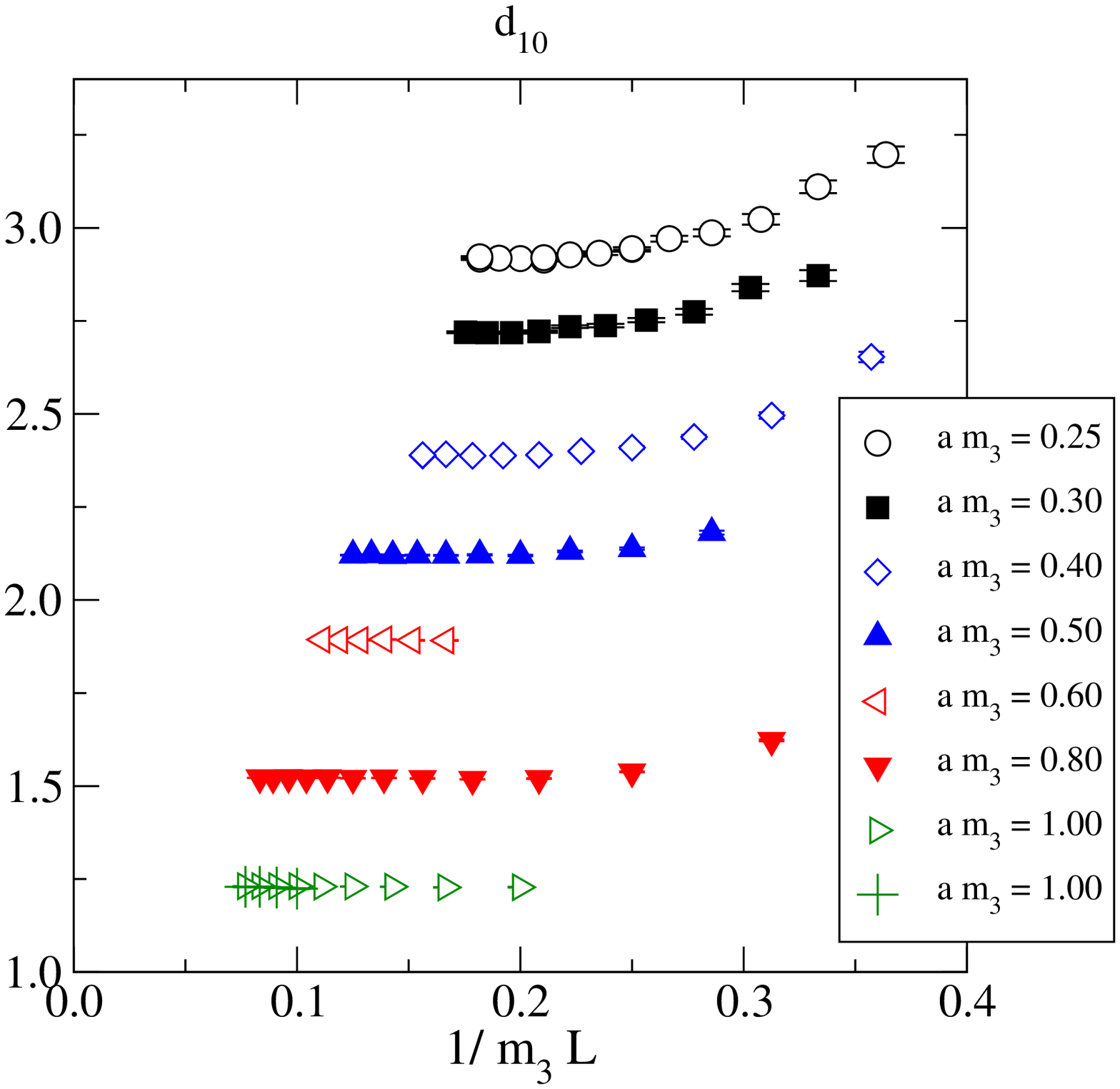}%
~~\epsfysize=5.0cm\epsfbox{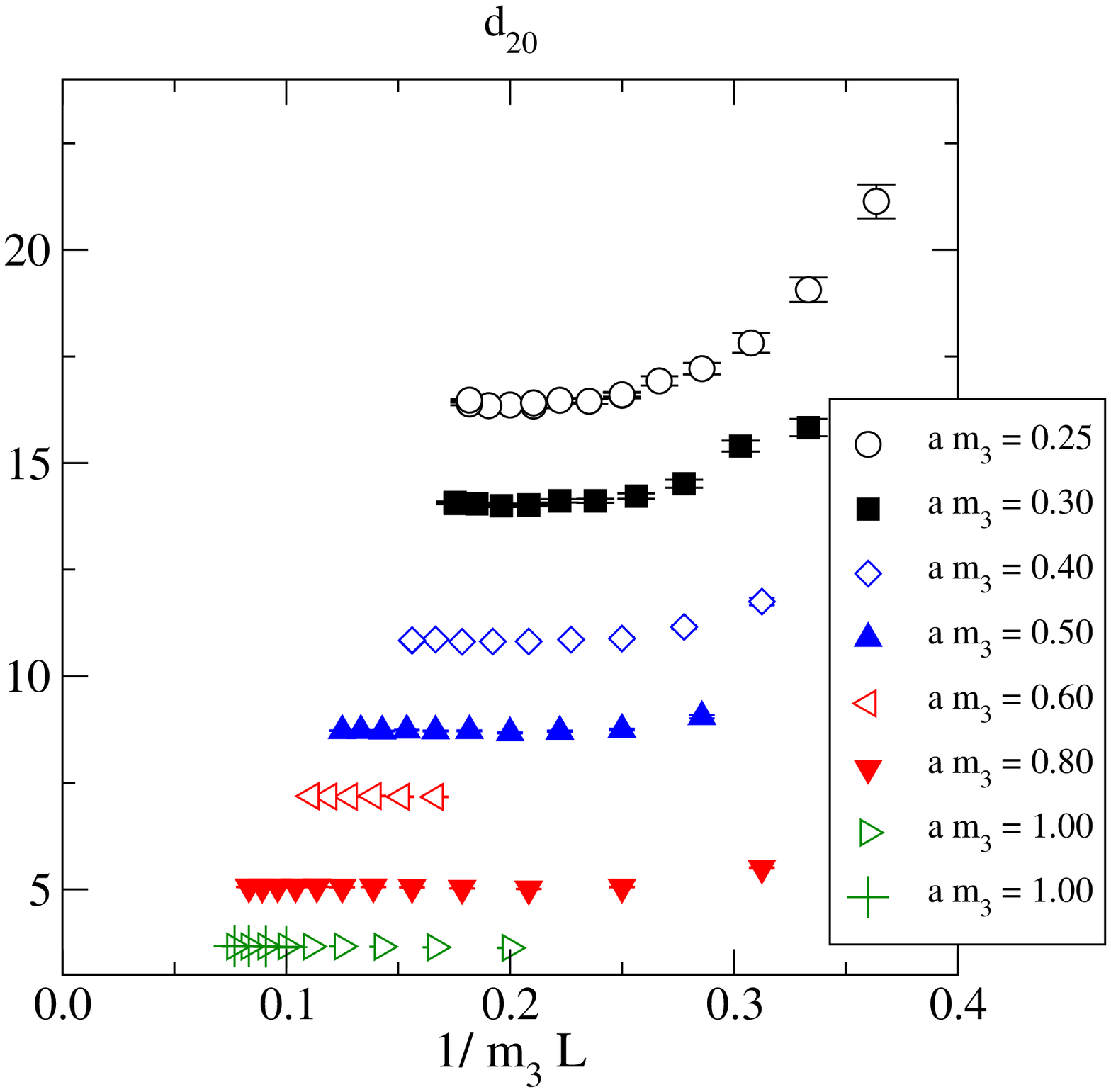}%
~~\epsfysize=5.0cm\epsfbox{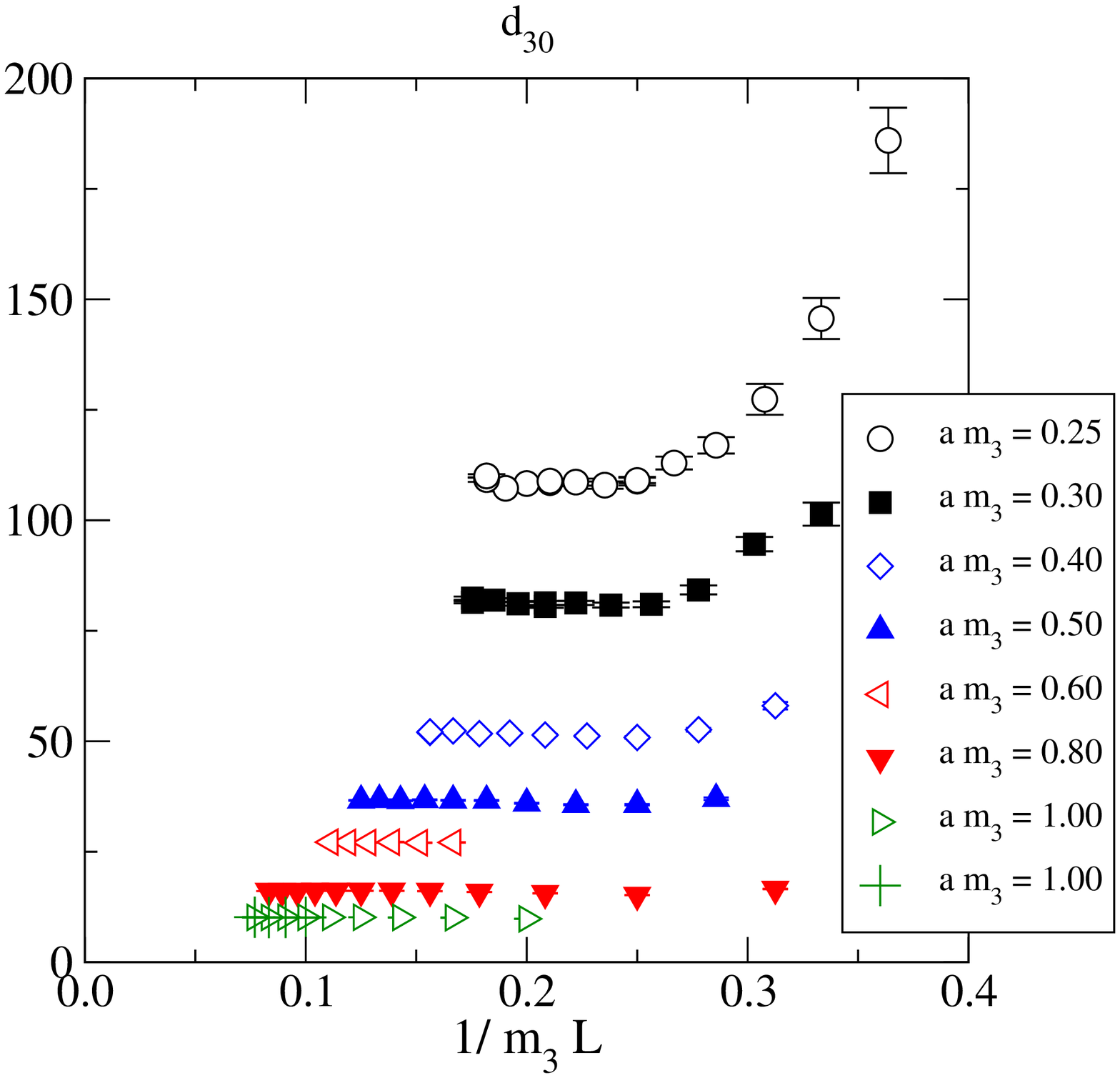}%
}

\vspace*{0.5cm}

\centerline{%
\epsfysize=5.0cm\epsfbox{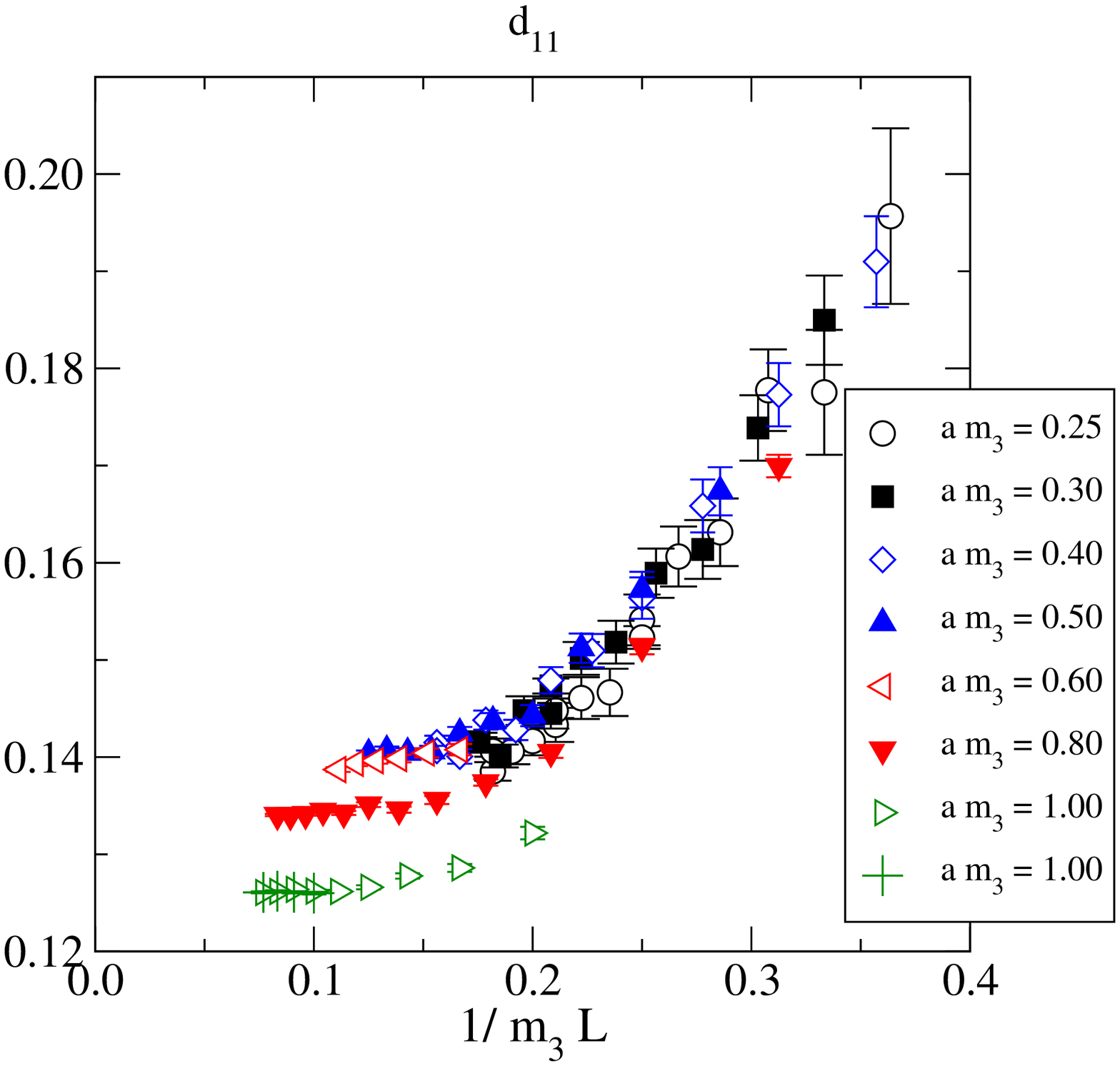}%
~~\epsfysize=5.0cm\epsfbox{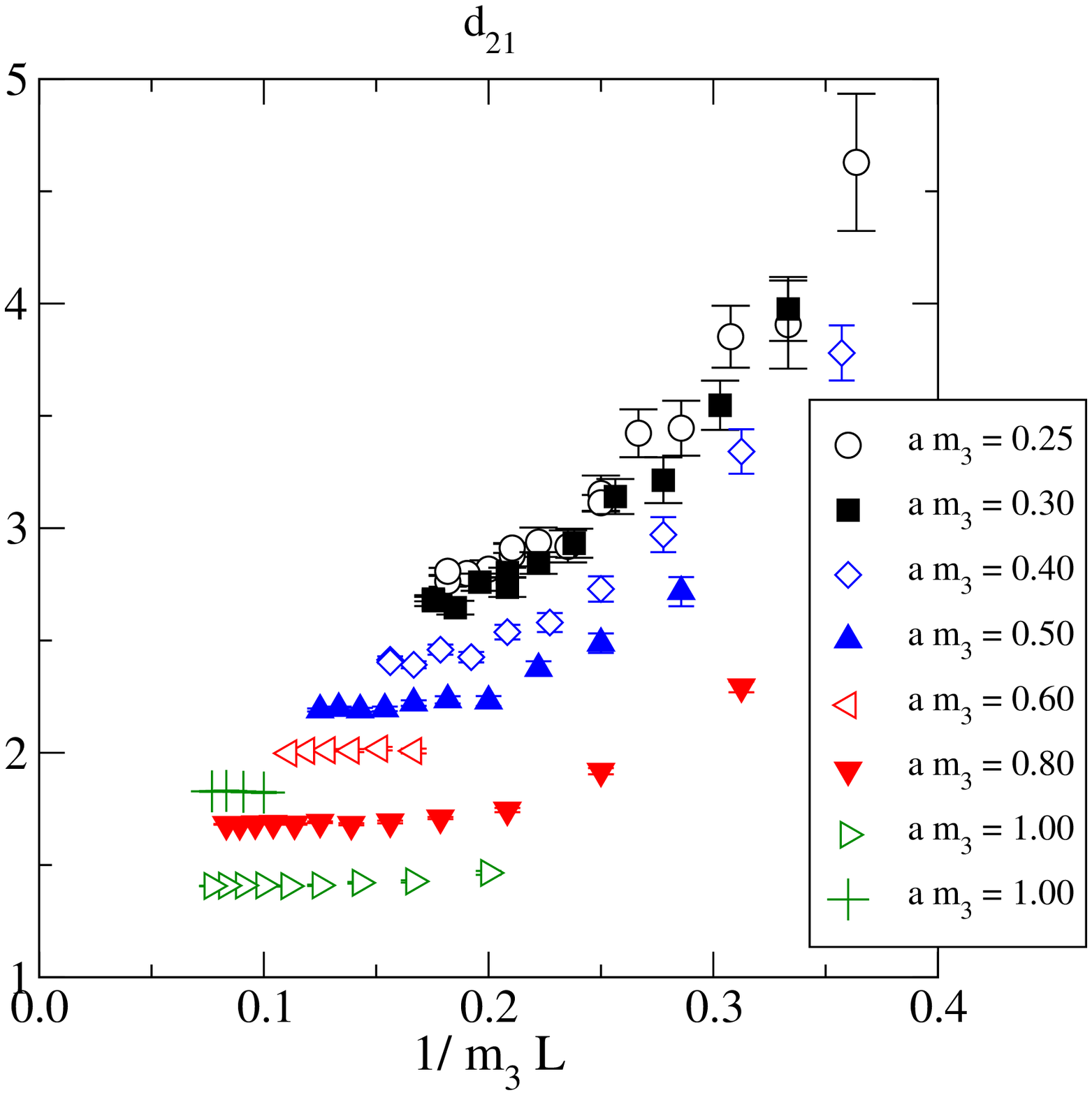}%
~~\epsfysize=5.0cm\epsfbox{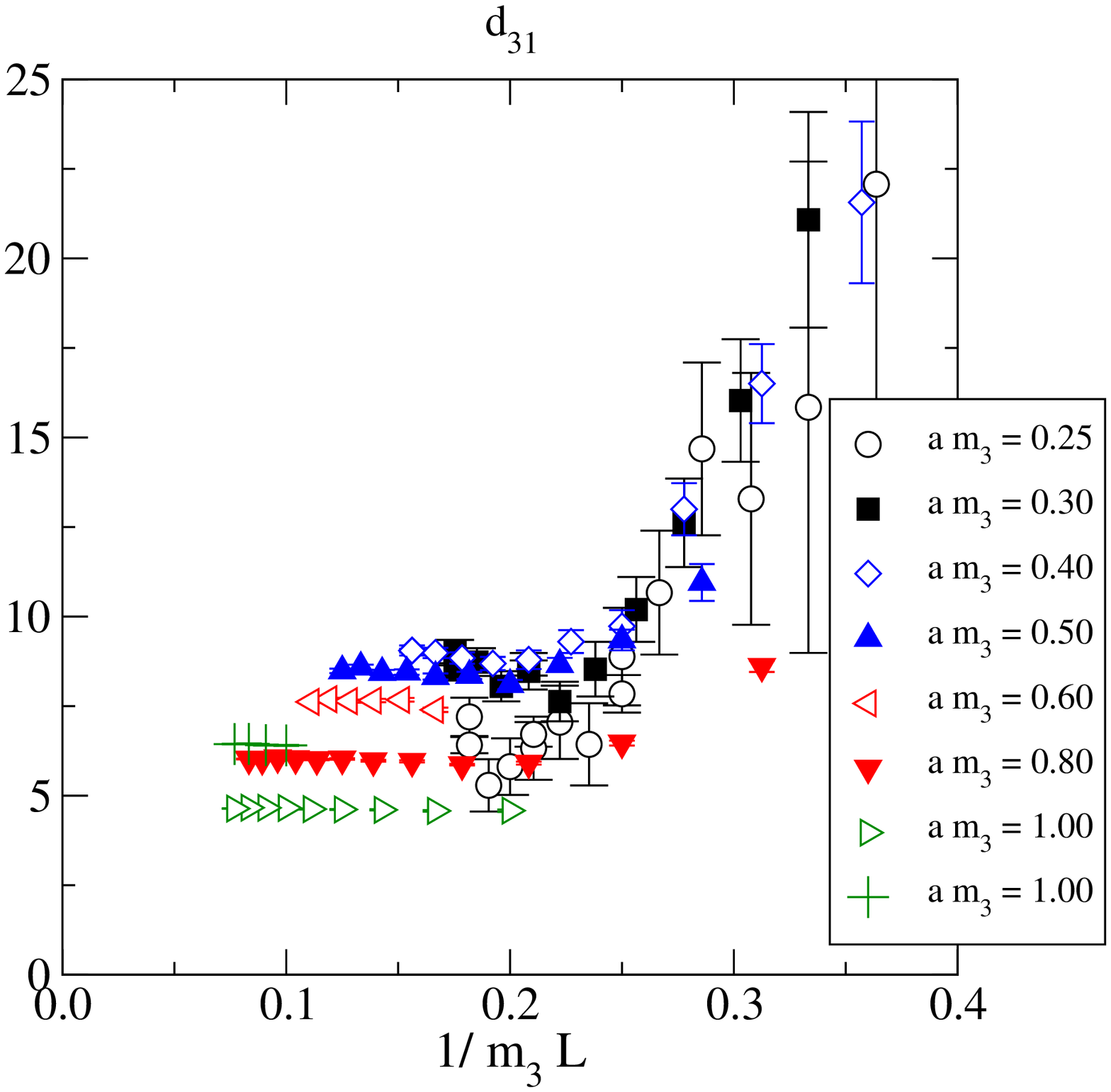}%
}

\vspace*{0.5cm}

\centerline{%
\epsfysize=5.0cm\epsfbox{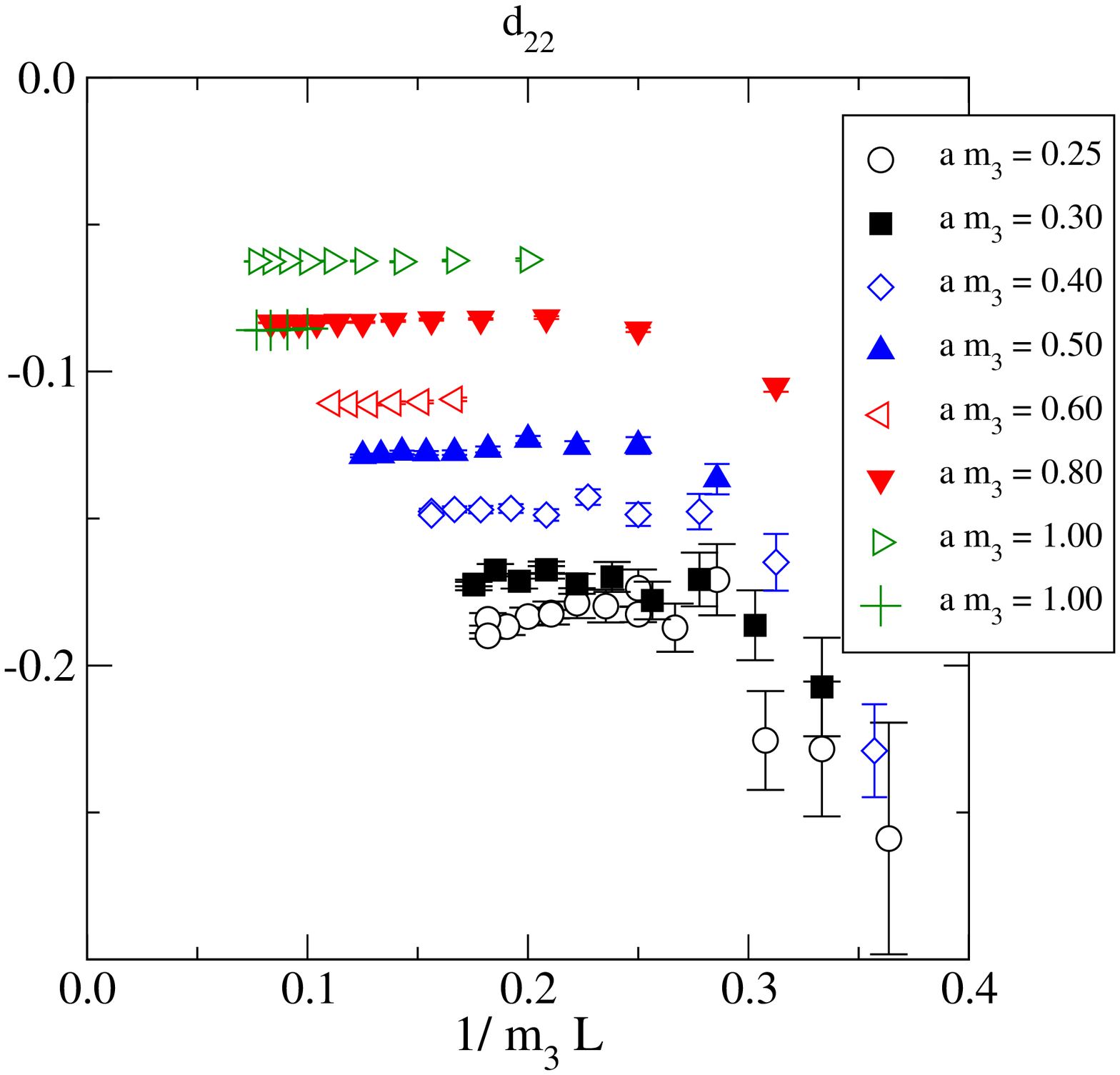}%
~~~~\epsfysize=5.0cm\epsfbox{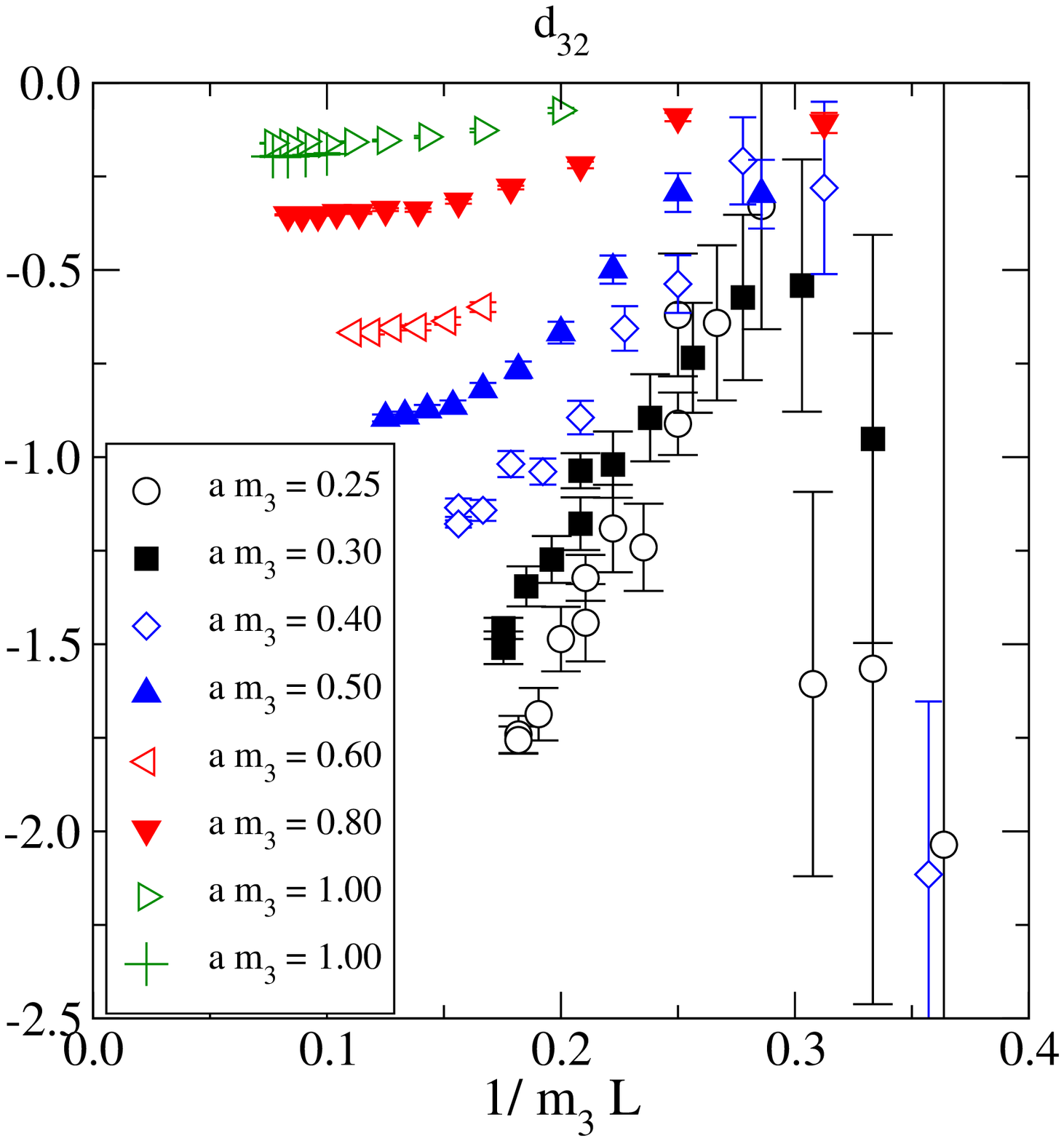}%
~~~~\epsfysize=5.0cm\epsfbox{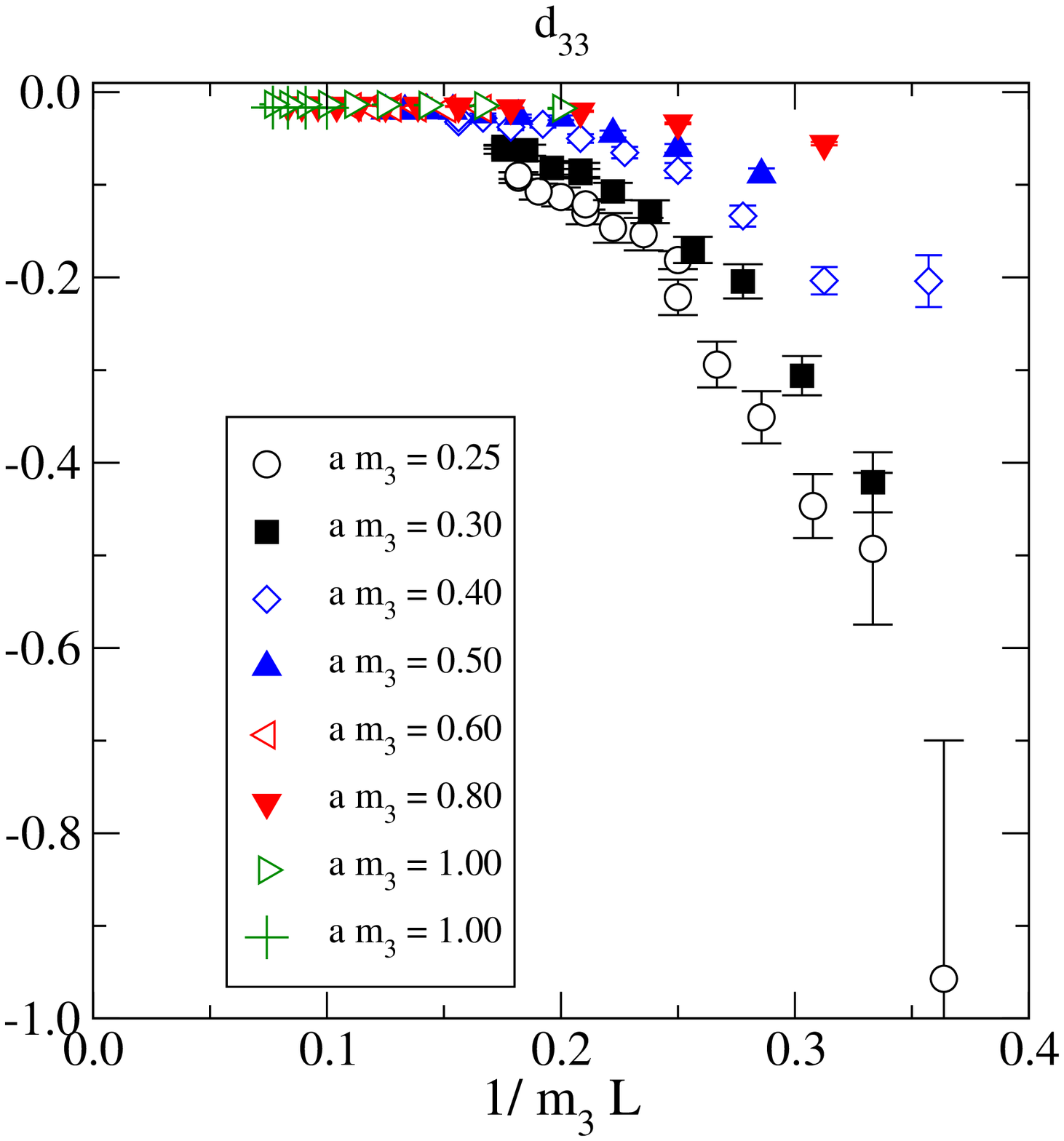}%
~~}

\vspace*{0.5cm}


\caption[a]{Shown are the $\tau\to 0$ limits for the coefficients $d_{ij}$, 
as a function of the inverse box size 
in ``physical'' units ($m_3 L = a m_3 N$). 
The data for $a m_3 = 1.0$ indicated with the pluses ($+$) is with
$\ln\beta = \ln 80$; the other cases are with 
$\ln \beta = \ln 24$.}

\la{fig:dij_all}
\end{figure}

Given the results of the $\tau\to 0$ extrapolations
(the complete data set, apart from $d_{00}$, 
is shown in \fig\ref{fig:dij_all}), 
the next step is to extrapolate to infinite volume.

As \fig\ref{fig:dij_all} shows, finite-volume effects 
become small at large volumes. However, the box size $N=L/a$ 
required for this grows as the mass $a m_3$ decreases
(the behaviour is more or less universal as a function of $m_3 L$). 
In addition, some of the coefficients appear to require larger
volumes than others. For the smallest masses, in particular, 
we are in many cases not yet in a region where all volume
dependence has died out. 

To be able to deal with this situation, some theoretical
knowledge about the functional dependence on the finite volume
is needed. The situation is complicated by the fact that
there are both massless and massive fields in the system; therefore
both powerlike and exponential volume dependences appear. 
However, an inspection of the known cases 
(\figs\ref{fig:precise}, \ref{fig:precise_2}) suggests
than in practice the magnitude of the power corrections is much
smaller than that of the exponential ones so that, strangely enough, 
the latter dominate in the volumes where our data lies. In this situation, 
we could then expect the dominant volume behaviour to be some exponential, 
$d_{ij}(m_3 L) - d_{ij}(\infty) \sim \exp(-m_3 L)/(m_3 L)^\alpha$.
Unfortunately, an inspection of some of the known cases
(particularly $d_{22}$) shows that, again because
of the fairly small volumes reached in practice, the behaviour
is not given by a simple exponential, but that there are at 
least two competing universal functions, 
because the position of the maximum in $m_3 L$ evolves
slightly with $a m_3$. At the same, allowing for too 
many free functional forms in the fits does not lead to good results
either, because a long extrapolation is needed, so that the ansatz 
needs to be fairly constrained. 

Having tested many procedures in the cases where exact results
are available, we have finally chosen the following strategy 
in order to deal with these challenges.
Let us consider a mass like $a m_3 = 0.8$. 
\fig\ref{fig:dij_all} shows that
this one has a reasonable plateau at affordable $N \le 15$ 
for all the coefficients $d_{ij}$,
but still the behaviour of the data 
is not too flat (i.e., some volume dependence is detectable).
One can then extract an infinite-volume value $d_{ij}(\infty)$ 
by fitting a constant to data in the range of the plateau, and
subtract it from the data in order to obtain the quantities 
\be
 g_{ij}(m_3L) \equiv d_{ij}(m_3L) - d_{ij}(\infty)
 \;. \la{gij}
\ee
Subsequently, one can try to obtain a reasonable interpolating 
fit $f_{ij}(m_3 L)$ for $g_{ij}(m_3 L)$, allowing to go also
to other values of $m_3 L$ than those simulated at $a m_3 = 0.8$.
In practice, we find that in the range $m_3 L > 2.5$ that 
we have considered (cf.\ Table~\ref{table:stats}), 
our data for $a m_3 = 0.8$ can be 
well modelled, for instance, by the ansatz
\be
 f_{ij}(x)
 \equiv 
 e^{-x} \biggl[ 
 \gamma_{ij}^{(1)} + 
 \gamma_{ij}^{(2)} \frac{1}{x} + 
 \gamma_{ij}^{(3)} \frac{1}{x^2} 
 \biggr]
 \;, \quad x = m_3 L \;, \la{fij}
\ee
where $\gamma_{ij}^{(n)}$ are fit parameters.
We have also experimented with other ans\"atze, but do not 
find a significant effect on our final results. 

After this empirical determination of the finite-volume effects
in one well-controlled case, we can go back to the other masses 
$a m'_3$, and use the fits $f_{ij}$ as a constrained ansatz. 
However, as already mentioned, there are cases, such as $d_{22}$, 
where the position of the maximum evolves with $am_3$; therefore
the results cannot be described by one universal function in 
our modest volumes. To incorporate this fact, 
we allow \eq\nr{fij} to in general split up into two functions, 
and take a finite-size scaling ansatz of the form
\be
 d_{ij}(x') = d_{ij}(\infty) + A_{ij}(a m'_3 ) \times 
 e^{-x'} 
 \Bigl[ \gamma_{ij}^{(1)} \Bigr] +
 B_{ij}(a m_3')  \times
 e^{-x'} 
 \biggl[  \gamma_{ij}^{(2)} \frac{1}{x'} + 
 \gamma_{ij}^{(3)} \frac{1}{(x')^2} 
 \biggr]
 \;, \la{dij_prime}
\ee
where $x' \equiv m_3' L$; 
$d_{ij}(x')$ are the direct measurements at the mass $a m'_3$
for various $N = L/a$; 
and $d_{ij}(\infty)$, $A_{ij}(a m'_3)$ and 
$B_{ij}(a m'_3)$ are volume-independent fit coefficients. 
Among the exactly known coefficients, 
the only case where the results change significantly
while going from \eq\nr{fij} to the more general \eq\nr{dij_prime}
is precisely $d_{22}(\infty)$; among the unknown ones, the ansatz does
systematically affect also $d_{20}(\infty)$, 
and particularly $d_{30}(\infty)$, 
at the smallest masses. For instance, in the last case, 
the values of $d_{30}(\infty)$ would 
be as much as $\sim 10\sigma$ higher at the two smallest masses
if we employed \eq\nr{fij} throughout. In the following, 
we cite results based on \eq\nr{dij_prime}, 
for reasons now to be explained. 

In \figs\ref{fig:precise}, \ref{fig:precise_2}, the results of 
such fits are compared with the exact results at the smallest masses. 
We do find compatibility within statistical errors ($\sim 2\sigma$) in all 
cases. The results for the coefficients $d_{ij} \equiv d_{ij}(\infty)$, 
both fitted and exact, are given in Table~\ref{table:dij} for all masses. 
Only one 
``failure'' can be detected, namely the coefficient $d_{10}$ at the 
largest masses $a m_3 = 0.80, 1.00$, where our (very small) NSPT error bars
appear to be underestimated (the difference is $\sim 10\sigma$).
Given that the largest masses are the least important ones for 
the subsequent steps, we have decided to let this problem ``pass''; 
let us stress, in any case, that the overall excellent agreement is 
a very non-trivial result, as a long extrapolation needs to be 
carried out, and could only be achieved after a considerable amount 
of experimenting with various procedures. 

Encouraged by these tests, as well as by the indication
in \figs\ref{fig:dij_tau}, \ref{fig:dij_all} that $d_{20}$, $d_{30}$
could more or less behave like $d_{10}$; $d_{31}$ like $d_{21}$ and 
$d_{11}$; and $d_{32}$ like $d_{22}$; we then apply the same procedure
to the remaining coefficients.  The results 
are given in the lower-most panel in Table~\ref{table:dij}
(errors are statistical only).
Finally, all the results, but with the normalization 
of \se\ref{se:setup}, are shown in \fig\ref{fig:phiij}
(the exactly known values of 
$\tilde \phi_{21}$,
$\tilde \phi_{22}$,
$\tilde \phi_{31}$,
$\tilde \phi_{32}$,
$\tilde \phi_{33}$ have been used as input to convert
$d_{21}$, 
$d_{22}$, 
$d_{31}$, 
$d_{32}$, 
$d_{33}$ to 
$\phi_{21}$, 
$\phi_{22}$, 
$\phi_{31}$, 
$\phi_{32}$, 
$\phi_{33}$, respectively; 
cf.\ \eqs\nr{def:d21}--%
\nr{def:d33}.
The functions  
$\tilde \phi_{21}$,
$\tilde \phi_{22}$,
$\tilde \phi_{31}$,
$\tilde \phi_{32}$,
$\tilde \phi_{33}$
have been numerically crosschecked only at $am_3 = 1.00$, 
where simulations with two different $\ln\beta$ were carried out; 
cf.\ Table~\ref{table:stats}).

%
\begin{table}[t]
\small

\begin{center}

exact

\vspace*{3mm}

\begin{tabular}{llllll}  \hline \\[-2mm]
 $a m_3$ & ~~$d_{00}$ & ~~$d_{10}$ & ~~$d_{11}$ & ~~$d_{21}$ & ~~$d_{22}$ 
\\[2mm] \hline \\[-1mm]
0.25 & 
0.92893 &
2.9249 & 
0.13737 & 
2.7649 &
-0.19780  \\
0.30 & 
0.91214 &
2.7230 &
0.13855 &
2.6428 &
-0.17972  \\
0.40 & 
0.87839 & 
2.3909 & 
0.13998 & 
2.4058 &
-0.15127  \\
0.50 & 
0.84462 &
2.1210  & 
0.14017 &
2.1949 &
-0.12936 \\
0.60 & 
0.81103 & 
1.8924  & 
0.13919 & 
2.0074 & 
-0.11150 \\
0.80 & 
0.74518  & 
1.5201  & 
0.13422 & 
1.6830 & 
-0.083553  \\
1.00 & 
0.68209 &
1.2285  & 
0.12612 & 
1.4072 & 
-0.062617 \\
1.00$^*$ &  
0.68209 &
1.2285  & 
0.12612 & 
1.8283 & 
-0.086011  \\[2mm] \hline
\end{tabular}

\vspace*{5mm}

fitted

\vspace*{3mm}

\begin{tabular}{llllll}  \hline \\[-2mm]
 $a m_3$ & ~~$d_{00}$ & ~~$d_{10}$ & ~~$d_{11}$ & ~~$d_{21}$ & ~~$d_{22}$ 
\\[2mm] \hline \\[-1mm]
0.25 & 
       0.9299(6) &
       2.923(3) & 
       0.1362(6) &
       2.74(4) &
      -0.199(2)   \\
0.30 & 
     0.9110(7) & 
       2.720(3) &
       0.1384(5) &
       2.63(3) &
      -0.177(2)  \\
0.40 & 
      0.8782(4) &
       2.392(2) &
       0.1392(3) &
       2.39(1) &
      -0.1509(7)  \\
0.50 & 
      0.8451(3) &
       2.122(1) &
       0.1403(4) &
       2.191(6) &
      -0.1292(4)  \\
0.60 & 
       0.8113(5) &
       1.893(1) &
       0.1384(4) &
       1.994(6) &
      -0.1110(4)  \\
0.80 & 
       0.7455(1) &
       1.5223(2) &
       0.13414(7) &
       1.683(1) &
      -0.08345(7)  \\
1.00 & 
      0.6823(1) &
       1.2309(2) &
       0.12614(5) &
       1.4077(7) &
      -0.06255(5)  \\
1.00$^*$ &  
       0.6822(3) &
       1.230(1) &
       0.1261(1) &
       1.829(3) &
      -0.08596(16)  \\[2mm] \hline
\end{tabular}

\vspace*{5mm}

fitted

\vspace*{3mm}

\begin{tabular}{llllll} \hline \\[-2mm]
 $a m_3$ &  ~~$d_{20}$ & ~~$d_{30}$ & ~~$d_{31}$ & ~~$d_{32}$ & ~~$d_{33}$ 
\\[2mm] \hline \\[-1mm]
0.25 &
      16.49(3) & 
     109.4(4) &
       7.2(7) &
      -2.16(6) &
      -0.060(19)  \\
0.30 &    
      14.09(3) &
      81.8(3) &
       9.6(3) &
      -1.73(5) &
      -0.038(10)  \\
0.40  &
     10.87(1) & 
      52.2(1) &
       9.4(1) &
      -1.27(1) &
      -0.029(3)  \\
0.50  &
       8.74(1) &
      36.7(1) &
       8.61(5) & 
      -0.916(7) &
      -0.0201(7) \\
0.60  &
      7.18(1) &
      27.2(1) &
       7.59(5) &
      -0.670(6) &
      -0.0155(5) \\
0.80  &
       5.058(2) &
      16.11(2) &
       6.035(7) &
      -0.3522(9) &
      -0.01423(6) \\
1.00  &
       3.664(1) & 
      10.16(1) &
       4.652(4) & 
      -0.1605(5) &
      -0.01327(3)  \\
1.00$^*$  &
       3.666(6) &
      10.19(6) &
       6.46(2) &
      -0.196(3) &
      -0.01677(11) \\[2mm] \hline
\end{tabular}
\end{center}
\normalsize

\caption[a]{The coefficients $d_{ij}$
(cf.\ \eq\nr{Phi_exp}) in the infinite-volume limit. 
The starred mass refers to $\ln\beta = \ln 80$; 
the others to $\ln \beta = \ln 24$. The numbers in 
parentheses indicate the statistical errors of the last
digits shown. 
}

\la{table:dij}
\end{table}
%

%
\subsection{Interpolation in $a m_3$}

\begin{figure}[p]


\centerline{%
\epsfysize=5.0cm\epsfbox{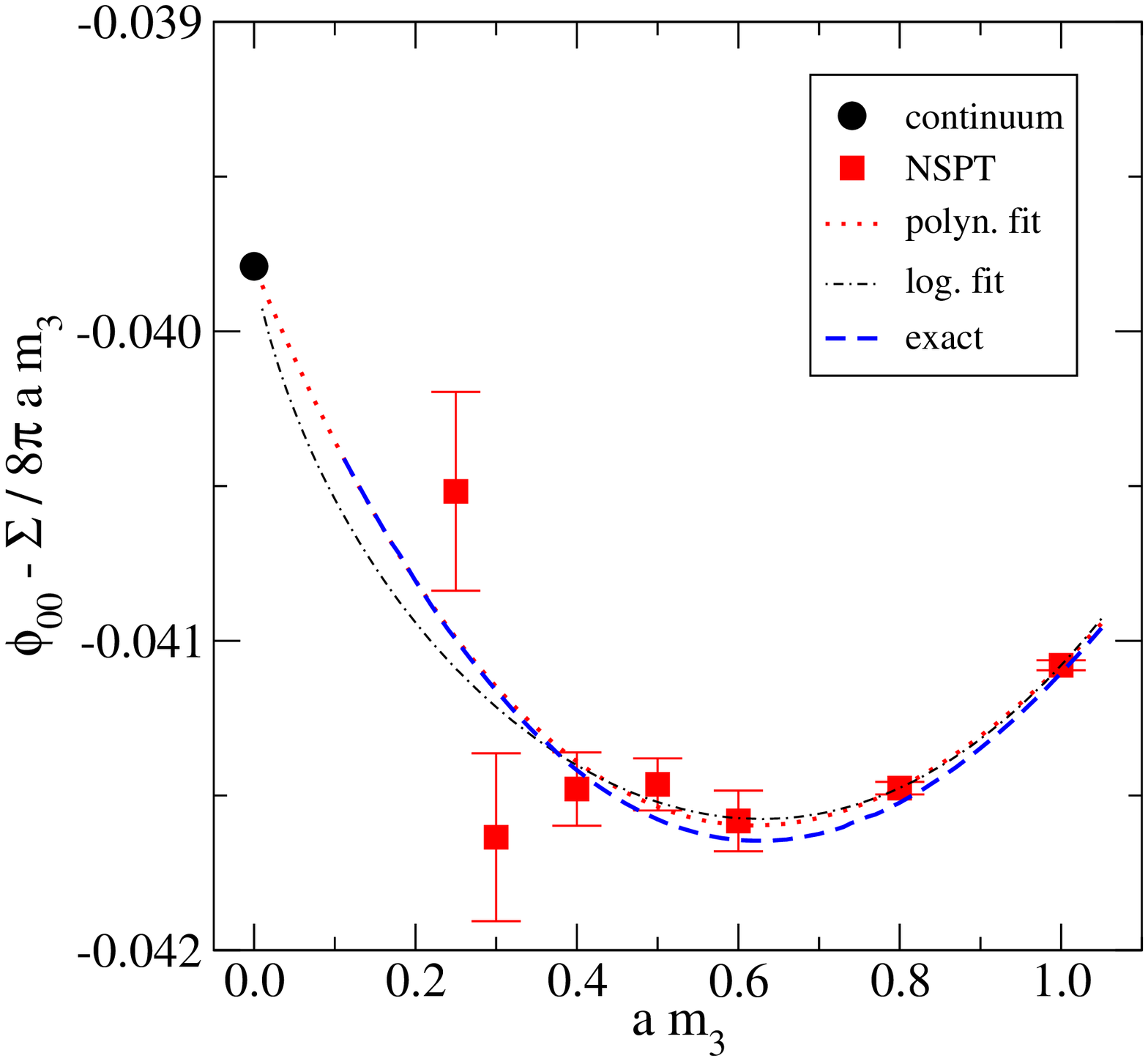}%
~~\epsfysize=5.0cm\epsfbox{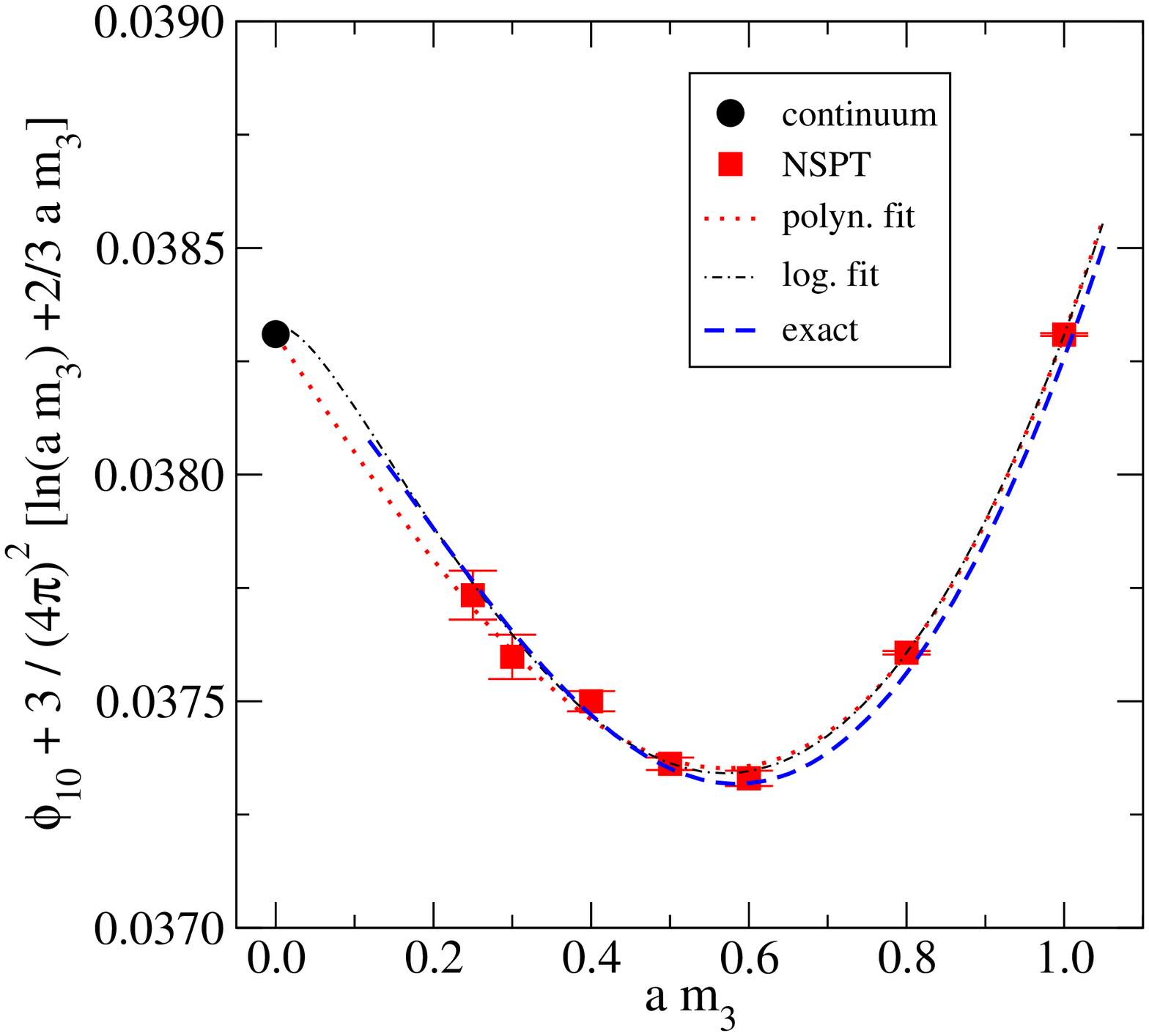}%
~~\epsfysize=5.0cm\epsfbox{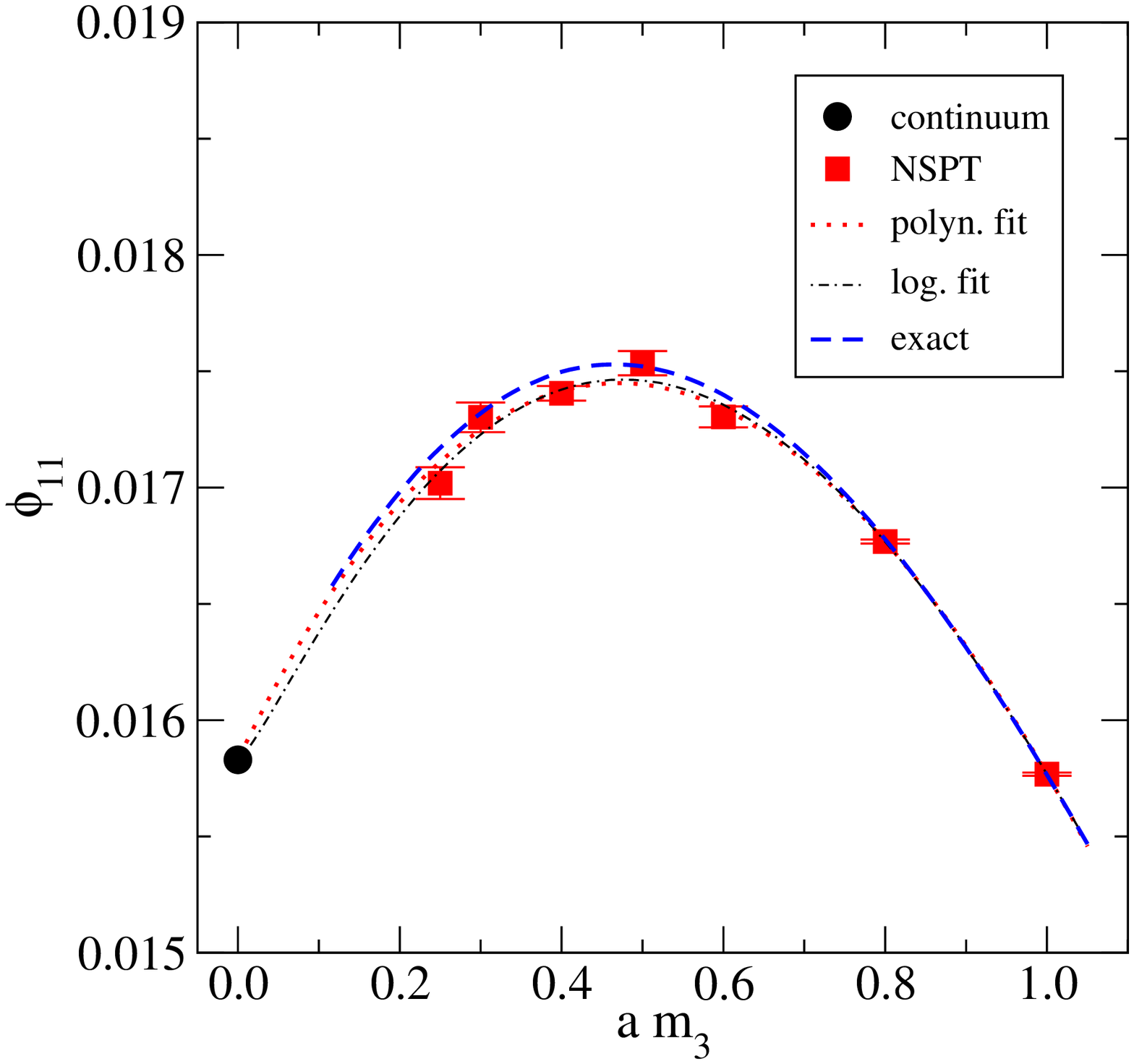}%
}

\vspace*{0.5cm}

\centerline{%
\epsfysize=5.0cm\epsfbox{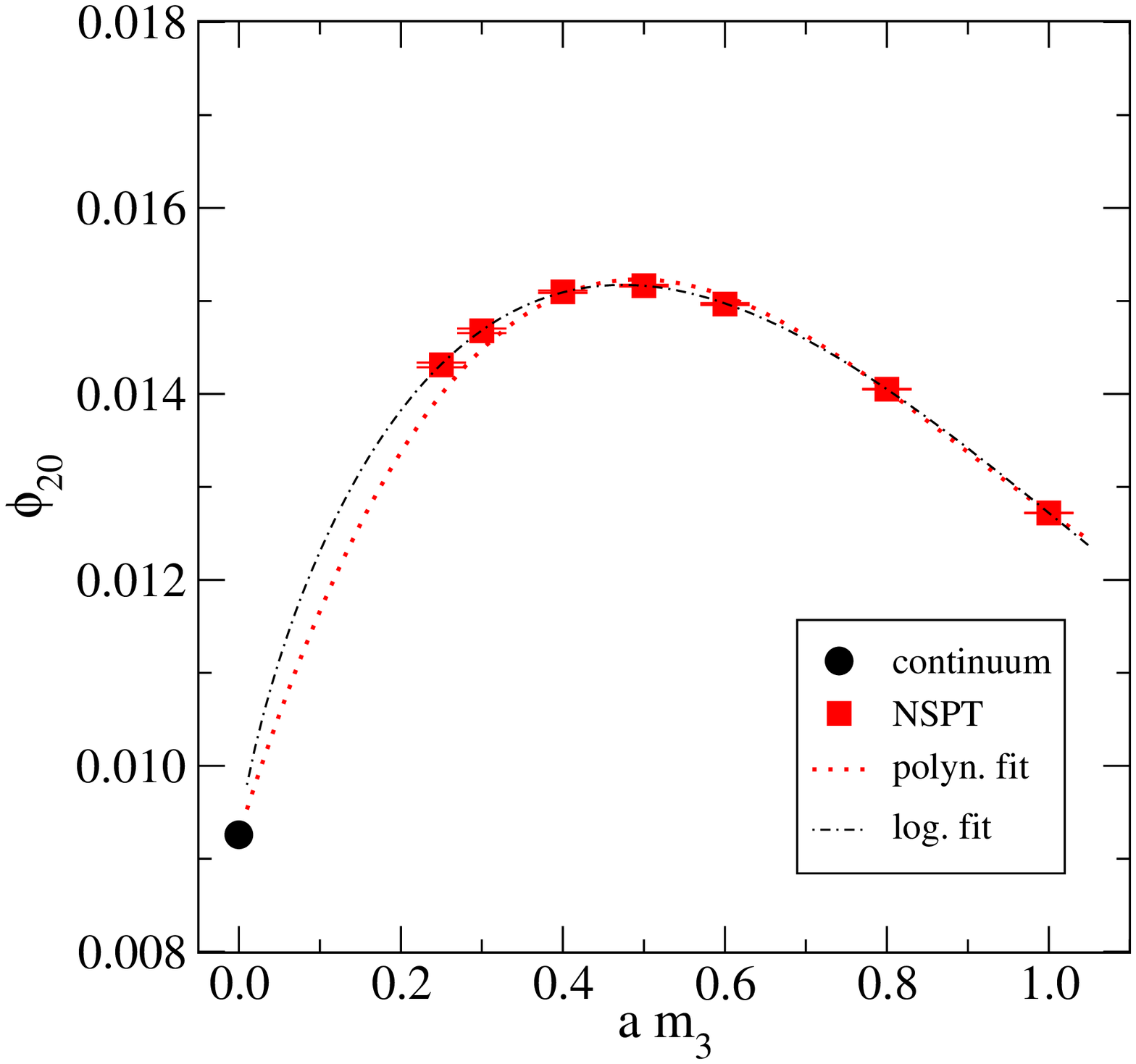}%
~~\epsfysize=5.0cm\epsfbox{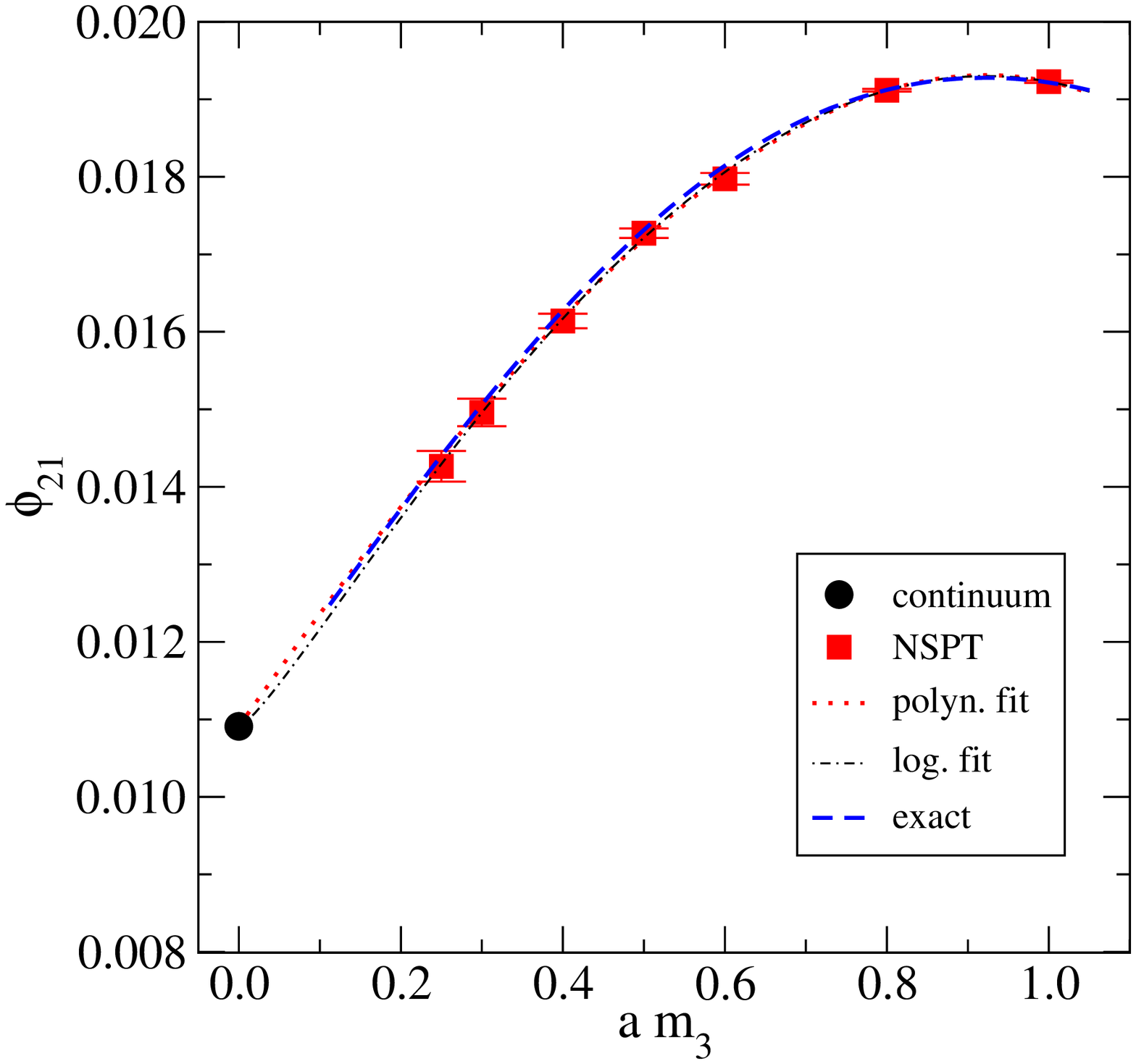}%
~~\epsfysize=5.0cm\epsfbox{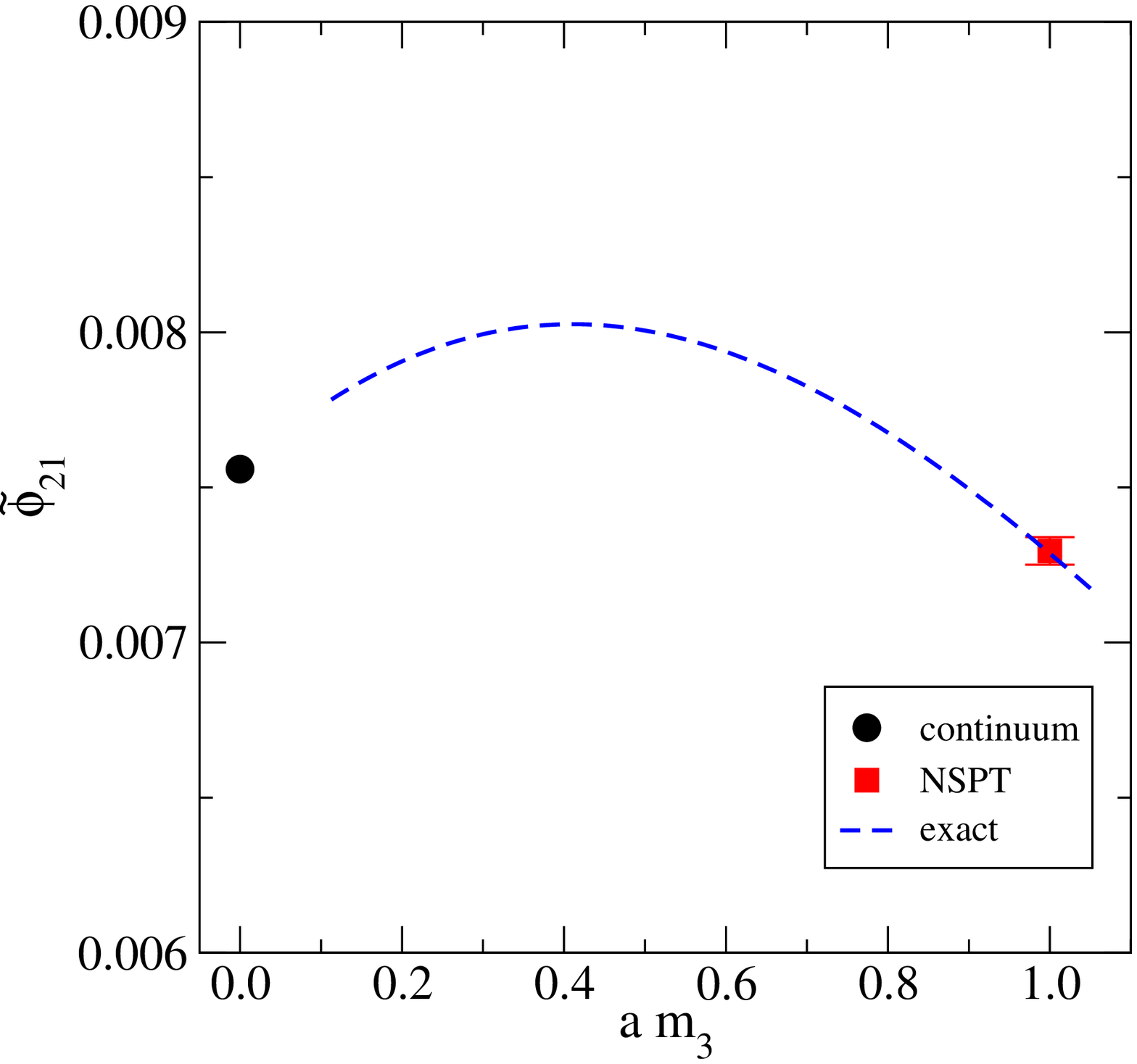}%
}

\vspace*{0.5cm}

\centerline{%
\epsfysize=5.0cm\epsfbox{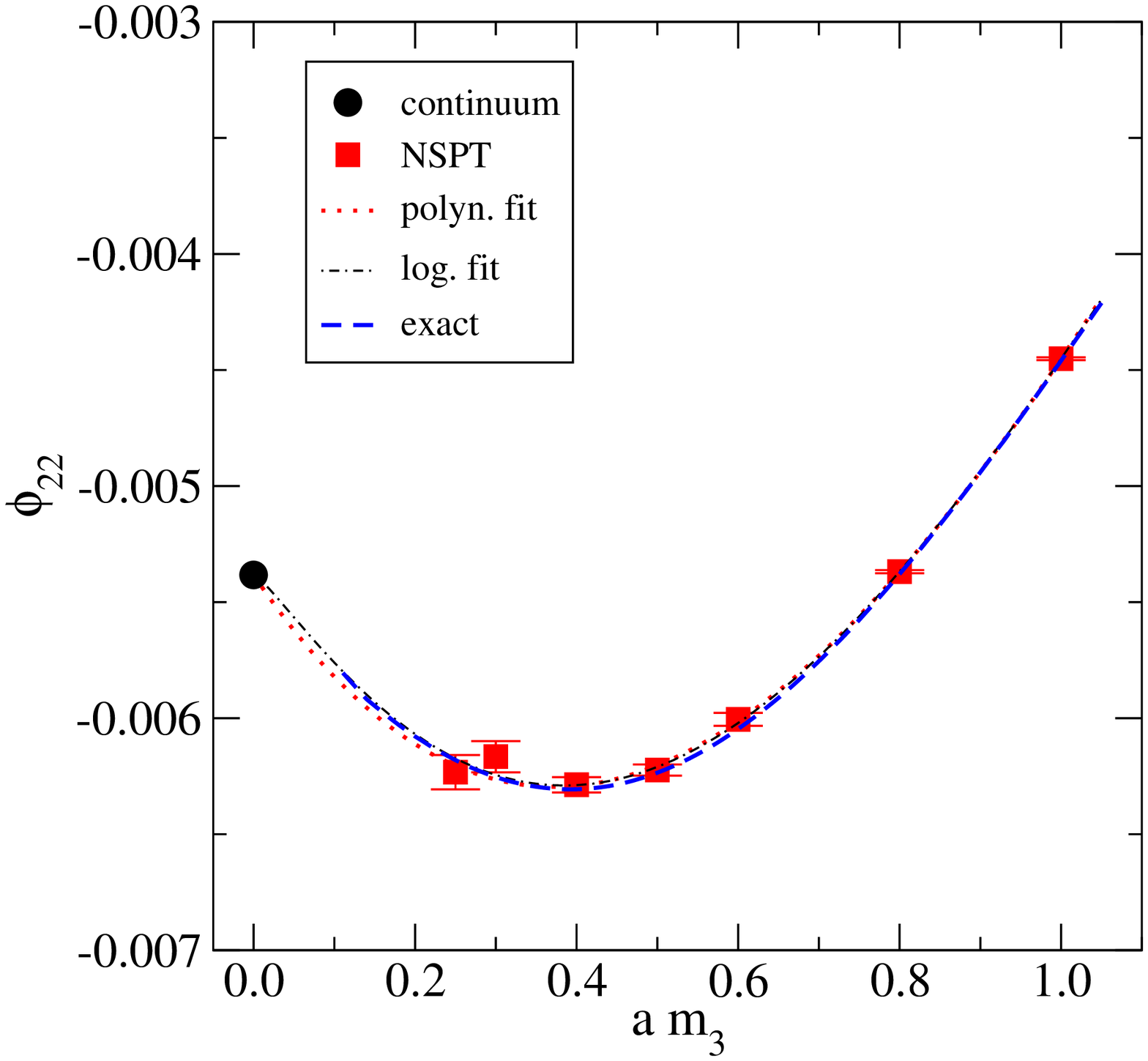}%
~~\epsfysize=5.0cm\epsfbox{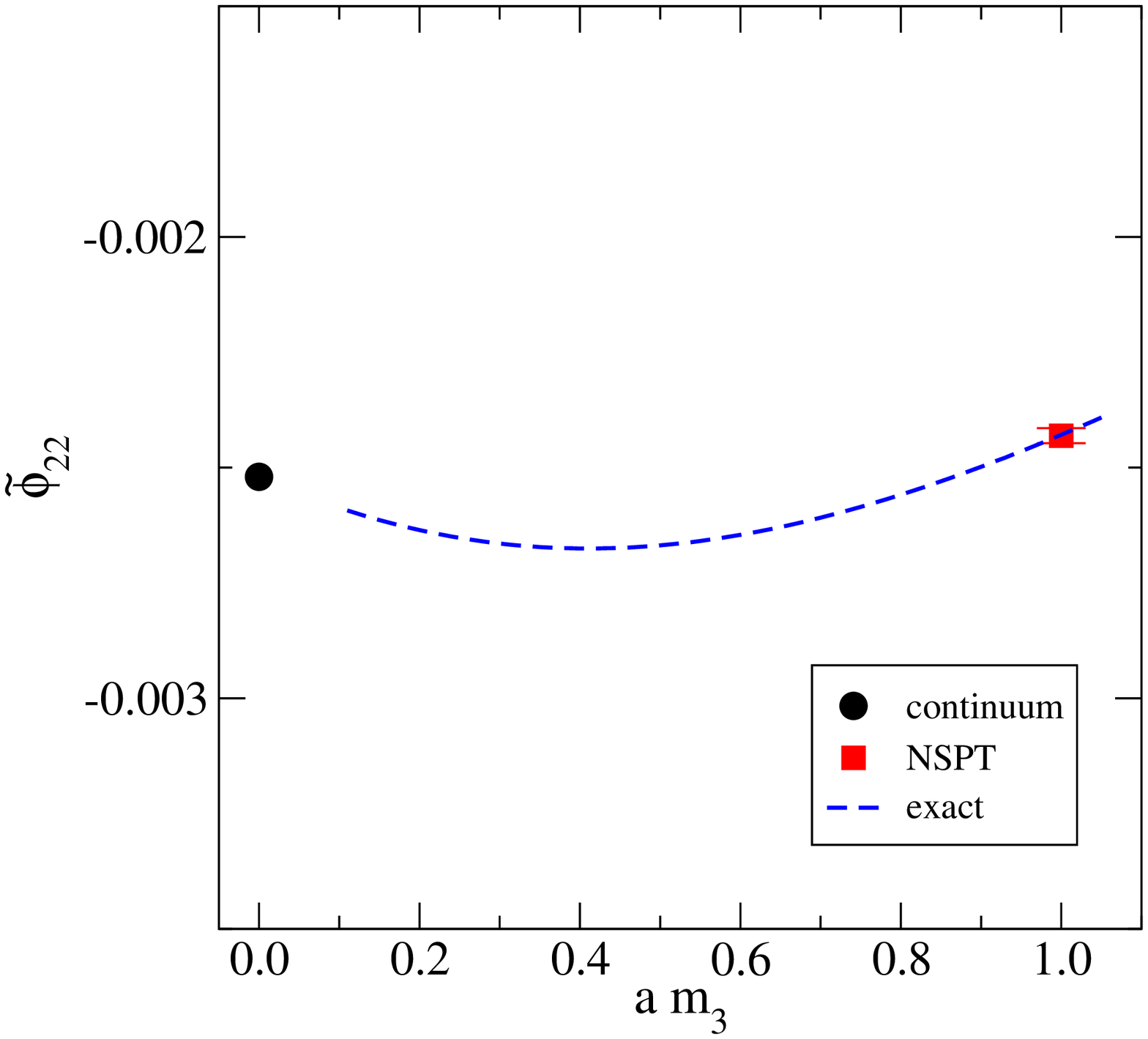}%
~~\epsfysize=5.0cm\epsfbox{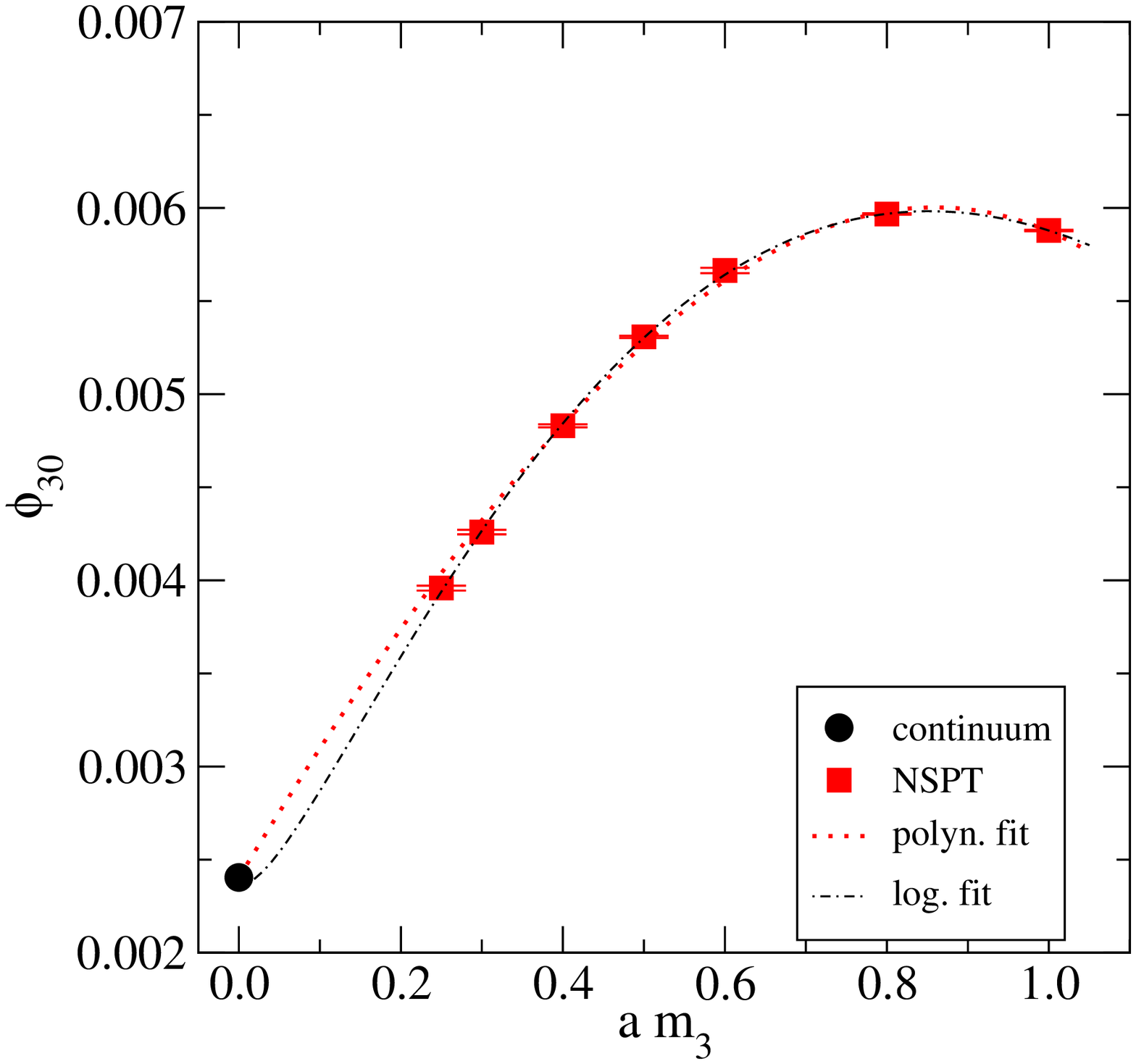}%
}

\vspace*{0.5cm}

\caption[a]{Shown are the coefficients 
in \eq\nr{phi}, as a function of $a m_3$. 
The polynomial fits are of third order in $a m_3$, 
and have been constrained to go through the continuum points; 
the logarithmic fits include the additional term $a m_3 \ln (1/a m_3)$, 
cf.\ \eq\nr{log_fit}.
}

\la{fig:phiij}
\end{figure}

\addtocounter{figure}{-1}

\begin{figure}[p]

\centerline{%
\epsfysize=5.0cm\epsfbox{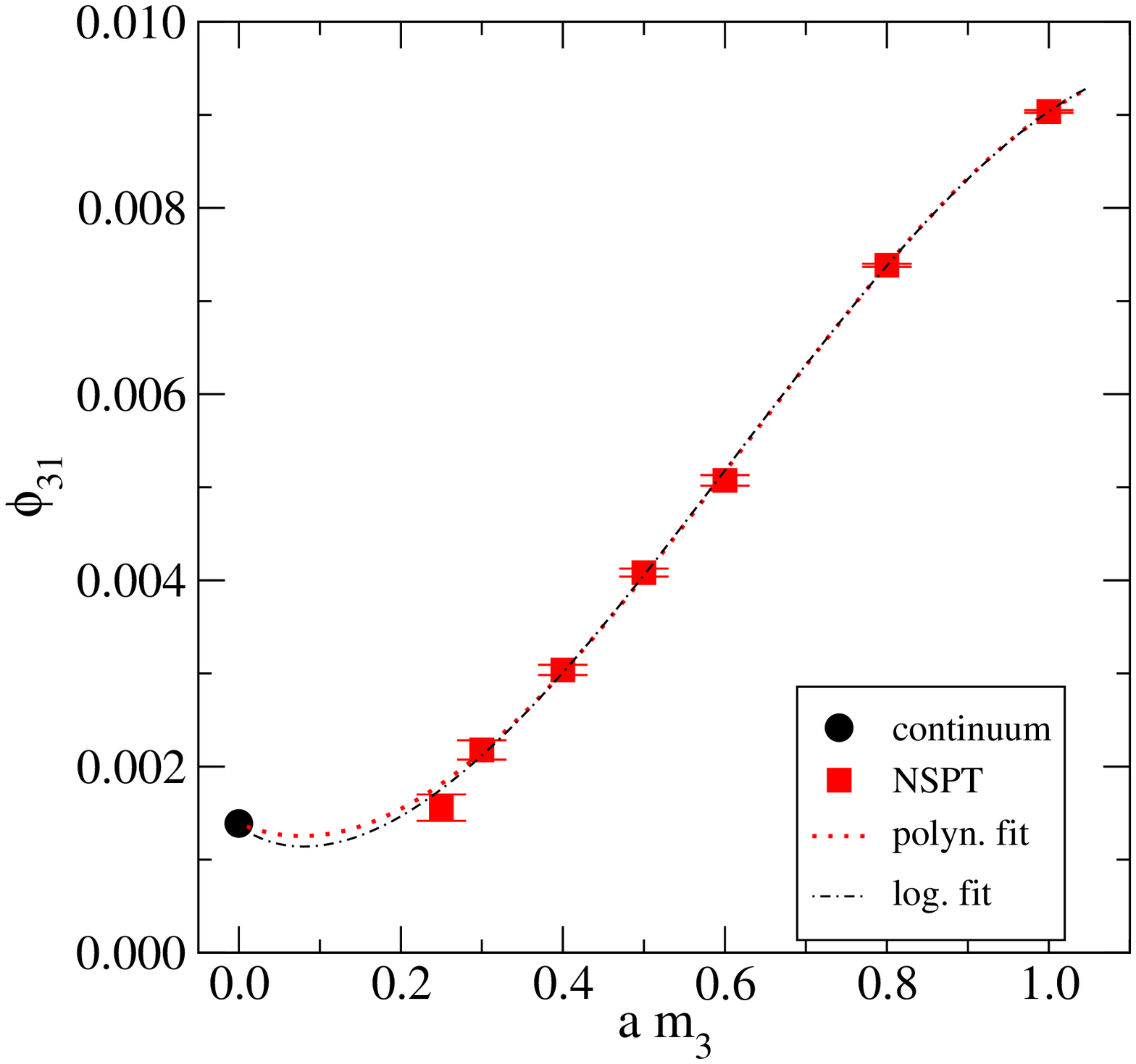}%
~~\epsfysize=5.0cm\epsfbox{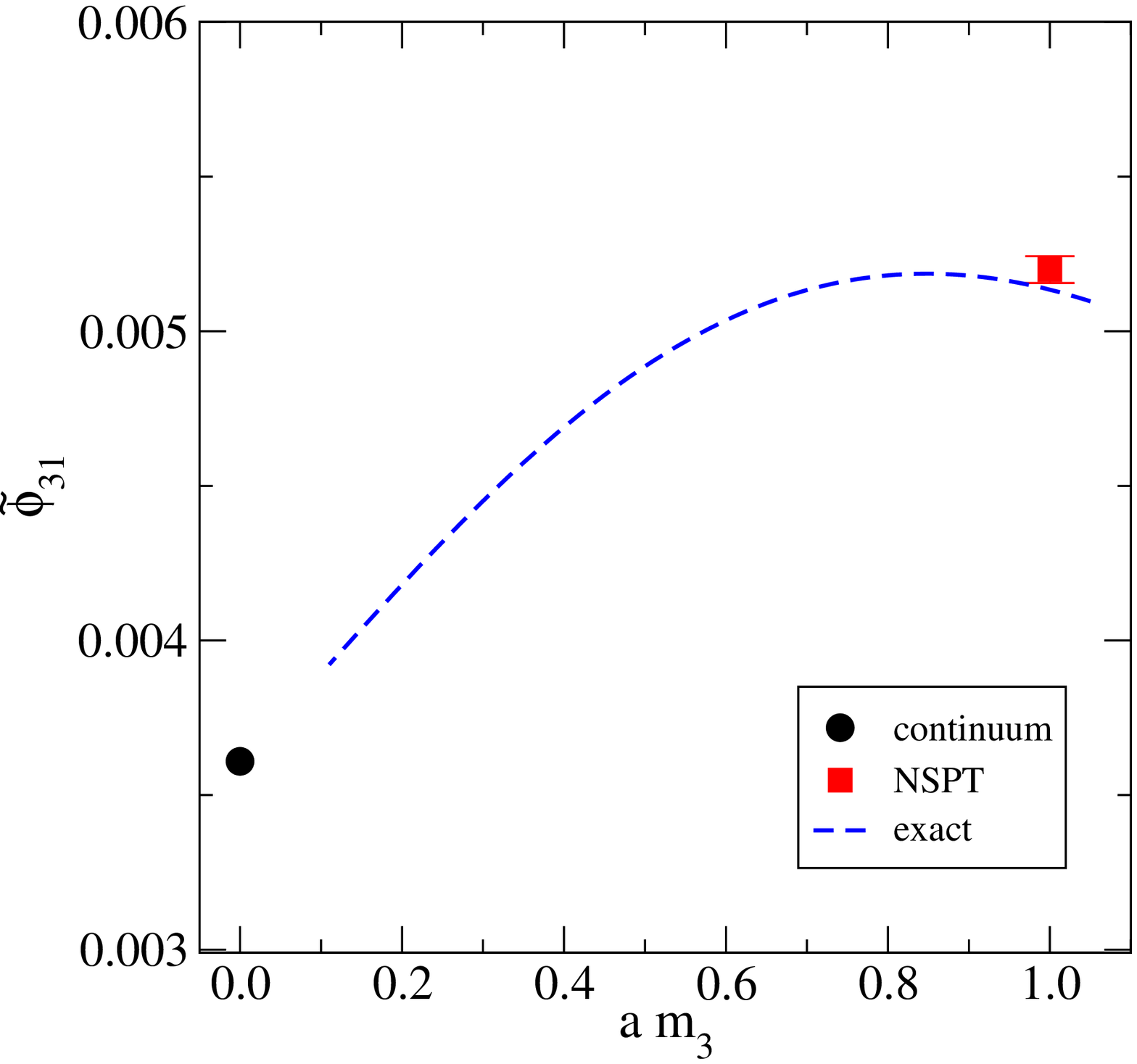}%
~~\epsfysize=5.0cm\epsfbox{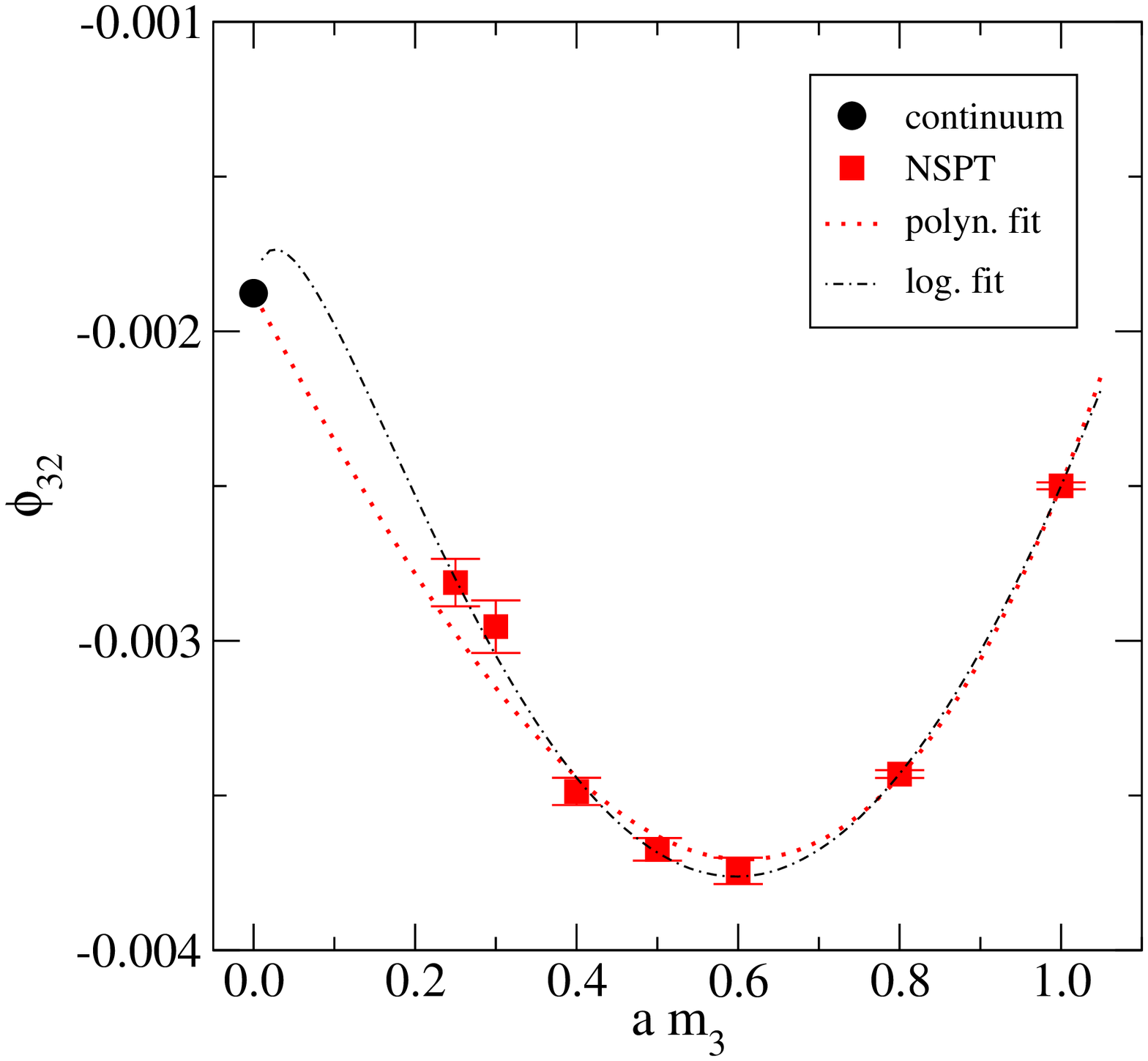}%
}

\vspace*{0.5cm}

\centerline{%
\epsfysize=5.0cm\epsfbox{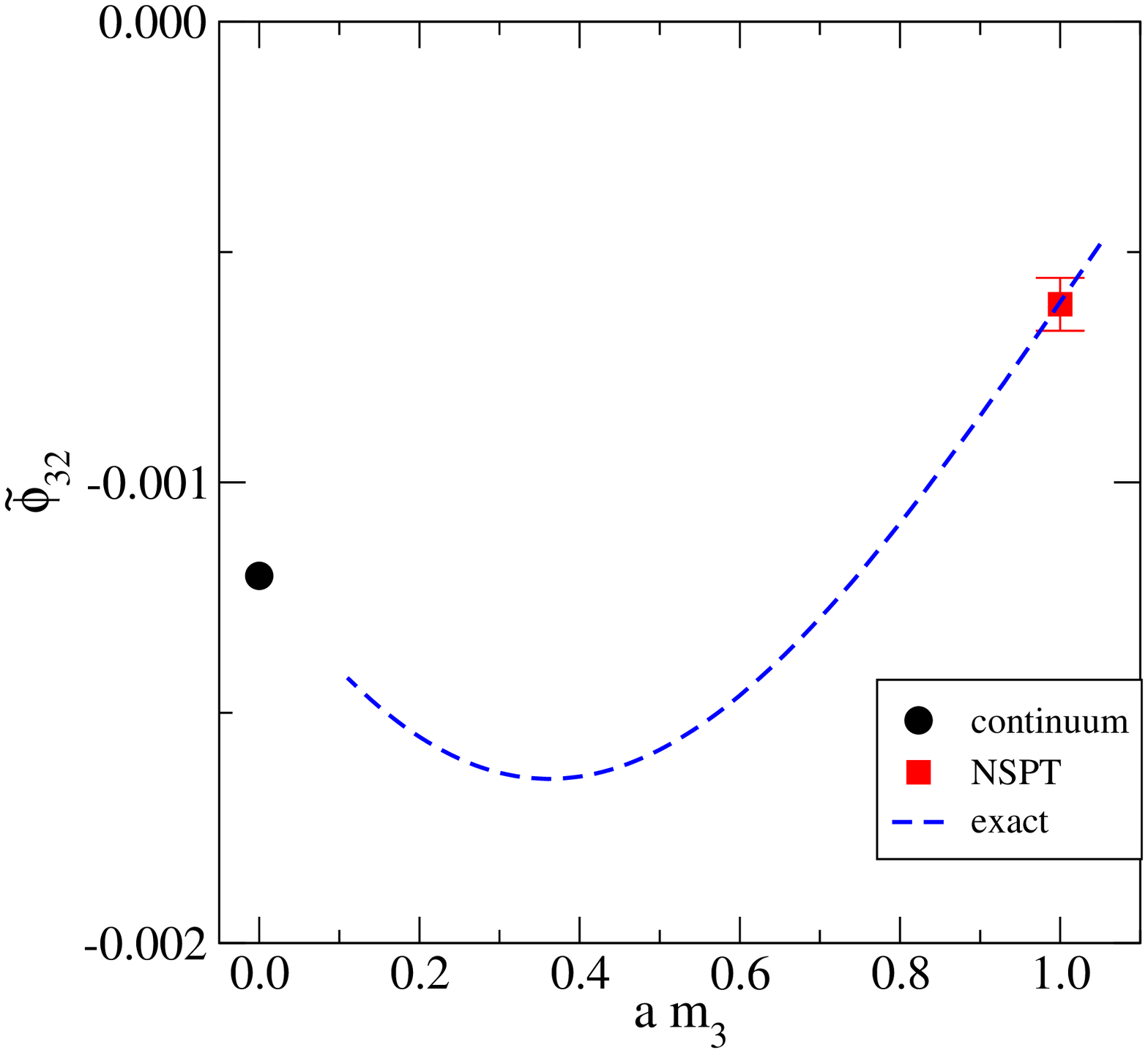}%
~~\epsfysize=5.0cm\epsfbox{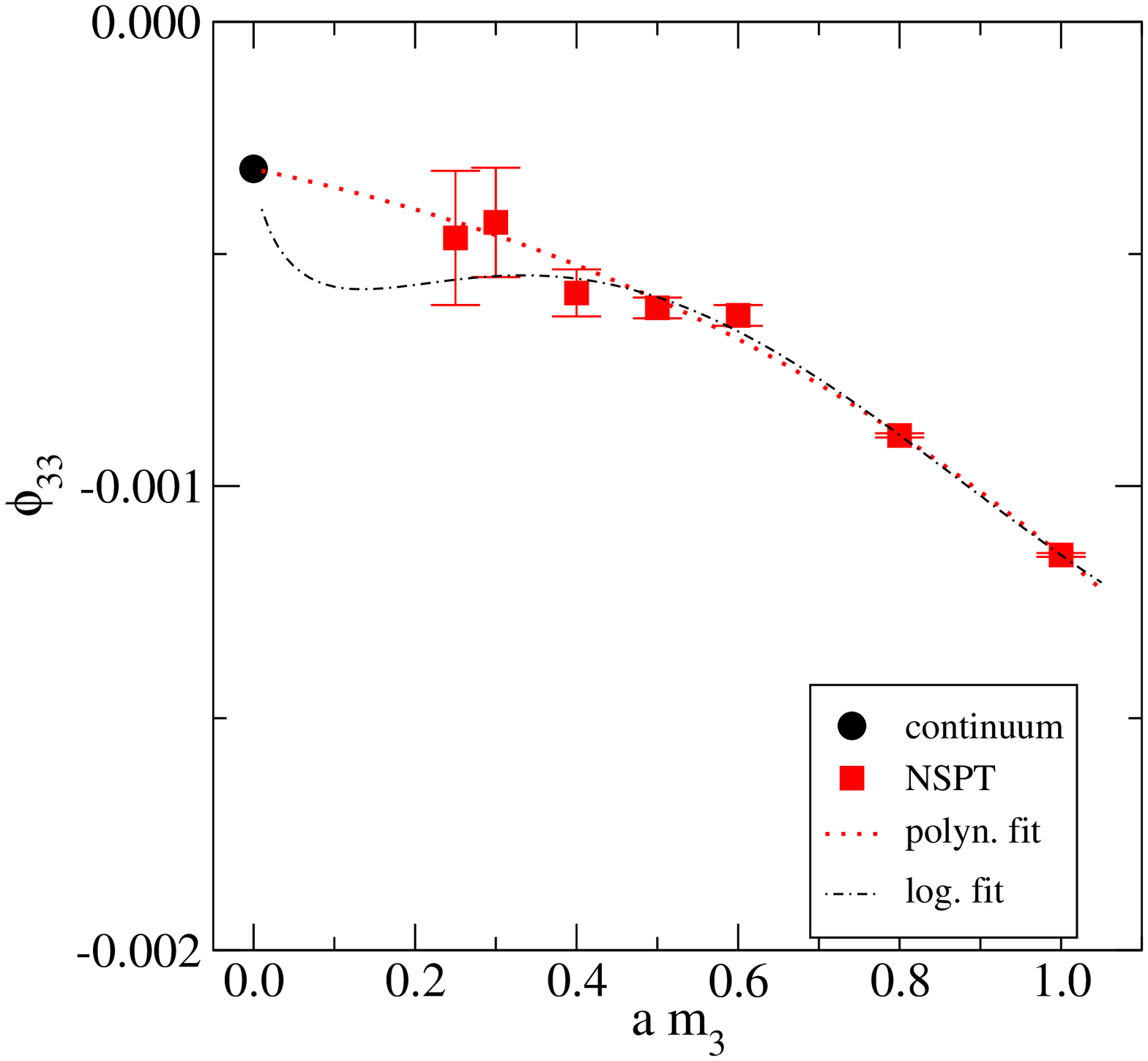}%
~~\epsfysize=5.0cm\epsfbox{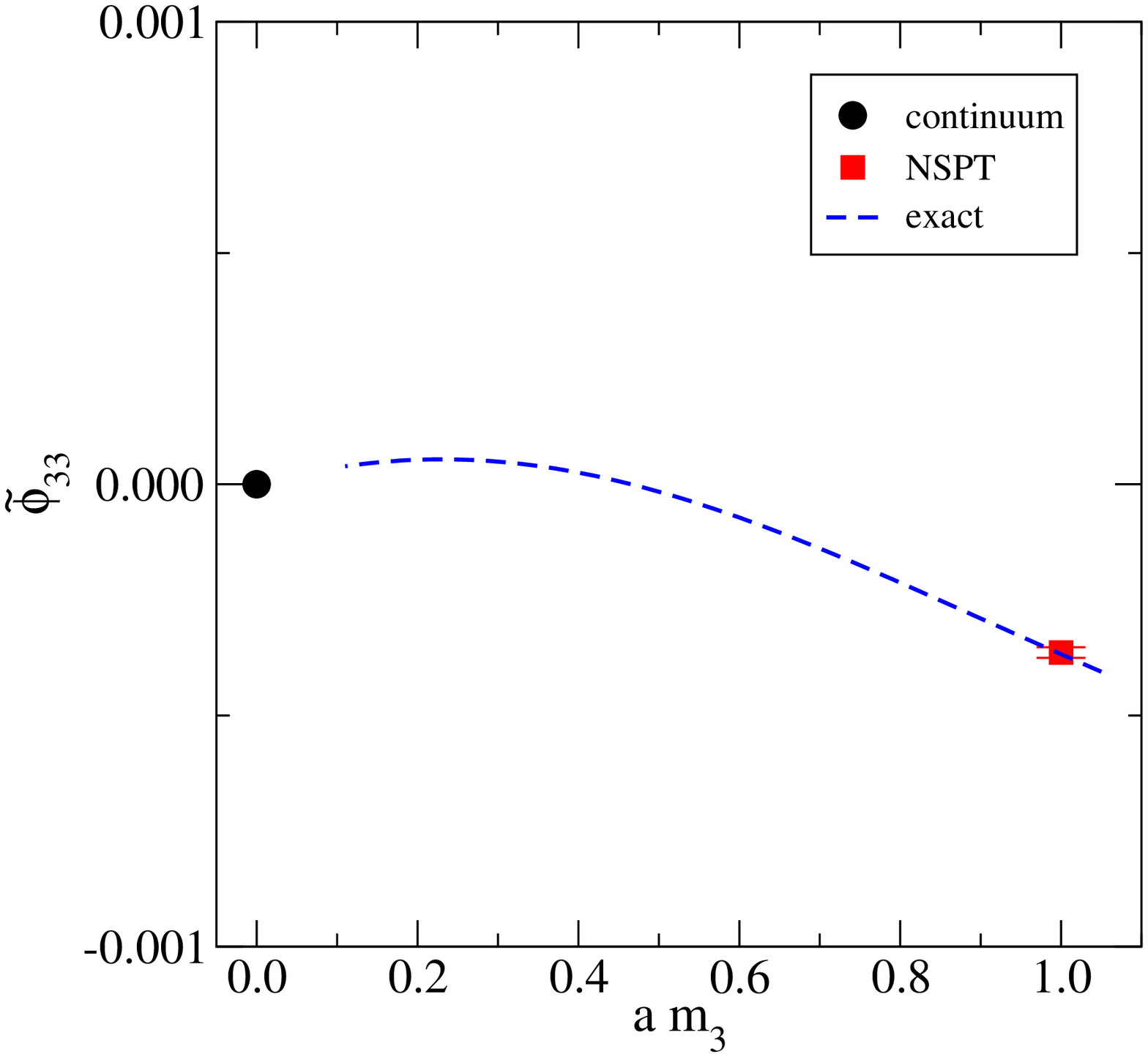}%
}

\vspace*{0.5cm}


\caption[a]{(Continued).
}

\end{figure}

The remaining task is to provide interpolating 
fits for our functions $\phi_{ij}(a m_3)$, 
$\tilde\phi_{ij}(a m_3)$. 
Indeed, lattice simulations 
such as those in ref.~\cite{ahkr} correspond to values
of $a m_3$, given by \eq\nr{amE},  as well as lattice spacings $\ln\beta$, 
which are in the range of our study, 
but seldom coincide exactly with our values.
The purpose of the interpolating fits is to nevertheless
make our results usable for the analysis of lattice simulations. 

In order to carry out the interpolating fits, a fit ansatz is again needed.
Since the continuum limit corresponds to $a m_3\to 0$, it may be
reasonable to use a finite-order polynomial in  $a m_3$ for this 
purpose. However, individual graphs do lead to other structures
as well, particularly logarithms like $a m_3 \ln (1/a m_3)$ 
(cf. \eqs\nr{hatH_exp}, \nr{hatB_exp}).  Even though these
logarithms cancel in all the analytically known terms, we are
not aware of a proof excluding them in general. In particular, 
the lattice simulations of ref.~\cite{ahkr} strongly suggest
the presence of such a logarithm, affecting the approach
to the continuum limit, and it would then be natural for 
it to appear in the coefficient $\phi_{20}$, which is 
numerically the most important unknown ingredient entering 
the analysis of ref.~\cite{ahkr}. 

%
\begin{table}
\small

\begin{center}
\begin{tabular}{rlllll} 
 coefficients & ~~$a^{(0)}_{ij}$ & ~~$a^{(1)}_{ij}$ & 
               ~~$a^{(2)}_{ij}$ & ~~$a^{(3)}_{ij}$
               & $\chi^2/$d.o.f. \\[2mm] \hline \\[0mm]
 $^*\phi_{00}^r$~~~ &
  -0.039789 &  -0.0062086 &   0.0056620 &  -0.00076693 & --- \\
 $^*\phi_{10}^r$~~~ &
   0.038310 &  -0.014881  &  -0.00082279 &   0.0029967 & --- \\
 $^*\phi_{11}$~~~ & 
   0.015831 &   0.0076464 &  -0.0094444 &   0.0017264 & --- \\
 $\phi_{20}$~~~ &
   0.0092578 &   0.027798 &  -0.039072 &   0.014739 &   55.07 \\
 $^*\phi_{21}$~~~ &
   0.010909 &   0.014947 &  -0.0022681 &  -0.0043976 & --- \\
 $^*\phi_{22}$~~~ &
  -0.0053827 &  -0.0050144 &   0.0074469 &  -0.0014995 & --- \\
 $\phi_{30}$~~~ &
   0.0024038 &   0.0071490 &  -0.0020206 &  -0.0016577 &   19.83 \\
 $\phi_{31}$~~~ &
   0.0013885 &  -0.0035354 &   0.024195 &  -0.013013 & 0.996 \\
 $\phi_{32}$~~~ &
  -0.0018767 &  -0.0049224 &    0.0013548 &   0.0029473 & 1.966 \\
 $\phi_{33}$~~~ &
  -0.00031675 &  -0.00036476 &  -0.00032547 &  -0.00014179 & 1.063 \\
 $^*\tilde\phi_{21}$~~~ &
   0.0075590 &   0.0023970 &  -0.0033645 &   0.00069377 & --- \\
 $^*\tilde\phi_{22}$~~~ &
  -0.0025197 &  -0.00079883 &   0.0011212 &  -0.00023111 & --- \\
 $^*\tilde\phi_{31}$~~~ &
   0.0036091 &   0.0031005 &   -0.00069559 &  -0.00088573 & --- \\
 $^*\tilde\phi_{32}$~~~ &
  -0.0012030 &  -0.0026312 &   0.0043532 &  -0.0011224 & --- \\
 $^*\tilde\phi_{33}$~~~ &
   0.0       &   0.00053248 &  -0.0013736 &   0.00047246 & --- \\[2mm] \hline
\end{tabular}
\end{center}

\caption[a]{The fit coefficients allowing to estimate the functions
$\phi_{ij}$, $\tilde\phi_{ij}$ in the range $0.0 \le a m_3 \le 1.0$, 
according to \eq\nr{polyn_fit}.
In the cases marked with a star, the fits have been carried out to 
the exact results rather than to NSPT data (we cite no
$\chi^2/$d.o.f.\ here because no error bars can be assigned
to the exact numbers); the accuracy of their 
description through a polynomial fit is on the 
per cent level. 
}

\la{table:a_ij}
\end{table}
%

%
\begin{table}
\small

\begin{center}
\begin{tabular}{rllllll} 
 coefficients & ~~$b^{(0)}_{ij}$ & ~~$b^{(1')}_{ij}$ & ~~$b^{(1)}_{ij}$ & 
               ~~$b^{(2)}_{ij}$ & ~~$b^{(3)}_{ij}$ 
               & $\chi^2/$d.o.f. \\[2mm] \hline \\[0mm]
 $\phi_{20}$~~~ &
   0.0092578 &   0.0097310 &   0.0090404 &  -0.010317 &   0.0047392 & 
   0.128
 \\
 $\phi_{30}$~~~ &
   0.0024038 &  -0.0036880 &   0.014462 &  -0.013444 &   0.0024569 &  
  1.185
 \\
 $\phi_{31}$~~~ &
   0.0013885 &  -0.0017147 &  -0.00022591 &   0.019122 &  -0.011248 &
 0.932
 \\
 $\phi_{32}$~~~ &
  -0.0018767 &   0.0058403 &  -0.016324 &   0.018943 &  -0.0032415 &
 0.361
 \\
 $\phi_{33}$~~~ &
  -0.00031675 &   -0.0029965 &   0.0052048 &  -0.0086842 &  0.0026476 & 
 0.737
 \\[2mm] \hline
\end{tabular}
\end{center}

\caption[a]{The fit coefficients allowing to estimate the functions
$\phi_{20}$, 
$\phi_{30}$,
$\phi_{31}$, 
$\phi_{32}$, 
$\phi_{33}$ 
in the range $0.0 \le a m_3 \le 1.0$, 
according to \eq\nr{log_fit}.
}

\la{table:b_ij}
\end{table}
%

Given these considerations, we have carried out 
fits of two types. Defining
\ba
 \phi_{00} & = & \frac{\Sigma}{8\pi a m_3} + \phi_{00}^r
 \;, \\ 
 \phi_{10} & = & \frac{\CA}{(4\pi)^2} \ln\frac{1}{a m_3} + \phi_{10}^r
 \;, \\ 
 \phi_{ij} & = & \phi_{ij}^r
 \;, \quad (ij) \neq (00), \; (10) 
 \;,
\ea
we have considered a polynomial fit, 
\be
 \phi_{ij}^r(a m_3)  =  
  a_{ij}^{(0)} + 
  a_{ij}^{(1)} \times a m_3   + 
  a_{ij}^{(2)} \times (a m_3)^2  + 
  a_{ij}^{(3)} \times (a m_3)^3 
 \;,  \la{polyn_fit}
\ee
and a logarithmic one, 
\be
 \phi_{ij}^r(a m_3) = 
  b_{ij}^{(0)} + 
  b_{ij}^{(1')} \times a m_3 \ln\frac{1}{a m_3}  + 
  b_{ij}^{(1)}  \times a m_3   + 
  b_{ij}^{(2)}  \times (a m_3)^2   + 
  b_{ij}^{(3)}  \times (a m_3)^3  
 \;.  \la{log_fit}
\ee
The fits have been constrained to have the correct continuum 
values $a_{ij}(0)$, $b_{ij}(0)$ which are known in all 
cases (cf. \se\ref{se:setup}). The fit functions are
illustrated in \fig\ref{fig:phiij}, and the results for 
the coefficients are given in Tables~\ref{table:a_ij}, \ref{table:b_ij}.

Two main observations can be made from the fits:
\begin{itemize}
\item
In all cases where exact results are available, the logarithmic 
fits agree reasonably well with the polynomial ones. 
This is expected in the sense that we know
that no logarithms exist in the exactly known functions. 

\item
For the most important unknown function, $\phi_{20}$, the logarithmic
fit {\em does} appear to produce a markedly better description of our
data than a polynomial fit; $\chi^2/${d.o.f.} decreases
dramatically, from $\sim 55$ to $\sim 0.13$. 
(This is the case also for the second-most
important unknown, $\phi_{30}$, where $\chi^2/${d.o.f.} decreases
from $\sim 20$ to $\sim 1.2$.) 
This observation would appear to be in 
accordance with the indications from lattice simulations~\cite{ahkr}.
In fact, the authors in ref.~\cite{ahkr} estimated 
the logarithmic term to be 
$
 \left\langle 
   \tr \bigl[ 
          {A_0^2}/{g_3^2}
       \bigr]
  \right\rangle_a \simeq ... + 
  (0.10 \,...\, 0.13) \times g_3^2 a \ln (1/ a)
$, 
which in our units converts to 
$
 b_{20}^{(1')} \simeq (0.10 \,...\, 0.13) / d_A 
 = 0.012 \,...\, 0.017
$.
Though the agreement with our value,
$
 b_{20}^{(1')} \approx 0.00973
$,
is not perfect, the order of magnitude is the same. 
It would be interesting to re-analyze the results of 
ref.~\cite{ahkr} with the coefficient of the logarithm
fixed to our value. 

\end{itemize} 
In summary, then, there appear to be good reasons to expect the 
presence of logarithms in $\phi_{20}$ and $\phi_{30}$. To unambiguously
confirm this expectation, it would obviously be very 
interesting to find a way to 
improve on the accuracy of the determination of
these coefficients, controlling in particular the 
difficult-to-estimate systematic errors that are 
related to the $\tau\to 0$ and $N\to\infty$ extrapolations
in the present method.
It would also be important to improve on the values of 
the coefficients associated with scalar self-couplings, 
$\phi_{31}$,
$\phi_{32}$,
$\phi_{33}$,
although this would mostly
serve as a theoretical consistency check, given that the values
of $\lambda_3/g_3^2$ corresponding to physical finite-temperature QCD 
are very small~\cite{adjoint}.

%
\section{Conclusions}
\la{se:concl}

The purpose of this paper has been to estimate the (Debye) mass
dependent part of the vacuum energy density of the three-dimensional
SU(3) + adjoint Higgs theory, up to 4-loop order in lattice perturbation
theory. The result can be parametrized in terms of 
the coefficients $\phi_{ij}$, $\tilde\phi_{ij}$, defined in \eq\nr{phi}. 
We have worked out the expressions for a number of these coefficients
analytically (\eqs\nr{def:f00}--\nr{def:tf33}), and estimated 
the remaining ones, 
$\phi_{20}$,
$\phi_{30}$,
$\phi_{31}$,
$\phi_{32}$,
$\phi_{33}$, 
for which only the continuum values are known analytically, 
with the help of Numerical Stochastic Perturbation Theory. The results
are illustrated in \fig\ref{fig:phiij}, and parametrized in terms of 
simple fits in \eqs\nr{polyn_fit}, \nr{log_fit}.

The main practical use of our results is that 
when combined with lattice Monte Carlo data, they 
should allow to improve  on the analysis 
of the sum (beyond the known 4-loop order) of infrared sensitive
contributions to the pressure~\cite{a0cond} and quark number
susceptibilities~\cite{ahkr} of hot QCD, 
given that discretization errors up to the 4-loop order 
can now be subtracted. However, our results might also have
some theoretical interest beyond these particular applications. 
For instance, they serve as a consistency check of the 
4-loop $\msbar$-scheme computation of ref.~\cite{aminusb}
(in the sense that our results appear to be consistent
with the continuum values indicated in \fig\ref{fig:phiij}), 
as well as of the super-renormalizability of the theory considered
and the power-counting arguments presented for it in ref.~\cite{framework}
(in the sense that no indications of ultraviolet divergences apart
from the known 1-loop and 2-loop ones in $\phi_{00}$, 
$\phi_{10}$ were seen). 

Concerning the new coefficients
$\phi_{20}$,
$\phi_{30}$,
$\phi_{31}$,
$\phi_{32}$ and
$\phi_{33}$, 
we unfortunately have to acknowledge that it appears difficult 
to improve significantly on the accuracy of our results with 
the present techniques. The problem is that two different
extrapolations, $\tau\to 0$ and $N\to\infty$, are needed in 
order to obtain the values at any fixed $a m_3$, and both of these
extrapolations introduce systematic and statistical errors.  
Therefore there is a need to crosscheck our results, 
and improve upon them, with standard techniques~\cite{HP}. 
Nevertheless, even in that approach, our study should 
serve as a basic framework. In particular, we would 
like to stress the insight that it is valuable to determine the 
coefficients $\phi_{ij}$, $\tilde\phi_{ij}$ as functions of $a m_3$, 
rather than to carry out an expansion in small $a m_3$, since 
realistic values of $a m_3$ ($\sim 0.1 \ldots 0.5$~\cite{ahkr}) are
in a region where the functions show more structure than 
just linear terms (cf.\ \fig\ref{fig:phiij}). The most 
important coefficients to determine are $\phi_{20}$ and 
$\phi_{30}$, which are independent of the scalar self-coupling, 
and for $a m_3 \ll 1$, 
a concrete challenge is to confirm or disprove the existence of the 
logarithmic term $\sim \rmO(a\ln(1/a))$ in $\phi_{20}$, 
for which independent indications
have been seen in ref.~\cite{ahkr} and in the present study.

%
\section*{Acknowledgments}
We thank V. Miccio for collaboration at preliminary stages 
of this work, and A.~Hietanen and K.~Rummukainen for useful discussions. 
The Parma group acknowledges support 
from I.N.F.N.\ under contract {\em i.s.~MI11}. 
We warmly thank {\em ECT*, Trento,} 
for providing computing time on the {\em BEN} system. 
The total amount of computing time used for this 
project corresponds to about 
$4 \times 10^{18}$
flop.
%


\appendix
\renewcommand{\thesection}{Appendix~\Alph{section}}
\renewcommand{\thesubsection}{\Alph{section}.\arabic{subsection}}
\renewcommand{\theequation}{\Alph{section}.\arabic{equation}}


%
\section{Graph-by-graph results for the vacuum energy density}
\la{app:graphs}

We list here results  
for the mass-dependent part of the 
vacuum energy density $\f_a$ of the theory 
in \eq\nr{SAH} (through \eq\nr{latt_cond} this produces the 
condensate that we are interested in).
At the 3-loop level we have only kept terms involving 
at least one $\lambda_3$. 
In the graphical notation to be used below, 
solid/wiggly lines represent tree-level $A_0$/$A_i$
lattice propagators, respectively. 

\paragraph{1-loop and 2-loop graphs.}

For brevity, we denote in the following $m\equiv m_3(\bmu)$.
The integrals appearing are defined in appendix~B, and 
in terms of these, the results read ($d=3$):
\ba
 \TopoVR(\Asc) &=& \frac{d_A}{2} J_a(m^2) 
 \;, \\
 \ToptVE(\Asc,\Asc) &=& \frac{\lambda_3}{4} d_A (d_A +2)[I_a(m^2)]^2
 \;, \\
 \pic{\Photon(0,15)(30,15){1.5}{6} \CArc(15,15)(15,0,360)}
 +
 \picc{\CArc(15,15)(15,0,360)%
 \PhotonArc(45,15)(15,0,360){1.5}{16}}
 &=&
 \frac{g_3^2}{4} d_A C_A \Bigl\{
 (2d-4) I_a(0) I_a(m^2) + [I_a(m^2)]^2 + 4 m^2 H_a(m^2) 
 + \nn &&{}
 +a^2\Bigl[ m^2 I_a(0) I_a(m^2) - I_a(0) / a^d + G_a(m^2) \Bigl] \Bigr\}
 \;. 
\ea

\paragraph{3-loop graphs involving $\lambda_3$.}

\ba
 \piccc{\CArc(15,15)(15,0,360) \CArc(45,15)(15,0,360) 
 \CArc(75,15)(15,0,360)}
 &=&
 \frac{\lambda_3^2}{4} d_A (d_A+2)^2 [I_a(m^2)]^2 I_a'(m^2) 
 \;, \\
 \ToprVB(\Asc,\Asc,\Asc,\Asc)
 &=&
 -\frac{\lambda_3^2}{4} d_A (d_A+2) B_a(m^2)
 \;, \\
 \picc{\CArc(15,15)(15,0,360) \CArc(45,15)(15,0,360)
 \Photon(15,0)(15,30){1.5}{5}}
 +
 \piccc{\CArc(15,15)(15,0,360) \CArc(45,15)(15,0,360)
 \PhotonArc(75,15)(15,0,360){1.5}{16}}
 &=& 
 \frac{g_3^2\lambda_3}{4} d_A C_A (d_A+2) I_a(m^2) \Bigl\{
 (2d-4) I_a(0) I_a'(m^2) 
 + \hspace*{1cm} \\ &&{}
 + 2 I_a(m^2) I_a'(m^2) + 4 H_a(m^2) 
 + 4 m^2 H_a'(m^2) 
 + \nn &&{}
 + a^2 \Bigl[ 
 I_a(0) I_a(m^2) + m^2 I_a(0) I_a'(m^2)
 + G_a'(m^2) \Bigr] \Bigr\}
 \;, \nn 
 \ToprVV(\Asc,\Asc,\Lgl,\Lsc,\Lsc)
 &=& 0
 \;. 
\ea

\paragraph{Mass counterterm contributions up to 3-loop order.}

Gauge field ``mass counterterms'' (arising from the Haar integration
measure) are not displayed, as they do not contribute to terms involving
at least one $\lambda_3$.
\ba
 \counterA &=& \frac{d_A}{2}  \delta m^2_a I_a(m^2)
 \;, \\
 \counterB &=& \frac{d_A}{4} (\delta m^2_a)^2 I_a'(m^2)
 \;, \\
 \counterC &=& \frac{\lambda_3}{2} d_A (d_A+2) \delta m^2_a I_a(m^2)
               I_a'(m^2)
 \;, \\
 \counterD + \counterE &=& 
 \frac{g_3^2}{4} d_A C_A \delta m^2_a \Bigl\{ 
 (2d-4) I_a(0) I_a'(m^2) 
 + \\ &&{}
 + 2 I_a(m^2) I_a'(m^2) + 4 H_a(m^2) 
 + 4 m^2 H_a'(m^2) 
 + \nn &&{}
 + a^2 \Bigl[ I_a(0) I_a(m^2) + m^2 I_a(0) I_a'(m^2)
 + G_a'(m^2) \Bigr] \Bigr\}
 \;. \nonumber 
\ea

%
\section{Basic lattice integrals}
\la{app:lattints}

We detail here the definitions
of the basic lattice integrals that appear in 
the expressions discussed in \se\ref{se:setup}.
The integration measure is 
\be
 \int\!{\rm d} p \equiv \int_{-\pi}^{\pi} \frac{{\rm d}^3 p}{(2\pi)^3}
 \;,
\ee
and we define the standard lattice momenta as 
\be
 \tilde p_i \equiv {2} \sin \frac{p_i}{2} \;, \quad
 \tilde p^2 \equiv \sum_{i=1}^3 \tilde p_i^2 
 \;. 
\ee
The integrals appearing then read (we use hats as a reminder of 
the use of lattice units in these expressions):
\ba
 \hat J_a(\amE^2) & \equiv & \int\! {\rm d}p \, \ln(\tilde p^2 + \amE^2 )  
 \;, \la{eq:Ja} \\
 \hat I_a(\amE^2) & \equiv & \int\! {\rm d}p \, \frac{1}{\tilde p^2 + \amE^2} 
 \;, \la{eq:Ia} \\
 \hat H_a(\amE^2) 
                       & \equiv  & \int\! {\rm d}p \, {\rm d}q \, 
                        \frac{1}{(\tilde p^2 + \amE^2)(\tilde q^2 + \amE^2)
                        \widetilde{(p+q)}^2} \la{eq:Ha}  
 \;,  \\
 \hat G_a(\amE^2) & \equiv  & \int\! {\rm d}p \, {\rm d}q \, 
                        \frac{\sum_i \tilde p_i^2 \tilde q_i^2}
                        {(\tilde p^2 + \amE^2)(\tilde q^2 + \amE^2)
                        \widetilde{(p+q)}^2}  
 \;. \la{eq:Ga}
\ea
Small-$\amE$ expansions for these functions have been worked
out in refs.~\cite{framework,contlatt} and are given, to the order
that was used for the small-$\amE$ expansions in \se\ref{se:setup}, by
\ba
  \hat J_a'(\amE^2) & = & 
  \hat I_a(\amE^2) \;, \\
  \hat I_a(\amE^2) & = & \frac{1}{4\pi} 
  \Bigl[ \Sigma - \amE + \rmO(\amE^2) \Bigr]
  \;, \la{eq:Iaexp} \\
  \hat H_a(\amE^2) & = & 
  \frac{1}{(4\pi)^2} \biggl[ 
    \ln\frac{3}{\amE} + \fr12 + \zeta + \rmO(\amE)
  \biggr]
 \;, \la{hatH_exp} \\
  \hat G_a(\amE^2) & = & 
  \frac{1}{(4\pi)^2}
  \Bigl[ 16 \kappa_1 - 4 \delta \, \amE^2 + \rmO(\amE^3) \Bigr]
 \;, 
\ea
where $\Sigma$,  $\zeta$, $\kappa_1$ and $\delta$ are numerical
coefficients mentioned below \eq\nr{dmassL}. 
Furthermore, we define
a 3-loop ``basketball'' integral through
\ba
 \hat B_a(\amE^2) 
                       & \equiv  & \int\! {\rm d}p \, {\rm d}q \, {\rm d}r \, 
                        \frac{1}{(\tilde p^2 + \amE^2)(\tilde q^2 + \amE^2)
                        (\tilde r^2 + \amE^2)[\widetilde{(p+q+r)}^2 + \amE^2]} 
 \;. \hspace*{0.5cm} \la{eq:Ba}  
\ea
In this case the small-$\amE$ expansion reads 
\be
 \hat B_a(\amE^2) = \frac{1}{(4\pi)^3}
 \biggl[
 \Sigma\times\theta - \amE \biggl(  
  4 \ln \frac{3}{2 \amE} + 4 \zeta + 6 
 \biggr) + \rmO(\amE^2)
 \biggr]
 \;. \la{hatB_exp}
\ee
The derivation of this result, which has to our knowledge
not appeared in the literature before, as well as a numerical estimate
for the new coefficient $\theta$, are given in appendix~C.

The expressions in appendix~A employ the functions
\ba 
 J_a(m^2) & \equiv & \frac{1}{a^3} \hat J_a (\amE^2)
 \;, \\ 
 I_a(m^2) & \equiv & \frac{1}{a} \hat I_a(\amE^2)
 \;, \\
 H_a(m^2) & \equiv & \hat H_a(\amE^2)
 \;, \\
 G_a(m^2) & \equiv & \frac{1}{a^4} \hat G_a(\amE^2)
 \;, \\
 B_a(m^2) & \equiv & \frac{1}{a} \hat B_a(\amE^2)
 \;,  \la{def:Ba}
\ea
where $a$ is the lattice spacing, and $\amE \equiv a m$.

In a finite volume, $V = (a N)^3$,  
the momentum integrations get replaced with
\be
 \int \! {\rm d}p\, f(\vec{p}) 
 \to 
 \frac{1}{N^3} 
 \sum_{n_1 = 0}^{N-1} 
 \sum_{n_2 = 0}^{N-1} 
 \sum_{n_3 = 0}^{N-1} 
 f\Bigl( \frac{2\pi \vec{n}}{N} \Bigr)
 \;, 
\ee
where $\vec{n} \equiv (n_1,n_2,n_3)$. In the case of massless
propagators, the zero-mode is left out.

%
\section{The 3-loop basketball in lattice regularization}

In order to work out the expansion in \eq\nr{hatB_exp}, 
we find it convenient to return from lattice units to physical 
units, considering then the function in \eq\nr{def:Ba}. 
Let us introduce the scalar propagator on the lattice, 
\be
 \g_a(x;m) \equiv \int_{-\pi/a}^{\pi/a} \! \frac{{\rm d}^3 p}{(2\pi)^3} \,
 \frac{ e^{i p\cdot x} }{\tilde p^2 + m^2}
 \;,
 \la{Ga}
\ee
where now 
$
  \tilde p^2 \equiv \sum_{i=1}^3 \tilde p_i^2 
$, 
$
  \tilde p_i \equiv \fr{2}{a} \sin \fr{a p_i}{2}
$.
For $x\neq 0$, the propagator remains finite in the continuum limit 
$a\to 0$, 
\be 
 \g_0(x;m) = \frac{\exp(- m |x|)}{4\pi |x|}
 \;, 
 \la{G0}
\ee
while for $x = 0$, it contains a linear divergence as shown 
in~\eq\nr{eq:Iaexp}, 
\be 
 \g_a(0;m) = \frac{1}{4 \pi a} \biggl[ 
 \Sigma - \amE + \rmO(\amE)^2
 \biggr]
 \;. \la{Sigma}
\ee

The integral we will be concerned with here is of the
``basketball'' type,
\be
 {B}^{(n)}_a(\{ m_i \}) \equiv
 \sum_x a^3 \, \prod_{i = 1}^n \g_a(x;m_i)
 \;. \la{bbn}
\ee
For $n=3$, this equals the integral $H_a$ 
defined in~\eq\nr{eq:Ha}, and for general masses,
the expansion close to the continuum 
limit is of the form~\cite{framework}
\be
 {B}^{(3)}_a(\{ m_i \}) = 
 \frac{1}{(4\pi)^2}
 \biggl[ 
 \ln\frac{6}{ a\sum_i m_i}  + \fr12 + \zeta + \rmO(a m_i) 
 \biggr]
 \;. \la{bb3}
\ee
The challenge now is to find 
the corresponding expansion for $n=4$.

The basic approach we follow is similar to the one used for the 
basketball integral in dimensional regularization in refs.~\cite{bn_phi4}.
The summation over configuration space in~\eq\nr{bbn} is divided
into two regions, $|x| \le r$ and $|x| > r$. We assume $a m_i \ll 1$, 
and can thus choose
\be 
 a \ll r \ll \frac{1}{m_i}
 \;. 
\ee
In the region $|x| \le r$, we now have $|x| m_i \ll 1$, and can 
expand in the masses; in the region $|x| > r$, we have $ a/|x| \ll 1$, 
and can use the continuum approximation for the propagators. 

\paragraph{The region $|x| > r$.}
The region of large $|x|$ gives a contribution which remains finite
in the limit $a\to 0$. It can thus be evaluated employing \eq\nr{G0}.
The integral is elementary, and we obtain
\ba
 \lim_{a\to 0} \Delta_{|x| > r} {B}_a^{(4)}(\{ m_i \})
 & = &  
 \int_r^\infty 4\pi |x|^2 {\rm d}|x|
 \,
 \prod_{i=1}^4 \g_0 (x;m_i)
 \nn & = &  
 \frac{1}{(4\pi)^3}
 \biggl[
 \frac{1}{r}  + M 
 \Bigl( 
 \ln M r + \gamma_\rmii{E} - 1
 \Bigr)
 \biggr] + \rmO(M^2 r) 
 \;,
 \la{larger}
\ea
where $M \equiv m_1 + m_2 + m_3 + m_4$.
Note in passing, for future reference, that
\be
 \int_r^\infty 4\pi |x|^2 {\rm d}|x|
 \,
 \prod_{i=1}^3 \g_0 (x;m_i)
 = \frac{1}{(4\pi)^2}
 \biggl(
 - \ln \tilde M r - \gamma_\rmii{E} 
 \biggr) + \rmO(\tilde M r)
 \;, \la{xtra}
\ee
where $\tilde M \equiv m_1 + m_2 + m_3$.

\paragraph{The region $|x| \le r$.}
In the region of small $|x|$, we rewrite the propagator of 
\eq\nr{Ga} in the equivalent form 
\be 
 \g_a(x;m) = \int_{-\pi/a}^{\pi/a} \! \frac{{\rm d}^3 p}{(2\pi)^3}  
 \biggl[ 
  \frac{e^{i p\cdot x}}{\tilde p^2}
  + \frac{1}{\tilde p^2 + m^2}
  - \frac{1}{\tilde p^2}
  -m^2 \frac{e^{i p\cdot x} - 1}{\tilde p^2 (\tilde p^2 + m^2)}
 \biggr]
 \;.
 \la{Ga1}
\ee
The point of this rewriting is that for $a \to 0$ and $x\neq 0$, the first
term behaves as $\sim 1/ 4\pi |x|$, the next two combine 
into a constant $\sim - m/ 4 \pi$, while the last term, which is both 
ultraviolet and infrared finite, behaves 
as $\rmO( m^2 |x|)$. Therefore, if we consider 
the limit $a\to 0$, $r m_i \ll 1$, 
the last term does not contribute in
$\Delta_{|x| \le r} {B}_a^{(4)}(\{ m_i \})$: 
\be
 \int_0^r \! |x|^2 {\rm d}|x| \frac{1}{(4\pi |x|)^3} \rmO(m_i^2 |x|)
 \sim \rmO(m_i^2 r) 
 \;.
\ee
The same is true for the case that there are two or more appearances
of the constant term: 
\be
 \int_0^r \! |x|^2 {\rm d}|x| \frac{1}{(4\pi |x|)^2} \rmO(m_i) \rmO(m_j)
 \sim \rmO(m_i m_j r) 
 \;.
\ee
Thus the only contributions come from four appearances of the 
first term in~\eq\nr{Ga1}, and three appearances of the first term as well as
one appearance of the middle terms: 
\ba 
 \lim_{a\to 0}  \Delta_{|x| \le r} {B}_a^{(4)}(\{ m_i \})
  & = &  \lim_{a\to 0} \sum_{|x|\le r} a^3 \, [\g_a(x;0)]^4 
 \la{smallr}
 \\ & + & 
 \lim_{a\to 0} \biggl\{ \sum_{i=1}^4 \Bigl[\g_a(0;m_i) - \g_a(0,0)\Bigr] 
 \sum_{|x| \le r } a^3 \, [\g_a(x;0)]^3 \biggr\} 
 + \rmO(m^2_i r)
 \;. \nonumber
\ea
Here and in the following, 
$
 \lim_{a\to 0}
$
is meant in a symbolic sense, 
since the sums actually diverge as $a\to 0$.

Given that $\g_a(0;m_i) - \g_a(0,0)$ is known (cf.\ \eq\nr{Sigma}), 
we are left with the evaluation of the sums on the first
and second rows in~\eq\nr{smallr}.
Since the propagators are massless, the outcomes only depend on 
the ratio $r/a$, where $a \ll r$. The first of the sums can be performed
by extending the sum to be over all space, and taking the continuum 
limit in the resulting subtraction, which is ultraviolet finite: 
\ba
 \lim_{a\to 0} \sum_{|x|\le r} a^3 \, [\g_a(x;0)]^4 & = &  
 \lim_{a\to 0} \sum_{x} a^3 \, [\g_a(x;0)]^4 - 
 \lim_{a\to 0} \sum_{|x| > r} a^3 \, [\g_a(x;0)]^4 
 \nn & = & 
 \lim_{a\to 0} \sum_{x} a^3 \, 
 [\g_a(x;0)]^4 - \frac{1}{(4\pi)^3} \frac{1}{r} 
 \;, 
 \la{sr1}
\ea
where we used \eq\nr{larger}.
The latter sum is slightly more difficult because it would be
infrared divergent at large distances, but it can be performed
as above, once we regulate  the infrared with 
mass terms, and use then~\eqs\nr{bb3}, \nr{xtra}:
\ba
 \lim_{a\to 0} \sum_{|x|\le r} a^3 \, [\g_a(x;0)]^3 & = &  
 \lim_{\tilde M\to 0} \biggl\{ \lim_{a\to 0} \sum_{x} a^3 \, 
 \prod_{i=1}^3 \g_a(x;m_i) - 
 \lim_{a\to 0} \sum_{|x| > r} a^3 \, \prod_{i=1}^3 \g_a(x;m_i) \biggr\} 
 \nn & = & 
 \lim_{a\to 0}
 \biggl\{
 \frac{1}{(4\pi)^2}\biggl( 
 \ln\frac{6}{a\tilde M} + \fr12 + \zeta + \ln \tilde Mr + \gamma_\rmii{E} 
 \biggr)
 \biggr\}
 \;. \la{sr2}
\ea
The infrared regulator $\tilde M$ is seen to 
cancel in \eq\nr{sr2}, as it should. 

Combining now 
\eqs\nr{Sigma}, \nr{larger}, 
\nr{smallr}, \nr{sr1} and \nr{sr2}, terms singular in $r$
cancel, and the limit ${m_i r \to 0}$ can be taken. 
The Euler gamma-constants $\gamma_\rmii{E}$ also cancel, and we finally obtain
\be
 {B}^{(4)}_a(\{ m_i \}) = \sum_{x} a^3 \, [\g_a(x;0)]^4 
 - \frac{M}{(4\pi)^3}
 \biggl(
 \ln\frac{6}{aM} + \fr32 + \zeta 
 \biggr)
 + \rmO(a m_i^2)
 \;.
 \la{bb4_res}
\ee

The task remains to evaluate the sum in~\eq\nr{bb4_res}. Employing 
the techniques introduced in ref.~\cite{lw} and worked out for
the three-dimensional case in ref.~\cite{ns}, we find that
\ba
 \sum_{x} a^3 \, [\g_a(x;0)]^4  & \equiv & 
 \frac{1}{a}\frac{\Sigma}{(4\pi)^3} \theta
 \;, \qquad 
 \theta =  3.0122(1)
 \;, \la{theta}
\ea
where the number in parentheses indicates the uncertainty of the last digit.
Combining \eqs\nr{bb4_res}, \nr{theta} leads to \eq\nr{hatB_exp}.

Finally, recall for completeness 
that in dimensional regularization (DR) at $d=3-2\epsilon$, 
the 2-loop~\cite{pert} and 3-loop~\cite{ar_3loop}
basketball integrals read 
($M = \sum_{i=1}^4m_i$, $\tilde M = \sum_{i=1}^3 m_i$)
\ba 
 B^{(3)}_\rmi{DR}(\{m_i\}) & = & \frac{\mu^{-4\epsilon}}{(4\pi)^2}
 \biggl(\frac{1}{4\epsilon} + \ln\frac{\bmu}{\tilde M} + \fr12 + 
 \mathcal{O}(\epsilon)  \biggr)
 \;, \\ 
 B^{(4)}_\rmi{DR}(\{m_i\}) & = & - \frac{\mu^{-6\epsilon}}{(4\pi)^3} M 
 \biggl(
 \frac{1}{4\epsilon} + \fr32\ln\frac{\bmu}{M} + 2 + 
 \fr12 \sum_{i = 1}^4 \frac{m_i}{M}\ln\frac{M}{2 m_i}   + 
 \mathcal{O}(\epsilon) 
 \biggr)
 \;.
\ea
For a complete discussion of basketball integrals in dimensional
regularization, see ref.~\cite{korner}.



\end{document}